\documentclass[useAMS,usenatbib]{mn2e}
\usepackage{longtable}
\usepackage{times}
\usepackage{inputenc}
\usepackage{mathrsfs}
\usepackage{color}
\usepackage{parskip}
\usepackage{graphicx,float}
\usepackage[german, english]{babel}
\usepackage{xspace}
\usepackage[dvips,ps2pdf]{thumbpdf}
\usepackage{textcomp}
\usepackage{url}
\usepackage{bm}
\usepackage{amsmath}
\usepackage{fixltx2e}
\usepackage{subcaption}
\usepackage{multirow}
\usepackage{multicol}
\usepackage{rotating}
\usepackage{pifont}
\usepackage{mathptmx}
\usepackage{caption}
\usepackage{capt-of}

\title{The Chandra Deep Group Survey -- cool core evolution in groups and clusters of galaxies}

\author[A. Pascut and T. J. Ponman]{A. Pascut and T. J. Ponman\\
\textit{School of Physics and Astronomy, The University of Birmingham, Birmingham B15 2TT}}

\begin{document}

\newcommand{\TcoolTage}{$t_{\rm cool}/t_{\rm Uni}$}
\newcommand{\TcoolTageb}{$\mathbf{t}_{\bf \rm cool}/\mathbf{t}_{\bf \rm Uni}$}
\newcommand{\csb}{$c_{\rm SB}$}
\newcommand{\csbb}{$\mathbf{c}_{\bf \rm SB}$}
\newcommand{\fcore}{$F_{\rm core}$}
\newcommand{\fcoreb}{$\mathbf{F}_{\bf \rm core}$}
\newcommand{\fc}{$f_{\rm c}$}
\newcommand{\fcb}{$\mathbf{f}_{\bf \rm c}$}
\newcommand{\tcool}{$t_{\rm cool}$}
\newcommand{\tcoolb}{$\mathbf{t}_{\bf \rm cool}$}
\newcommand{\R}{\ensuremath {\mathrm {R_{500}}}}
\newcommand\ltsim{\ifmmode\stackrel{<}{_{\sim}}\else$\stackrel{<}{_{\sim}}$\fi}
\newcommand{\tick}{\ding{51}} 
\newcommand{\cross}{\ding{53}}
\maketitle

\begin{abstract}
We report the results of a study which assembles deep observations
with the ACIS-I instrument on the {\it Chandra Observatory} to study the
evolution in the core properties of a sample of galaxy groups and 
clusters out to redshifts $z\approx 1.3$. A search for extended objects within
these fields yields a total of 62 systems for which redshifts are available,
and we added a further 24 non-X-ray-selected clusters, to 
investigate the impact of selection effects and improve our statistics
at high redshift. Six different estimators of cool core strength are applied to these data: the entropy ($K$) and cooling time (\tcool) 
within the cluster core, the cooling time as a fraction of the age of the 
Universe (\TcoolTage), and three estimators based
on the cuspiness of the X-ray surface brightness profile. 
A variety of statistical tests are used to quantify evolutionary trends
in these cool core indicators. In agreement with some previous
studies, we find that there is significant evolution in \TcoolTage,
but little evolution in \tcool, suggesting that gas is accumulating
within the core, but that the cooling time deep in the core
is controlled by AGN feedback. We show that this result
extends down to the group regime and appears to be robust
against a variety of selection biases (detection bias, archival biases
and biases due to the presence of central X-ray AGN) which we consider.\\
\end{abstract}


\section{Introduction}

The hot ionized gas in clusters of galaxies, also known as
intra-cluster medium (ICM), loses its thermal energy through X-ray
radiation. The time scale on which an isothermal parcel of gas with
uniform density can radiate away its thermal energy is inversely
proportional to its density. As a result, cooling times at the centre of
the clusters, where the density is high, are shorter than in the outer
regions. Observations of low redshift clusters show that
clusters with central cooling time shorter than their age are common in
the local Universe, and they represent $\sim 50\%-90\%$ of the population
\citep{Peres1998, Sanderson2006, Chen2007, Hudson2010, Santos2010}. In
the light of this, clusters have been divided into two classes: cool core
(CC) systems, which have a short central cooling time, a cuspy central
surface brightness and usually manifest a drop in their central
temperature, and non cool core (NCC) clusters, with the opposite
properties.

Evidence for the existence of two distinct cluster populations came from
the observation of bimodality in the distribution of the cooling time
\citep{Cavagnolo2009} or the closely related gas entropy
\citep{Cavagnolo2009,Sanderson2009,Mahdavi2013} in the central regions of
clusters. On the other hand, other studies have found no clear evidence for
bimodality in cluster properties, and some authors,
e.g. \citet{Santos2008}, have split core properties into {\it three}
classes, with an intermediate weak cool core (WCC) class between strong
cool cores (SCC) and NCC clusters. Whether the observed
distribution is representative for the cluster population depends on the
sample used for the study. Biases in sample selection can affect the
observed distribution and lead to misinterpretation of the results. For
example, the study of \cite{Cavagnolo2009}, which is based on an X-ray
selected archival sample, might have a bias against WCC clusters if
observations of strong CCs and/or disturbed clusters (i.  e. generally
NCCs) are preferred over the regular, WCC clusters. 

Different models have been put forward to explain the observed
distribution in core properties in terms of the dynamical and/or
thermal history of clusters. In the model of \citet{Burns2008},
cluster merging is the mechanism which creates NCC clusters by destroying 
the cooling core in CC clusters. The natural state of a cluster is the CC
one since most clusters have central cooling times which are less than
their age. This model agrees with the high fraction of CC at low
redshift and the observed bimodality in the central cooling state. The
simulations of \cite{Burns2008} predict no evolution in the CC
fraction up to a redshift of 1. Moreover they show that the
probability of mergers increases with the system mass and therefore CC
are more common in low mass systems. It is not yet clear whether this
prediction is borne out observationally due to the
substantial variation in CC fraction found by different methods
used for CC/NCC classification, and the lack of statistically selected
samples of galaxy groups. However, there is observational evidence in
favour of this merger-driven model from the fact that most cool core
clusters have a regular surface brightness, whilst many NCC clusters
are disturbed \citep{O'Hara2006,Maughan2012}. Also,
\cite{Rossetti2010} showed that none of the clusters classified as
cool cores in their sample have detected radio relics, which are a
sign of mergers. On the other hand, some simulations \citep{Poole2006}
suggest that CCs cannot be destroyed by mergers. If the main effect
of mergers is to redistribute the core gas, rather than to raise its
entropy, then the core is reassembled quite rapidly, and even the most
massive mergers would only temporarily disrupt it.

Another class of models assumes that the observed thermal
state of the cluster core was established early, as a result of the
entropy level established in the intergalactic gas before
cluster formation \citep{McCarthy2004a}. NCC clusters will then be those for
which the entropy of the intergalactic gas has been raised to a
sufficiently high value that the cluster has not had enough time to radiate
away its thermal energy and develop a cool core. Conversely, CC clusters
experienced a lower level of entropy injection.

Irrespective of the mechanism which generates the
distribution of core properties, there is an observed tendency for cool
core clusters to host a central active galactic nucleus
(AGN)\citep{Dong2010}. Moreover, it has been shown that there is a
correlation between the strength of the cool core and the radio power of
the central AGN \citep{Mittal2009}.  The coexistence of an AGN and CC
plays an important role in the thermal evolution of ICM. AGN, through
their feedback, are thought to represent the main heating
source for the ICM, whilst the cool gas in the cluster core 
constitutes the reservoir for black hole accretion \citep{Croston2005,Rafferty2006,McNamara2007,McNamara2012,Ma2013,Russell2013}.  

One way in which AGN interact with the ICM is 
through relatvistic plasma jets, which can push aside the ICM, creating 
lower density regions detectable in X-ray images of clusters as `cavities' with
reduced surface brightness. Cavities have been detecetd
in clusters at low \citep{Boehringer1993,Fabian2000,McNamara2000,Blanton2011,Gitti2011} 
and high redshift \citep{Hlavacek-Larrondo2012}, while evidence for cavities
in groups is currently limited to low redshift systems
\citep{Morita2006,Gastaldello2009,Randall2009,Gitti2010,OSullivan2011}
due to to groups' lower surface brightness compared to clusters. Based on
the volume and pressure of these cavities, the energy input from the AGN
can be estimated. Studies of cavities in clusters have shown that AGN
can typically
provide the necessary power to balance the energy lost through
cooling in clusters \citep{Birzan2004,Rafferty2006}, whilst in
galaxy groups their impact is even more significant, and they may be able to 
provide more energy than is lost through cooling \citep{OSullivan2011a}.  

These results demonstrate that the contribution of AGN 
to the thermal state of the ICM cannot be ignored, and  
\cite{McCarthy2008} introduced a model which combines
pre-heating at high redshifts and AGN feedback to explain the existence
of CC and NCC systems. More recently, Voit and collaborators \citep{Voit2011b,Voit2014}
have explored the relationship between cooling, thermal conduction, thermal
instability and AGN feedback within cluster cores. They find that many
properties of the gas in cluster cores can be explained in terms of
the balance between these processes. We will return to this below, in the light
of our results.

Studies of the evolution of cool cores face two major problems: the
construction of an unbiased sample with the necessary statistics at high
redshift to be able to draw any conclusion about any evolutionary trends,
and the definition of a parameter that can separate a CC cluster from a NCC
one for a variety of systems at different redshifts and for data with
different quality.

One parameter frequently used to characterize the thermal state of a
cluster core is the central cooling time
\citep{Edge1992,Peres1998,Bauer2005,Mittal2009}, which is directly related
to the physical definition of a cool core as one in which cooling is
significant. Central entropy, which is closely related to cooling time, is
another physical parameter used to characterize CCs
\citep{Cavagnolo2009}. Other cool core estimators have been defined based
on the observed X-ray properties associated with CC clusters, such as the
central temperature drop \citep{Maughan2012} and central surface brightness
excess \citep{Vikhlinin2007,Santos2008,Maughan2012}.

How well do these various parameters perform in separating CC and NCC
systems? \cite{Hudson2010} applied 16 cool core estimators to the HIFLUGCS
(HIghest X-ray FLUx Galaxy Cluster Sample) sample of low redshift clusters
and found that cooling time and entropy are the quantities which show the
most pronounced bimodality in their distribution.

Studies of the evolution of cool cores, using X-ray
selected samples, have shown that CC are common at low redshift
\citep{Peres1998}. \cite{Bauer2005} showed that their fraction in X-ray
luminous clusters does not change strongly up to a redshift of 0.4 when
the central cooling time is used as a CC estimator.  The investigation of
how this fraction changes with redshift has been extended beyond
redshift 0.5, mainly by studies which use CC estimators based on the
surface brightness excess
\citep{Vikhlinin2007,Santos2008,Maughan2012}. These studies found
that the fraction of cool core clusters drops significantly, resulting in
a lack of strong cool cores at high redshift. In contrast, the study of 
\cite{Alshino2010}, which used
a CC estimator based on central surface brightness excess to examine
a sample of groups and clusters from the XMM-LSS survey, confirmed
the lack of strong CCs in {\it clusters} at high redshift, but 
reported an {\it increase} in the strength of cool cores in cooler groups.
Further evidence on the evolution of core properties
comes from optical studies, since CC clusters have associated
H$_{\alpha}$ \citep{Bauer2005} and other optical line
emission. \cite{Samuele2011} studied a sample of 77 clusters up to a
redshift of 0.7 and found a lack of cool core clusters at redshifts
greater than 0.5.

Recent results \citep{Semler2012,McDonald2013}
based on samples of clusters selected by the Sunyaev-Zeldovich (SZ)
effect, with Chandra follow-up, demonstrate that CC clusters do exist
at redshifts greater than 0.5. Moreover,
\cite{McDonald2013} found that there is no evolution in central
cooling time out to redshifts $\sim1$. There are also studies on
individual clusters, although not very numerous, which show that
there are strong cool cores at high redshift. The WARPS cluster studied
by \cite{Santos2012} is a CC cluster at redshift 1.03. Another
interesting system is 3C188, studied by \cite{Siemiginowska2010},
which is a strong CC system at z=1.03 with a powerful radio AGN at its
centre. Signs of cooling at the centre of the cluster surrounding the
$z=1.04$ powerful quasar PKS1229-021 have also been reported by
\cite{Russell2012}.

While most of these evolutionary studies have concentrated
on rich clusters, and show a reduction in the incidence of strong
CCs at high redshift, the one study
\citep{Alshino2010} which covers groups, finds a conflicting trend
in less massive systems, whereby the CC strength tends to increase
at high redshift.  This study is based
on XMM data, which has limited spatial resolution. The aim of
the present paper is to present the results of a study of the evolution of
CCs across the full mass range from groups to clusters using the deepest
available high spatial resolution data, which we extract from the Chandra
archive. This X-ray selected sample constitutes the {\it Chandra Deep Group
Survey} (CDGS).  The CDGS sample and our selection criteria are presented
in Section 2. In Section 3 we describe the methods adopted to extract X-ray
properties for each system, and we examine a number of cool core
estimators. Our main results are presented in Section 4. Section 5 contains
a discussion of possible selection biases which might have an impact on our
results, and the addition of a set of high redshift non X-ray selected
systems with which we enlarge our sample. Finally, in Section 6 we discuss
the conclusions from this work. A $\Lambda$ cold dark matter cosmology with
H$_{0}= 100 h = 70$ km s$^{-1}$ Mpc$^{-1}$ $\Omega_{M}=0.3$ and
$\Omega_{\Lambda}=0.7$ is adopted throughout the paper.

\section{Sample Selection and data reduction}
\label{secData}
Our study is based on a Chandra archival sample of 62 systems with
temperatures between $\sim1$ and $\sim12$ keV and redshifts that span the
range between 0.07 and 1.3, with means in temperature and redshift of 4.0
keV and 0.55 respectively. The sky coordinates of the systems in our sample
together with the X-ray properties derived from our analysis are listed in
Table \ref{table:table1} and Table~\ref{table:table2}.

The strategy adopted for our sample selection has a twofold motivation:
firstly, the necessity of a large sample, with enough statistics to allow
the study of cool core evolutionary trends in groups and clusters, and
secondly, the requirement for data of sufficient quality to permit spectral
and spatial analysis for all systems in the
sample.

The use of Chandra data is crucial for our study because of the high
resolution required to resolve the cores in our systems out to high
redshifts, in order to apply different cool core estimators and also to
resolve and exclude contaminating point sources. Chandra's
advantage over all other X-ray telescopes is its high
angular resolution of $\sim$0.5 arcsecond (FWHM), which corresponds to 4 kpc at
a redshift of 1. 

The observations used by CDGS to search for extended sources,
have been selected from the Chandra archive using the following criteria:

\begin{itemize}
\item[--] Only ACIS-I observations are used. Chandra has two 
detectors which can be used for spectral imaging: ACIS-I and ACIS-S. We
use only ACIS-I observations due to their larger field of view compared
to ACIS-S. This allows us to maximize the number of {\it serendipitous}
clusters in our sample (i.e. systems which were not the target of the
Chandra observation, and are therefore free from observer selection
bias). To construct a sample as large as possible we made use of all
ACIS-I observations available in the archive as of September 2009 (when
the analysis commenced) which meet certain criteria. 
 
\item[--] Only high galactic latitude 
($|b|>20^{\circ}$) pointings were included, to avoid heavy galactic absorption. 
 
\item[--] Observations for which the target 
    is a low redshift extended system that occupies most of the field of
    view were excluded. A consequence of this requirement is that our sample lacks very
    low redshift systems. This can be seen in Table~\ref{table:table1} --
    with the exception of one system, all sources lie at redshifts greater
    than 0.1.
 \end{itemize} 

All individual observations from the archive with the 
  above mentioned properties have been grouped into fields (i.e. a single
  observation, or a group of observations with similar pointings). In order
  to provide data of adequate quality for our analysis out to high
  redshift, we considered only fields with a total exposure time of at least
  70 ks, though individual areas within a field can have shorter exposures
  than this. These selection criteria result in a total of 66 fields, covering
  an area of $\sim 10 $ degree$^2$.

Each observation was reprocessed starting from level 1 event files in order
to use the latest calibration files for the charge transfer inefficiency
and time dependent gain corrections and to create new bad pixel files with
hot pixels and those affected by cosmic ray events flagged. Calibration
files are taken from the Calibration Database (version 4.5) and data
reprocessing and all subsequent data analysis has been performed with the
Chandra software package CIAO (version 4.4). Three types of filters have
been applied to the corrected level 1 events file to create a corrected and
filtered level 2 events file for use in our data analysis. The first filter
is for bad event grades (we used ASCA grades 0,2,4,6) and for `clean'
status column. The other two filters are for background cleaning. The first
removes background flares, which seriously affect only a few
observations. Flaring periods were removed from the eventfile by extracting
a lightcurve from the whole chip, excluding sources, and eliminating
periods of time in which the count rate is 20$\%$ higher than the median
rate. The second background filter was applied only to observations taken
in VFAINT mode. The VFAINT cleaning procedure removes events generated by
high energy particles and is applied in order to reduce the level of
particle background.

After reprocessing and cleaning the event file, observations with similar
pointings were merged to create a single event file (field) for all
overlapping observations. This file was used for all our spatial analysis,
whilst individual observations were used for spectral analysis.

We searched all fields for sources using a source searching algorithm
based on the Voronoi tessellation algorithm implemented in CIAO. All
detected sources were tested for extension using a Bayesian extension
test developed by \cite{Slack2014} which checks for a significant
difference in fit statistic between a point source model and a beta
model blurred with the point spread function. Our final candidate list
includes only extended sources with at least 100 counts in the
soft band (0.5-2.0 keV). 
This threshold is motivated by the fact that our subsequent
analysis requires enough counts to construct a useful spectrum and
surface brightness profile. This restriction also has the advantage of
greatly simplifying selection biases, as we will see in
Section~\ref{subsection:detbias}. The flux corresponding to the 100
count limit varies with the exposure time of the source.
Assuming a spectrum corresponding to a thermal plasma with a
temperature of 3~keV and abundance 0.3~solar, at redshift 0.5,
the 0.5-2.0 keV flux limit is approximately $8 \times 10^{-15} t_{100}^{-1}$
 $\rm erg~cm^{-2}~ s^{-1}$, where $t_{100}$ is the exposure time in units
of 100~ksec, which varies from 0.1 to 40 for our sources.

A number of sources which, although extended, were found to be dominated by
a bright central point source (presumably an AGN) were excluded, as described
in Section~\ref{subsection:SBP}, and four apparently bona-fide extended
sources were also dropped from our list because no redshift was available
for them. Our total X-ray selected sample of 62 groups and clusters
is listed in Table~\ref{table:table1}. 33 are serendipitous detections, whilst
the remaining 29 were the main target of the Chandra observation in which
they were detected. The redshift value quoted in the Table for each system is 
derived from the literature. Note that some of these redshifts are photometric. 
The position given for each system corresponds to the R.A. and 
Declination (J2000) of the X-ray peak.

\begin{table*}
\centering
\normalsize
\begin{minipage}{180mm}

    \caption{Catalogue of groups and clusters used. Columns represent: source ID (increasing with redshift),  Right Ascension (R.A.), Declination (Dec.), redshift, reference for redshift and the number of galaxies used to derive the cited redshift (when available), a source flag, alternative names given in the literature for the system and any other notes. R.A and Dec are given for J2000 and represent the position of the X-ray peak. All redshifts are spectroscopic except those marked with an asterisk which are photometric. For each source, the Flag column contains a 't' if the source is the target of the observation, an 'a' if is contaminated by a central AGN and a 'c' if the beta model fit to the surface brightness profile has been adjusted (see \protect\ref{subsection:SBP}).}
    
\scriptsize
   \centering
  \begin{tabular}{@{}lrrlrll}  
  \hline
  ID  &  R.A. (deg) & Dec. (deg) &  z & Ngal & Flag & Literature names\\
\hline
  CDGS1 & 214.4486 & +52.6954 & 0.066 & 23[1] & \textendash a\textendash & EGSXG J1417.7+5241  \\ 
  CDGS2 & 149.8517 & +01.7736 & 0.12$^{*}$ & ---[2] & \textendash\textendash \textendash  & \\
  CDGS3 & 150.4316 & +02.4281 & 0.12$^{*}$ & ---[2] & \textendash\textendash \textendash & \\
  CDGS4 & 26.2022 & -04.5494 & 0.17$^{*}$ & ---[3] & \textendash\textendash\textendash& \\
  CDGS5 & 215.003 & +53.1122 & 0.200 & 19[1] & \textendash\textendash\textendash & EGSXG J1420.0+5306 \\
  CDGS6 & 221.6679 & +09.3385 & 0.204$^{*}$& ---[6] & \textendash\textendash\textendash & \\
  CDGS7 & 212.907 & +52.3147 & 0.21$^{*}$ & ---[4] & \textendash\textendash\textendash &  \\
  CDGS8 & 150.1967 & +01.6537 & 0.220 & 14[2] & \textendash\textendash\textendash &  \\
  CDGS9 & 8.4430 & -43.2917 & 0.223 & 1[5] & \textendash\textendash\textendash & XMMES1\_145 \\
  CDGS10 & 255.1737 & +64.2167 & 0.225 & 1[7] & \textendash \textendash c & RXJ1700.7+6413;Abell2246; \\
  CDGS11 & 214.3371 & +52.5964 & 0.236 & 9[1] & \textendash\textendash\textendash & EGSXG J1417.3+5235 \\
  CDGS12 & 210.31717 & +02.7534 & 0.245 & ---[8] & \textendash\textendash\textendash &  \\
  CDGS13 & 235.3019 & +66.4410 & 0.245 & ---[9] & \textendash\textendash\textendash & \\
  CDGS14 & 222.6074 & +58.2201 & 0.28$^{*}$ & ---[10] & \textendash\textendash\textendash & \\
  CDGS15 & 150.1798 & +01.7689 & 0.346 & 14[2] & \textendash\textendash\textendash & \\
  CDGS16 & 170.0304 & -12.0864 & 0.352 & 13[11] & t \textendash\textendash & \\
  CDGS17 & 292.9568 & -26.5761 & 0.352 & 35[12] & tac & MACSJ1931.8-2634\\

  CDGS18 & 161.9225 & +59.1156 & 0.36$^{*}$ & ---[10] & \textendash \textendash \textendash &  \\
  CDGS19 & 170.0416 & -12.1476 & 0.369 & 22[11] & t \textendash\textendash & \\
  CDGS20 &  8.6137 & -43.3168 & 0.3925& 1[5] & \textendash\textendash\textendash & XMMES1\_224 \\
  CDGS21 & 29.9557 & -08.8331 & 0.406 & 31[12] & tac & MACS0159\\
  CDGS22 & 29.9637 & -08.9219 & 0.407$^{*}$ & ---[13] & \textendash\textendash\textendash &  \\
  CDGS23 & 249.1566 & +41.1337 & 0.423 & 3[14] & \textendash\textendash\textendash &\\
  CDGS24 & 327.672 & -05.6853 & 0.439 & 30[15] & \textendash\textendash\textendash &  \\
  CDGS25 & 138.4395 & +40.9412 & 0.442 & 1[16] & ta \textendash& MACSJ0913.7+4056; CL09104+4109\\
  CDGS26 & 52.4231 & -02.1960 & 0.450 & ---[17] & t \textendash c &   MACSJ0329.6-0211\\
  CDGS27 & 255.3481 & +64.2366 & 0.453 & ---[18] & t \textendash c & RXJ1701.3+6414 \\
  CDGS28& 212.8357 & +52.2027 & 0.460 & 21[19] & tac & Cl 1409+524 \\
  CDGS29 & 245.3532 & +38.1691 & 0.461 & ---[20] & tac & MACSJ1621.3+3810\\
  CDGS30 & 169.9805 & -12.0402 & 0.479 & 17[21] & t \textendash\textendash & \\
  CDGS31 & 197.7571 & -03.1768 & 0.494 & ---[22] & t \textendash\textendash & MACS1311.0-0311\\
  CDGS32 & 158.8557 & +57.8484 & 0.5$^{*}$ & ---[23] & \textendash\textendash\textendash & \\
  CDGS33 & 158.8076 & +57.8387 & 0.5$^{*}$ & ---[23] & \textendash\textendash\textendash & \\
  CDGS34 & 109.3822 & +37.7581 & 0.546 & 142[24] &t \textendash c& MACSJ0717.5+3745\\
  CDGS35 & 170.2387 & +23.4462 & 0.562 & ---[25] & t \textendash\textendash & RXJ1120.9+2326; V1121+2327\\
  CDGS36 & 132.1985 & +44.9380 & 0.570 & 11[26] & t \textendash\textendash & RX J0848+4456; CL0848.6+4453\\
  CDGS37 & 6.3736 & -12.3761 & 0.586 & 108[27] & t \textendash\textendash &  MACS0025.4-1222  \\
  CDGS38 & 314.0887 & -04.6307 & 0.587& 149[28] & t \textendash c &  MS2053.7-0449\\
  CDGS39 & 314.0721 & -04.6988 & 0.600 & ---[29] & \textendash\textendash\textendash & \\
  CDGS40 & 222.5374 & +09.0802 & 0.644 & 9[30] & \textendash\textendash\textendash & \\
  CDGS41 & 52.9582 & -27.8274 & 0.679 & 2[31] & \textendash\textendash\textendash & \\
  CDGS42 & 214.4736 & 52.5795 & 0.683 & 11[1] & \textendash a \textendash & EGSXG J1417.9+5235 \\
  CDGS43 & 61.352 & -41.0057 & 0.686 & ---[32] & t \textendash\textendash &  \\
  CDGS44 & 185.3565 & +49.3092 & 0.700 & ---[25] & t \textendash\textendash & RXJ1221.4+4918; V1221+4918\\
  CDGS45 & 345.6999 & +08.7307 & 0.722 & 1[33] & t \textendash\textendash & WARPJ2302.8+0843; CLJ2302.8+0844\\
  CDGS46 & 168.2731 & -26.2612 & 0.725 & 2[33] & t \textendash\textendash & WARPS1113.0-2615  CLJ1113.1-2615\\
  CDGS47 & 149.9211 & +02.5229 & 0.730 & 12[2] & \textendash\textendash\textendash &\\
  CDGS48 & 53.0401 & -27.7099 & 0.734 & 4[31] & \textendash\textendash\textendash&  \\
  CDGS49 & 215.1388 & +53.1392 & 0.734 & 17[1] & \textendash\textendash\textendash & EGSXG J1420.5+5308 \\
  CDGS50 & 349.6286 & +00.5661 & 0.756 & 8[34] & t \textendash\textendash & RCS2318+0034 \\
  CDGS51 & 175.0927 & +66.1374 & 0.784 & 22[35] & t \textendash\textendash &  MS1137.5+6625\\
  CDGS52 & 199.3407 & +29.1889 & 0.805 & 6[36] & t \textendash\textendash & RDCS 1317+2911\\
  CDGS53 & 214.0694 & +52.0995 & 0.832 & 1[1] & \textendash\textendash\textendash & EGSXG J1416.2+5205\\
  CDGS54 & 150.504 & +02.2246 & 0.9$^{*}$  & ---[2] & \textendash\textendash\textendash &\\
  CDGS55 & 53.0803 & -27.9017 & 0.964 & 2[31] & \textendash\textendash c & \\
  CDGS56 & 355.3011 & -51.3285 & 1.00 & 15[37] & t \textendash\textendash & SPT-CLJ2341-5119 \\
  CDGS57 & 213.7967 & +36.2008 & 1.026 & 25[38] & t \textendash c & WARPS J1415.1+3612 \\
  CDGS58 & 137.6857 & +54.3697 & 1.101 & 20[39] & t \textendash\textendash & \\
  CDGS59 & 137.5357 & +54.3163 & 1.103 & 17[40] & \textendash \textendash \textendash &  RXJ 0910+5419 \\
  CDGS60 & 193.2273 & -29.4546 & 1.237 & 36[41] &t \textendash\textendash & RDCS1252-29\\
  CDGS61& 132.2435 & +44.8664 & 1.261 & 6[42] & t \textendash\textendash & RXJ0848.9+4452; RDCS0848.9+4452 \\
  CDGS62 & 132.1507 & +44.8975 & 1.273 & 8[43] & t \textendash\textendash & RXJ0848.6+4453; RDCS0848.6+4453; CLG J0848+4453\\

\hline
\label{table:table1}
\end{tabular}

\medskip
\raggedright
Redshift References: 1:\citealt{Finoguenov2007}; 2:\citealt{Knobel2012}; 3:\citealt{Mehrtens2012}; 4:\citealt{Wen2011}; 5:\citealt{Feruglio2008}; 6:\citealt{Hsieh2005}; 7:\citealt{Struble1987}; 8:\citealt{Bonamente2012}; 9:\citealt{Romer2000}; 10:\citealt{Wen2012}; 11:\citealt{Tran2009a}; 12:\citealt{Ebeling2010}; 13:\citealt{Hao2010}; 14:\citealt{Manners2003}; 15:\citealt{Finoguenov2009}; 16:\citealt{Kleinmann1988}; 17:\citealt{Kotov2006}; 18:\citealt{Vikhlinin1998a}; 19:\citealt{Dressler1992}; 20:\citealt{Allen2008}; 21:\citealt{Gonzalez2005}; 22:\citealt{Schmidt2007}; 23:\citealt{Yang2004}; 24:\citealt{Ebeling2007}; 25:\citealt{Mullis2003}; 26:\citealt{Holden2001}; 27:\citealt{Bradavc2008}; 28:\citealt{Tran2005a}; 29:\citealt{Barkhouse2006}; 30:\citealt{Finoguenov2009}; 31:\citealt{Szokoly2004}; 32:\citealt{Burenin2007}; 33:\citealt{Perlman2002}; 34:\citealt{Stern2010}; 35:\citealt{Donahue1999}; 36:\citealt{Holden2002}; 37:\citealt{Song2012}; 38:\citealt{Huang2009}; 39:\citealt{Tanaka2008};
40:\citealt{Rumbaugh2013}; 41:\citealt{Rosati2004}; 42:\citealt{Rosati1999}; 43:\citealt{Stanford1997}

\end{minipage}
\end{table*}

\section{Data Analysis}

Our aim is to study the evolution of CCs in groups and clusters of
galaxies and compare evolutionary trends between these two classes of
objects. Therefore an X-ray spectral and spatial analysis has been
performed on each system in our sample in order to characterize the
gas properties and 
derive parameters which can be used as CC estimators. We use
mean gas temperature estimated from our spectral fits to distinguish
between groups and clusters by applying a temperature cut of 3
keV. There is, of course, a degree of arbitrariness in this choice,
and previous studies have adopted temperature thresholds
between groups and clusters ranging
from 1 keV to 3 keV \citep{Sun2009,Finoguenov2001,Gastaldello2007}.

\subsection{X-ray derived parameters}

\subsubsection{\textbf{\R}}

\label{subsection:r500}

\R, the radius enclosing a mean density of 500 times the critical density
at the system's redshift, is estimated iteratively using the observed
relation between radius and temperature derived by \cite{Sun2009} for a
sample of 57 low redshift groups and clusters of galaxies:

\begin{equation}
 hE(z)R_{500}=0.602\left(\dfrac{T_{500}}{3 {\rm keV}}\right)^{0.53} ~~,
\end{equation}

where the evolution factor is

\begin{equation}
 E(z)=\sqrt{\Omega_{M}(1+z)^{3}+\Omega_{\Lambda}}~~,
\end{equation}

with $h=0.7$ for our cosmology, $z$ is the system redshift, and $T_{500}$
the gas temperature within $R_{500}$.

\cite{Sun2009} evaluate $T_{500}$ by creating a three-dimensional temperature profile and integrating it between $0.15R_{500}$ and $R_{500}$. They exclude the inner region of the system in order to remove the contribution of a CC or a central AGN which would bias the mean temperature towards lower or higher (respectively) values.
In our case we lack the data quality required to create a temperature profile, so our $T_{500}$ is derived by fitting a spectrum extracted from within a circle of radius $R_{500}$, and is therefore the projected mean temperature within $R_{500}$, including the central region. The only case in which we exclude a central region is when we find evidence for the existence of an X-ray AGN, which can be detected as a point like source in the hard band (2.0-7.0 keV) image of the system. In that situation, we remove data within a circle enclosing 95$\%$ of the counts from a point spread function at the position of the AGN. Since we include the central region in our spectrum, the contribution from a CC, if it is present, will bias our temperature downwards. However, the magnitude of this bias has been shown to be at the 4-5\% level for both groups and clusters \citep{Osmond2004,Pratt2009}, which is much smaller that our statistical errors of $\sim$20\%.

Evaluation of  $R_{500}$ involves an iterative procedure. A first estimate of $T_{500}$ is derived by fitting a spectrum extracted from a region equivalent to the source detection region. This temperature is used to calculate $R_{500}$ which provides the extraction radius for a new spectrum, from which we derive a new temperature. The process is then repeated until convergence.

\subsubsection{\textbf{Gas temperature}}

The mean temperature of the gas within $R_{500}$ was obtained by
fitting a spectrum extracted within a circular region of radius equal
to $R_{500}$ with a model composed of two main components: one for
cluster emission and the other for particle and photon background. The
cluster contribution was modelled with an
absorbed thermal plasma (APEC) model. The free parameters are
temperature and normalization, while we fixed the redshift at the
known value, the abundance at 0.3 solar \citep{Mushotzky1997}
using the abundance table from \cite{Anders1989}, and the
absorbing column at the Galactic value \citep{Dickey1990}.

We model the background emission, instead of subtracting it, because
this allows us to use the Cash statistic \citep{Cash1979} in our fitting
procedure, which is less biased for sparse data compared to the $\chi^{2}$
statistic \citep{Humphrey2009}. However, it can only be applied to
Poisson distributed data, a condition which would not be valid after
background subtraction. Our background model includes components for
cosmic X-ray background (galactic and extragalactic), particle and
instrumental background. Galactic emission is modelled by two thermal
plasma models: one for the Galactic Halo
\citep{Snowden1998,Henley2010a} and one for the Local Hot
Bubble. Cosmic background is modelled as a power law with a fixed slope
of 1.4 \citep{DeLuca2004}, while to model the quiescent particle
background we use a broken power law \citep{Snowden2008}. Instrumental
background due to fluorescence of material in the telescope and focal
plane is modelled by five Gaussians to account for the most prominent
lines in the spectrum.

As we are dealing with multiple observations for each system, we have
extracted background spectra from the entire field of view of each
individual observation in which the system is present after excluding all
sources. All extracted spectra were merged and our background model fitted
to this merged spectrum. The same approach was used for the source spectra.

\subsubsection{\textbf{Surface brightness profiles}}
\label{subsection:SBP}

To characterize the spatial distribution of X-ray emission from the cluster
gas we constructed azimuthally-averaged surface brightness profiles using
concentric circular annuli centred on the X-ray peak, within an outer
radius of $2.5R_{500}$. These profiles were fitted with a single beta-model
\citep{Cavaliere1976} to which we add a constant to allow for the
background contribution:

\begin{equation}
\label{equation:beta}
 S(r)=S_{0}(1+(r/r_{c})^2)^{-3\beta+0.5}+C ~~,
\end{equation} 

where $S_{0}$, $r_{c}$ and $C$ are the central surface brightness, core
radius and the background constant, respectively. Blurring by the Chandra
point spread function (PSF) is allowed for during the fitting process using a
model generated with the Chandra MARX simulator for each source, at the
appropriate off-axis angle.

While the single beta model can describe well the surface brightness
distribution of NCC clusters \citep{Mohr1999,Henning2009},
it represents a poor approximation for CCs because of their central surface
brightness excess above the model \citep{Neumann1999,Vikhlinin2006}. \cite{Chen2007} showed
that a significant improvement in the fit of CC clusters can be obtained by
adding a second component to the model to account for the central excess
emission. The quality of our data do not permit a more complex model to be
fitted, and in practice our main aim will be to use the fitted profile to
estimate the gas density in the core of each system (at r=0.01$R_{500}$)
using geometrical deprojection, which has a relatively straightforward
analytical form for the case of a single beta model (see
Section \ref{subsection:cooling_time} for details of the geometrical
deprojection and the choice of r=0.01$R_{500}$). Because we need to
obtain the density at a particular radius, our primary requirement is a
good match of the model to the data around that radius. We checked the
adequacy of our fit for each system and found that for most cases it
matches the data well into 0.01$R_{500}$. In a few cases with strongly
peaked profiles, the default fit underestimates the data at small radii.
For these cases, we first fit the central region using a beta-model with a
small core radius, and then fix the amplitude whilst relaxing other
parameters, to achieve the best fit possible at larger radii, subject to
providing a good match near the centre. Systems for which such adjustment
was needed are flagged with a `c' in column 6 of Table \ref{table:table1}.

\begin{figure}
\centering
 \includegraphics[width=0.5\textwidth,keepaspectratio]{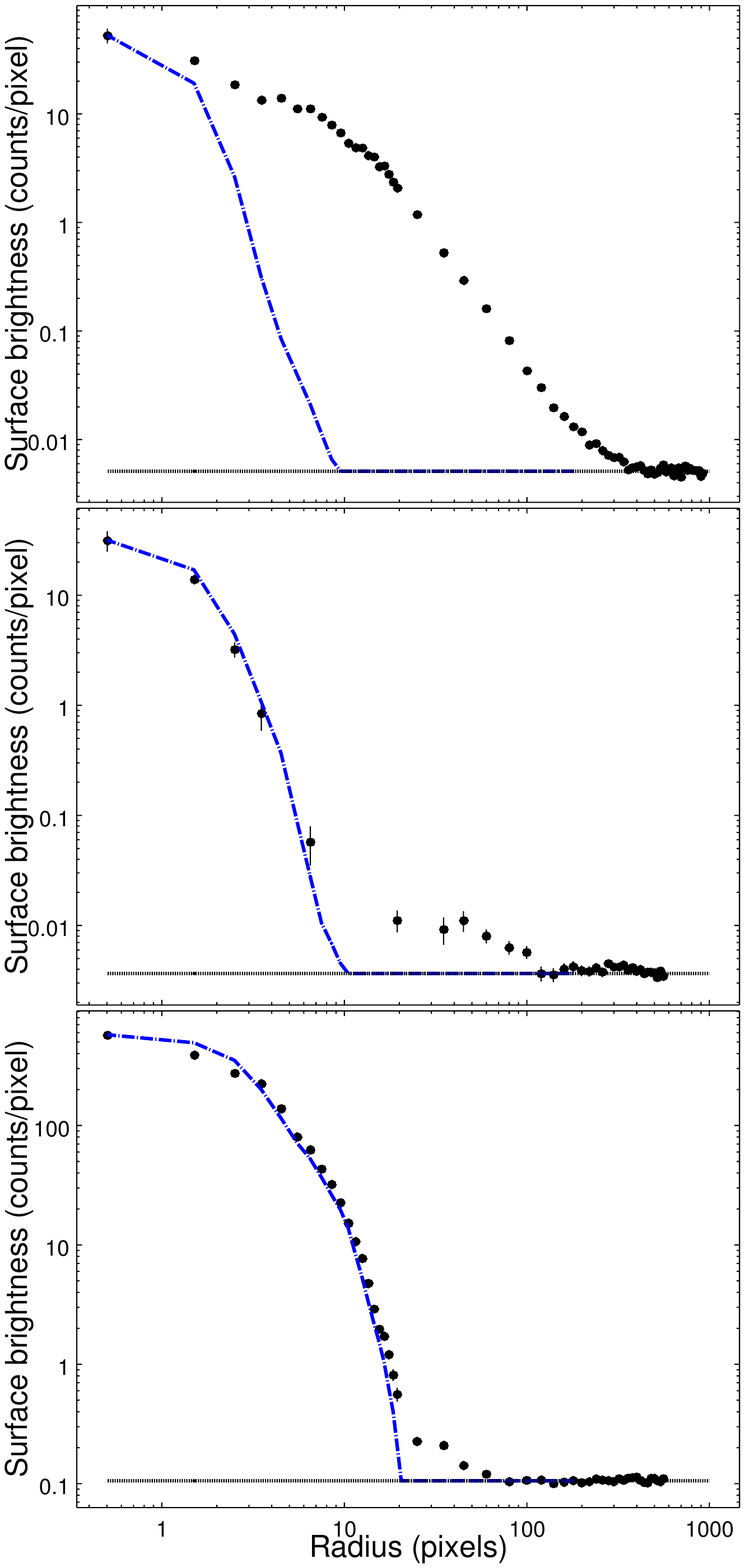}
 \caption{Three cases of AGN contamination. Top panel: the central AGN is strong but the cluster's flux dominates; source is kept in the sample. Middle panel: AGN dominates over the clusters' flux; the source is excluded from our sample. Bottom panel: The AGN is dominant and there is some evidence for the presence of extended emission; source is excluded from the sample. Black filled symbols represent the data while the modelled PSF is represented by the blue dashed line. The horizontal dotted line marks the background level. }
  \label{AGN_contamination}
\end{figure}

It is well-established that the central galaxies in many low redshift groups and clusters display nuclear activity. Such AGN can be bright X-ray point sources, which may contaminate the cluster X-ray flux. We checked for the existence of a central AGN in three different ways: by looking for the presence of a central point source at the position of the cluster candidate in the hard band image, by comparing the surface brightness profile of the source with the point spread function, and by comparing the fit statistics of a thermal plasma plus power law model fit (to model the cluster emission plus the AGN) with a thermal plasma only model applied to the source spectrum. Cluster candidates in which we found evidence of AGN contamination were divided into three classes: (1) Sources with clear spatial extension in which the central AGN does not dominate the total flux -- in this case the source was retained in the cluster list and the central AGN excised during data analysis. (2) Sources with clear extension but 
with a dominant central AGN. (3) Sources with only marginal extension, but with clear evidence for the presence of an AGN. In cases (2) and (3) the source was excluded from our catalogue. An example of each case is presented in Figure \ref{AGN_contamination}.

\subsubsection{\textbf{Cooling Time}}

\label{subsection:cooling_time}

The mechanism by which gas in clusters of galaxies cools is radiation of
its thermal energy through X-ray emission. One simple parameter which can
characterize the thermal state of the gas is the cooling time, which is
defined as the characteristic timescale on which the gas radiates away its
thermal energy.  The cooling time at a radius $r$ is
\begin{equation}
 t_{cool}(r)=\dfrac{3}{2}\,\dfrac{\mu_{e}\,n_{e}\,V\,kT}{\mu\,L_{x}}~,
 \label{equation:tcool}
\end{equation} 
where $kT$ and $n_{\rmn{e}}$ are the gas temperature and electron
number density in a spherical shell of volume $V$ at radius $r$, and
$L_{\rmn{x}}$ is the luminosity radiated by the shell. The mean mass
per electron ($\mu_{\rmn{e}}$) and mean mass per particle ($\mu$) have
values of 1.15 and 0.597, respectively, corresponding to a fully
ionized thermal plasma with metallicity 0.3 Z$_\odot$
\citep{Sutherland1993}.

The gas density at radius $r$ is derived from the normalization of the
thermal plasma fit to the source spectrum and derived counts emissivity
using the following equation:

\begin{equation}
 n_{e}(r)=\sqrt{\dfrac{(n_{e}/n_{H})\, N_{\rm spec} \,4\pi [D_{a}(1+z)]^2}{V\,10^{-14}}\dfrac{\epsilon(r)}{C}}~,
\end{equation} 

where D$_{a}$ is the angular diameter distance , $\epsilon(r)$ is the
counts emissivity integrated over the volume of the shell (i.e. the total
count/s from the shell) and C is the total number of counts from the source
within \R. $N_{\rm spec}$ is the normalization of the thermal plasma model
fitted to the spectrum extracted within \R, with all point sources
excluded, which for the APEC model is related to the emission measure by

$N_{\rm spec}=\dfrac{10^{-14}}{4\pi[D_a(1+z)]^2}\int n_{e}n_{H}dV $.

The analytical expression for the counts emissivity profile

\begin{equation}
 \epsilon(r)=\epsilon_{0}(1+(r/r_{c})^2)^{-\eta}~,
\end{equation} 

can be obtained from the surface brightness profile of the form given in
Equation \ref{equation:beta} by geometrical deprojection, assuming a
spherically symmetric distribution. Since surface brightness represents the
projection on the sky of emissivity, the surface brightness profile can be
written as a integral along the line of sight of emissivity:

\begin{equation}
 S(b)=2 \int_0^{\infty} \epsilon(r) dl~,
\end{equation} 

where $r^2=b^2+l^2$ and $l$ is the direction along the line of
sight. Solving the integral, we can obtain the slope and the normalization
of the emissivity profile as a function of the beta-model parameters. Hence
$\eta=3\beta$ and $\epsilon_{0}=S_{0}/(2r_{c} \int_0^{\pi/2} \cos \alpha
^{2(\eta -1)} d\alpha)$ .

The temperature of the gas is required to derive gas density from the
emissivity, and hence to calculate entropy and cooling time. We use the
global temperature, as our data quality does not allow us to construct
temperature profiles. For CC systems, the temperature drops in the core, by
a factor of up to 2 or 3 from its peak value (or a smaller factor compared
to the {\it mean} global temperature). As a result, we will somewhat
overestimate the central cooling time in CC systems, by a factor of
approximately $\sqrt{2}$.

Clearly, the cooling time rises progressively with radius, as the density
drops, so we need to pick a scale radius at which to extract the cooling
time which will characterise the cluster core. We would like this radius to
be as small as possible, subject to it being resolved in our
observations. However, we do want the derived gas properties to represent
the group/cluster core. \citet{Sun2007} has pointed out that some galaxy
groups contain dense gas within the central galaxy, which he refers to as a
`compact corona'. These small gas halos are distinct from classic cool
cores and are more closely associated with the central galaxy itself. These
compact coronae have sizes typically between 1-4 kpc
\citep{Vikhlinin2001,Sun2007}, though they can be as large as 10 kpc. On
the basis of these considerations, we pick our scale radius for calculation
of the cooling time to be $0.01R_{500}$, which is deep inside the CC region
even for low mass systems but generally outside the inner 4 kpc.
Our surface brightness profiles have a radial resolution of $0.49\arcsec$,
which is similar to the FWHM of the Chandra on-axis PSF. This corresponds
to a physical scale of 4 kpc at z=1, and is smaller than $0.01R_{500}$ for
all our systems apart from CDGS62 at z=1.27, for which
0.01\R\ lies just inside the innermost bin. Although our cooling
time is derived from the analytical emissivity profile fitted
(allowing for PSF blurring) to the radial surface brightness profile, the
value for CDGS62 should be regarded as slightly less robust than the others, 
since it involves a small extrapolation inwards from the innermost data bin.

\subsubsection{\textbf{Entropy}}

Another parameter which can be used to characterize the thermal state of
the gas is its entropy, which we define here as
$K=\dfrac{kT}{n_{e}^{2/3}}$. This definition is widely adopted in X-ray
studies of clusters, and the standard thermodynamic definition of entropy
can be obtained from it by applying a logarithm and adding a constant
\citep{Voit2005}. To characterise the cluster core properties, we evaluate
the entropy at a scale radius of $0.01R_{500}$.

\subsubsection{\textbf{Error calculation}}
\label{subsec:errors}
Uncertainties in the values of \tcool\ and $K$ are estimated using Monte
Carlo simulations. The density profile parameters and the temperature
are perturbed in a Gaussian fashion based on their derived 
fitting errors. For each newly created 
set of parameters, a value for \tcool\ and $K$ at 0.01\R\ is calculated.
1000 such random realisations are generated and the required
errors are derived from the distribution of \tcool\ and $K$ which result. For the poorest quality datasets, the derived errors can be very large,
as can be seen in Table 2.

\subsection{Quantifying cool core status}
In order to study the evolution of cooling in cluster cores, we need to
choose an indicator of cool core strength. Ideally, this indicator should
be able to distinguish CC and NCC systems in a way which is minimally
affected by variations in redshift, temperature and data quality. As
discussed earlier, several CC estimators have been used in the literature:
some are based on the central temperature drop \citep{Maughan2012}, some
quantify the central surface brightness excess
\citep{Maughan2012,Alshino2010,Santos2008}, whilst others are based on
physical characteristics like central cooling time or entropy
\citep{Peres1998,Bauer2005,Mittal2009}.

Parameters that define the CC strength based on the amplitude of the
central temperature drop observed in the temperature profile of the system
are not accessible to us here because of the high quality data required to
construct temperature profiles. As many of our systems lie not far above
our 100 count lower limit, even calculating the ratio of central to outer
temperature is not feasible. Central cooling leads to increased gas
density, resulting in a sharp central cusp in surface brightness. This has
been used to define a number of different CC diagnostics. These approaches
have the advantage that they require only imaging data and can therefore be
applied over a wide range in data quality. When defining these parameters, generally a size for the CC is assumed in order to separate the emission
coming from the core from the larger scale emission.

\onecolumn
\small
\captionsetup[longtable]{labelformat=simple, labelsep=period, labelfont=bf}
\begin{longtable}{c c@{\hskip 0.4cm } cc@{\vspace{1.5 mm} \hskip 0.4cm }cccccc} 

\caption{X-ray derived properties. Columns represent: 1) Source ID, which is the same as in Table \ref{table:table1}; 2) Number of soft source counts (0.5-2.0 keV); 3) \R estimated iteratively as explained in \ref{subsection:r500}; 4) Gas temperature estimated from a thermal plasma model fit to a spectrum extracted within \R; 5) Cooling time; 6) Cooling time normalized by the age of the cluster which is the age of the Universe at the cluster's redshift; 7) Entropy; 8)-10) Three cuspiness cool core indicators (see \ref{subsection:cuspiness}). All errors are $1 \sigma$ errors. Errors for cooling time and entropy are calculated using Monte Carlo simulations (see \ref{subsec:errors}), while errors in cuspiness cool core indicators are estimated based on error propagation. Unconstrained errors are marked with asterisks. }\\
  \hline

ID & Counts & $R_{500}$ & kT & $t_{cool}$ & $t_{cool}/t_{Uni}$ & K & Csb & Fratio & Fc  \\ [-0.4ex]
& & (Mpc) & (keV) & (Gyr) & & (keV cm$^{2}$) & & & \\ [-3.0ex]\\ \hline
\endfirsthead

\caption{\emph{continued}}\\
\hline
ID & Counts & $R_{500}$ & kT & $t_{cool}$ & $t_{cool}/t_{Uni}$ & K & Csb & Fratio & Fc  \\ [-0.4ex]
& & (Mpc) & (keV) & (Gyr) & & (keV cm$^{2}$) & & & \\ [-3.0ex]\\ \hline
\endhead
\hline
\multicolumn{10}{r}{\emph{continued on next page}}
\endfoot
\hline
\endlastfoot

CDGS1 &  2522 & 0.470 & $ 0.98^{+0.18}_{-0.06}$ &$ 1.12 \pm 0.57$ & $ 0.09 \pm 0.04$ & $ 22.95 \pm 7.31$ & $ 0.393 \pm 0.606$ & $0.402 \pm 0.374$ & $3.87  \pm 0.26$\\
CDGS2 &  1008 & 0.519 & $ 1.30^{+0.21}_{-0.06}$ &$ 3.27 \pm 6.12$ & $ 0.27 \pm 0.50$ & $ 46.94 \pm 42.62$ & $ 0.139 \pm 0.143$ & $0.278 \pm 0.208$ & $1.62  \pm 0.20$\\
CDGS3 &  1982 & 0.557 & $ 2.08^{+1.83}_{-0.52}$ &$ 16.81 \pm 22.72$ & $ 1.36 \pm 1.85$ & $ 156.47 \pm 184.10$ & $ 0.055 \pm 0.008$ & $0.173 \pm 0.017$ & $0.52  \pm 0.08$\\
CDGS4&  357 & 0.562 & $ 1.54^{+0.46}_{-0.30}$ &$ 1.57 \pm 1.08$ & $ 0.13 \pm 0.09$ & $ 29.48 \pm 13.96$ & $ 0.100 \pm 0.024$ & $0.152 \pm 0.032$ & $2.27  \pm 0.64$\\
CDGS5 &  1047 & 0.492 & $ 1.25^{+0.11}_{-0.14}$ &$ 1.96 \pm 0.91$ & $ 0.17 \pm 0.08$ & $ 32.95 \pm 9.24$ & $ 0.153 \pm 0.079$ & $0.140 \pm 0.105$ & $1.79  \pm 0.41$\\
CDGS6 &  2132 & 0.864 & $ 3.42^{+0.80}_{-0.45}$ &$ 22.19 \pm 14.99$ & $ 1.95 \pm 1.32$ & $ 244.52 \pm 123.01$ & $ 0.057 \pm 0.008$ & $0.195 \pm 0.017$ & $1.20  \pm 0.19$\\
CDGS7 &  173 & 0.437 & $ 1.01^{+0.19}_{-0.11}$ &$ 4.26 \pm 6.54$ & $ 0.38 \pm 0.58$ & $ 55.70 \pm 45.33$ & $ 0.173 \pm 0.055$ & $0.360 \pm 0.114$ & $1.44  \pm 0.50$\\
CDGS8&  2413 & 0.708 & $ 2.54^{+0.50}_{-0.49}$ &$ 8.27 \pm 1.99$ & $ 0.74 \pm 0.18$ & $ 107.51 \pm 28.49$ & $ 0.048 \pm 0.005$ & $0.225 \pm 0.013$ & $0.61  \pm 0.06$\\
CDGS9 &  910 & 0.778 & $ 2.93^{+1.19}_{-0.62}$ &$ 1.47 \pm 1.25$ & $ 0.13 \pm 0.11$ & $ 36.71 \pm 24.66$ & $ 0.157 \pm 0.020$ & $0.302 \pm 0.031$ & $1.88  \pm 0.25$\\
CDGS10&  17428 & 0.833 & $ 3.42^{+0.20}_{-0.22}$ &$ 1.67 \pm 1.26$ & $ 0.15 \pm 0.11$ & $ 43.69 \pm 18.27$ & $ 0.200 \pm 0.004$ & $0.509 \pm 0.009$ & $1.49  \pm 0.04$\\
CDGS11 &  324 & 0.630 & $ 1.83^{+1.00}_{-0.36}$ &$ 3.80 \pm 16.67$ & $ 0.34 \pm 1.51$ & $ 56.18 \pm 95.19$ & $ 0.215 \pm 0.297$ & $0.343 \pm 0.107$ & $0.90  \pm 0.21$\\
CDGS12 &  1528 & 0.660 & $ 2.11^{+0.45}_{-0.27}$ &$ 1.27 \pm 0.62$ & $ 0.12 \pm 0.06$ & $ 28.07 \pm 9.81$ & $ 0.137 \pm 0.015$ & $0.207 \pm 0.018$ & $2.39  \pm 0.32$\\
CDGS13 &  1254 & 0.819 & $ 3.37^{+0.95}_{-0.57}$ &$ 9.96 \pm 4.13$ & $ 0.91 \pm 0.38$ & $ 142.31 \pm 64.28$ & $ 0.043 \pm 0.008$ & $0.167 \pm 0.014$ & $0.86  \pm 0.10$\\
CDGS14 &  440 & 0.727 & $ 2.77^{+2.64}_{-0.86}$ &$ 1.27 \pm 2.21$ & $ 0.12 \pm 0.21$ & $ 32.36 \pm 46.18$ & $ 0.186 \pm 0.028$ & $0.357 \pm 0.054$ & $1.85  \pm 0.27$\\
CDGS15 &  499 & 0.557 & $ 1.76^{+0.55}_{-0.20}$ &$ 0.71 \pm 0.32$ & $ 0.07 \pm 0.03$ & $ 17.79 \pm 6.89$ & $ 0.173 \pm 0.029$ & $0.250 \pm 0.034$ & $2.23  \pm 0.42$\\
CDGS16&  368 & 0.720 & $ 2.99^{+1.22}_{-0.86}$ &$ 10.07 \pm 12.81$ & $ 1.01 \pm 1.29$ & $ 133.72 \pm 118.12$ & $ 0.088 \pm 0.021$ & $0.295 \pm 0.062$ & $1.09  \pm 0.36$\\
CDGS17 &  48672 & 1.107 & $ 6.57^{+0.35}_{-0.22}$ &$ 0.54 \pm 0.06$ & $ 0.05 \pm 0.01$ & $ 30.08 \pm 2.79$ & $ 0.216 \pm 0.003$ & $0.635 \pm 0.005$ & $2.17  \pm 0.03$\\
CDGS18 &  733 & 0.662 & $ 2.53^{+0.42}_{-0.29}$ &$ 5.25 \pm 3.00$ & $ 0.53 \pm 0.30$ & $ 79.10 \pm 31.32$ & $ 0.085 \pm 0.013$ & $0.263 \pm 0.027$ & $0.86  \pm 0.16$\\
CDGS19&  516 & 0.735 & $ 3.10^{+1.30}_{-0.73}$ &$ 14.42 \pm 9.86$ & $ 1.47 \pm 1.01$ & $ 173.22 \pm 101.88$ & $ 0.055 \pm 0.014$ & $0.208 \pm 0.034$ & $0.77  \pm 0.23$\\
CDGS20 &  1345 & 0.803 & $ 3.52^{+0.76}_{-0.57}$ &$ 6.23 \pm 4.24$ & $ 0.65 \pm 0.44$ & $ 106.58 \pm 53.60$ & $ 0.085 \pm 0.010$ & $0.234 \pm 0.017$ & $1.29  \pm 0.19$\\
CDGS21&  22738 & 1.221 & $ 8.38^{+0.87}_{-0.41}$ &$ 0.66 \pm 0.05$ & $ 0.07 \pm 0.01$ & $ 39.75 \pm 4.10$ & $ 0.224 \pm 0.004$ & $0.566 \pm 0.006$ & $1.94  \pm 0.04$\\
CDGS22 &  639 & 0.668 & $ 2.73^{+0.68}_{-0.51}$ &$ 11.38 \pm 5.50$ & $ 1.20 \pm 0.58$ & $ 137.77 \pm 54.19$ & $ 0.050 \pm 0.011$ & $0.149 \pm 0.021$ & $0.86  \pm 0.25$\\
CDGS23 &  203 & 0.558 & $ 1.92^{+1.41}_{-0.32}$ &$ 7.71 \pm 9.43$ & $ 0.83 \pm 1.01$ & $ 90.18 \pm 88.01$ & $ 0.084 \pm 0.027$ & $0.325 \pm 0.073$ & $0.96  \pm 0.38$\\
CDGS24&  675 & 0.553 & $ 1.94^{+0.28}_{-0.22}$ &$ 5.26 \pm 3.66$ & $ 0.57 \pm 0.40$ & $ 70.11 \pm 33.75$ & $ 0.081 \pm 0.014$ & $0.190 \pm 0.023$ & $1.25  \pm 0.32$\\
CDGS25 &  14141 & 0.956 & $ 5.59^{+0.25}_{-0.21}$ &$ 0.39 \pm 0.03$ & $ 0.042 \pm 0.003$ & $ 22.06 \pm 1.38$ & $ 0.364 \pm 0.006$ & $0.715 \pm 0.010$ & $1.57  \pm 0.03$\\
CDGS26&  14546 & 0.956 & $ 5.59^{+0.48}_{-0.36}$ &$ 0.49 \pm 0.03$ & $ 0.053 \pm 0.003$ & $ 25.68 \pm 2.22$ & $ 0.272 \pm 0.005$ & $0.565 \pm 0.008$ & $1.94  \pm 0.05$\\
CDGS27&  10852 & 0.865 & $ 4.59^{+0.43}_{-0.42}$ &$ 0.76 \pm 0.27$ & $ 0.08 \pm 0.03$ & $ 30.68 \pm 7.47$ & $ 0.163 \pm 0.005$ & $0.324 \pm 0.007$ & $2.04  \pm 0.09$\\
CDGS28&  9837 & 0.897 & $ 4.89^{+0.48}_{-0.25}$ &$ 0.34 \pm 0.03$ & $ 0.038 \pm 0.003$ & $ 18.59 \pm 1.91$ & $ 0.330 \pm 0.007$ & $0.654 \pm 0.011$ & $1.91  \pm 0.06$\\
CDGS29 &  17937 & 1.090 & $ 7.10^{+0.58}_{-0.53}$ &$ 0.93 \pm 0.08$ & $ 0.10 \pm 0.01$ & $ 45.50 \pm 4.23$ & $ 0.242 \pm 0.005$ & $0.554 \pm 0.007$ & $2.10  \pm 0.05$\\
CDGS30 &  654 & 0.660 & $ 2.66^{+0.45}_{-0.35}$ &$ 2.84 \pm 1.65$ & $ 0.32 \pm 0.19$ & $ 54.02 \pm 22.41$ & $ 0.109 \pm 0.016$ & $0.311 \pm 0.030$ & $1.21  \pm 0.26$\\
CDGS31 &  11194 & 1.013 & $ 6.44^{+0.39}_{-0.55}$ &$ 1.59 \pm 0.17$ & $ 0.18 \pm 0.02$ & $ 61.31 \pm 6.64$ & $ 0.170 \pm 0.005$ & $0.619 \pm 0.010$ & $1.27  \pm 0.04$\\
CDGS32 &  205 & 0.476 & $ 1.57^{+0.24}_{-0.20}$ &$ 2.48 \pm 2.80$ & $ 0.28 \pm 0.32$ & $ 40.00 \pm 25.63$ & $ 0.095 \pm 0.027$ & $0.259 \pm 0.054$ & $1.00  \pm 0.26$\\
CDGS33&  428 & 0.664 & $ 3.01^{+0.93}_{-0.94}$ &$ 12.20 \pm 7.38$ & $ 1.40 \pm 0.84$ & $ 152.60 \pm 74.42$ & $ 0.078 \pm 0.019$ & $0.208 \pm 0.034$ & $0.99  \pm 0.32$\\
CDGS34 &  20101 & 1.392 & $ 12.36^{+0.73}_{-0.63}$ &$ 7.60 \pm 0.40$ & $ 0.90 \pm 0.05$ & $ 248.07 \pm 15.62$ & $ 0.032 \pm 0.002$ & $0.229 \pm 0.004$ & $0.70  \pm 0.03$\\
CDGS35&  1745 & 0.763 & $ 4.18^{+0.39}_{-0.37}$ &$ 12.29 \pm 1.06$ & $ 1.48 \pm 0.13$ & $ 185.34 \pm 19.19$ & $ 0.029 \pm 0.005$ & $0.128 \pm 0.010$ & $0.51  \pm 0.11$\\
CDGS36 &  1142 & 0.625 & $ 2.87^{+0.41}_{-0.58}$ &$ 5.05 \pm 1.65$ & $ 0.61 \pm 0.20$ & $ 82.47 \pm 23.62$ & $ 0.097 \pm 0.011$ & $0.282 \pm 0.021$ & $1.01  \pm 0.17$\\
CDGS37 &  11370 & 1.077 & $ 7.96^{+0.35}_{-0.36}$ &$ 9.76 \pm 0.32$ & $ 1.20 \pm 0.04$ & $ 233.19 \pm 10.84$ & $ 0.032 \pm 0.002$ & $0.248 \pm 0.006$ & $0.59  \pm 0.04$\\
CDGS38&  1850 & 0.821 & $ 4.83^{+0.85}_{-0.95}$ &$ 1.18 \pm 0.55$ & $ 0.14 \pm 0.07$ & $ 42.45 \pm 14.67$ & $ 0.104 \pm 0.009$ & $0.339 \pm 0.018$ & $1.17  \pm 0.14$\\
CDGS39&  193 & 0.751 & $ 3.58^{+9.00}_{-1.53}$ &$ 0.63 \pm 1.18$ & $ 0.08 \pm 0.15$ & $ 23.35 \pm 54.95$ & $ 0.328 \pm 0.059$ & $0.713 \pm 0.167$ & $1.17  \pm 0.19$\\
CDGS40 &  127 & 0.685 & $ 3.58^{+21.56}_{-1.75}$ &$ 4.31 \pm 15.85$ & $ 0.55 \pm 2.04$ & $ 84.13 \pm 429.50$ & $ 0.103 \pm 0.037$ & $0.345 \pm 0.084$ & $1.05  \pm 0.47$\\
CDGS41&  2725 & 0.507 & $ 2.16^{+1.11}_{-0.35}$ &$ 16.00 \pm 18.03$ & $ 2.12 \pm 2.38$ & $ 153.01 \pm 134.27$ & $ 0.046 \pm 0.009$ & $0.104 \pm 0.072$ & $0.84  \pm 0.19$\\
CDGS42&  330 & 0.450 & $ 1.67^{+1.65}_{-0.93}$ &$ 6.57 \pm 49.42$ & $ 0.87 \pm 6.55$ & $ 77.21 \pm 164.94$ & $ 0.119 \pm 0.035$ & $0.228 \pm 0.059$ & $1.60  \pm 0.59$\\
CDGS43 &  1478 & 0.791 & $ 4.92^{+0.48}_{-0.40}$ &$ 3.90 \pm 0.78$ & $ 0.52 \pm 0.10$ & $ 95.14 \pm 15.49$ & $ 0.074 \pm 0.008$ & $0.397 \pm 0.022$ & $0.75  \pm 0.10$\\
CDGS44 &  2526 & 0.972 & $ 7.37^{+1.78}_{-1.46}$ &$ 11.80 \pm 4.08$ & $ 1.58 \pm 0.55$ & $ 253.74 \pm 83.36$ & $ 0.035 \pm 0.005$ & $0.194 \pm 0.010$ & $0.59  \pm 0.08$\\
CDGS45 &  1334 & 0.777 & $ 4.98^{+0.86}_{-0.47}$ &$ 5.76 \pm 2.14$ & $ 0.79 \pm 0.29$ & $ 124.31 \pm 36.34$ & $ 0.063 \pm 0.008$ & $0.292 \pm 0.020$ & $0.80  \pm 0.13$\\
CDGS46 &  1033 & 0.695 & $ 4.13^{+0.86}_{-0.74}$ &$ 4.06 \pm 1.33$ & $ 0.56 \pm 0.18$ & $ 87.89 \pm 25.12$ & $ 0.099 \pm 0.012$ & $0.341 \pm 0.025$ & $0.80  \pm 0.14$\\
CDGS47 &  1262 & 0.757 & $ 5.00^{+1.56}_{-1.68}$ &$ 16.36 \pm 4.52$ & $ 2.25 \pm 0.62$ & $ 249.86 \pm 93.94$ & $ 0.028 \pm 0.006$ & $0.122 \pm 0.013$ & $0.53  \pm 0.13$\\
CDGS48 &  1496 & 0.610 & $ 3.31^{+1.19}_{-0.75}$ &$ 16.40 \pm 12.87$ & $ 2.25 \pm 1.77$ & $ 195.69 \pm 121.22$ & $ 0.080 \pm 0.089$ & $0.110 \pm 0.034$ & $0.95  \pm 0.16$\\
CDGS49 &  542 & 0.550 & $ 2.53^{+0.88}_{-0.60}$ &$ 1.71 \pm 2.29$ & $ 0.24 \pm 0.32$ & $ 37.34 \pm 32.01$ & $ 0.160 \pm 0.026$ & $0.292 \pm 0.051$ & $1.53  \pm 0.29$\\
CDGS50&  1531 & 1.063 & $ 9.31^{+7.81}_{-4.54}$ &$ 8.68 \pm 4.91$ & $ 1.22 \pm 0.69$ & $ 237.14 \pm 171.73$ & $ 0.057 \pm 0.007$ & $0.512 \pm 0.033$ & $0.73  \pm 0.08$\\
CDGS51 &  3730 & 0.882 & $ 6.68^{+1.00}_{-0.75}$ &$ 3.02 \pm 0.48$ & $ 0.43 \pm 0.07$ & $ 96.74 \pm 16.53$ & $ 0.097 \pm 0.006$ & $0.432 \pm 0.014$ & $0.97  \pm 0.07$\\
CDGS52 &  321 & 0.627 & $ 3.34^{+1.69}_{-0.80}$ &$ 3.18 \pm 3.09$ & $ 0.46 \pm 0.45$ & $ 65.97 \pm 52.96$ & $ 0.126 \pm 0.027$ & $0.224 \pm 0.038$ & $1.49  \pm 0.49$\\
CDGS53&  340 & 0.566 & $ 2.85^{+1.14}_{-0.75}$ &$ 5.66 \pm 45.89$ & $ 0.84 \pm 6.80$ & $ 88.40 \pm 163.53$ & $ 0.120 \pm 0.027$ & $0.346 \pm 0.083$ & $0.78  \pm 0.22$\\
CDGS54 &  594 & 0.666 & $ 4.36^{+2.27}_{-2.15}$ &$ 3.87 \pm 4.79$ & $ 0.60 \pm 0.75$ & $ 87.90 \pm 84.47$ & $ 0.093 \pm 0.015$ & $0.252 \pm 0.028$ & $1.07  \pm 0.20$\\
CDGS55&  620 & 0.679 & $ 5.44^{+8.15}_{-2.48}$ &$ 1.70 \pm 10.42$ & $ 0.28 \pm 1.69$ & $ 58.40 \pm 216.32$ & $ 0.242 \pm 0.063$ & $1.061 \pm 0.367$ & $1.03  \pm 0.13$\\
CDGS56 &  1781 & 0.980 & $ 10.43^{+5.87}_{-3.32}$ &$ 3.41 \pm 1.78$ & $ 0.57 \pm 0.30$ & $ 136.92 \pm 80.37$ & $ 0.105 \pm 0.009$ & $0.406 \pm 0.021$ & $1.21  \pm 0.12$\\
CDGS57 &  1200 & 0.722 & $ 5.98^{+2.13}_{-1.03}$ &$ 0.58 \pm 0.86$ & $ 0.10 \pm 0.15$ & $ 30.05 \pm 27.71$ & $ 0.139 \pm 0.013$ & $0.352 \pm 0.022$ & $1.39  \pm 0.18$\\
CDGS58 &  385 & 0.577 & $ 4.75^{+1.82}_{-1.69}$ &$ 7.43 \pm 8.63$ & $ 1.33 \pm 1.54$ & $ 143.06 \pm 102.50$ & $ 0.098 \pm 0.019$ & $0.300 \pm 0.042$ & $0.83  \pm 0.23$\\
CDGS59 &  313 & 0.444 & $ 2.57^{+0.37}_{-0.33}$ &$ 0.40 \pm 8.35$ & $ 0.07 \pm 1.50$ & $ 14.18 \pm 57.47$ & $ 0.073 \pm 0.018$ & $0.156 \pm 0.029$ & $1.88  \pm 0.61$\\
CDGS60&  757 & 0.664 & $ 6.07^{+2.87}_{-1.27}$ &$ 4.38 \pm 2.59$ & $ 0.86 \pm 0.51$ & $ 116.94 \pm 67.07$ & $ 0.090 \pm 0.014$ & $0.268 \pm 0.026$ & $0.98  \pm 0.21$\\
CDGS61&  351 & 0.621 & $ 6.50^{+4.10}_{-3.05}$ &$ 7.61 \pm 23.68$ & $ 1.51 \pm 4.70$ & $ 176.47 \pm 251.38$ & $ 0.109 \pm 0.021$ & $0.288 \pm 0.045$ & $1.00  \pm 0.28$\\
CDGS62 &  124 & 0.331 & $ 1.81^{+0.68}_{-0.59}$ &$ 3.64 \pm *** $ & $ 0.73 \pm ***$ & $ 53.11 \pm ***$ & $ 0.118 \pm 0.044$ & $0.147 \pm 0.049$ & $0.74  \pm 0.55$\\
\label{table:table2}
\end{longtable}
\normalsize
\nopagebreak
\twocolumn

It is not {\it a
  priori} clear what scale should be chosen to separate core from cluster
emission. \cite{Maughan2012} use a fraction of \R, whilst \cite{
  Santos2008} argue that cluster cores cannot be expected to evolve in a
self-similar fashion and so use a fixed metric radius of 40~kpc.

Given the wide mass and redshift ranges spanned by our sample,
the choice of core radius has a significant impact, and is
therefore a disadvantage for these methods. We therefore prefer to base
the bulk of our analysis on more physically motivated 
CC indicators. 
However, in Section~\ref{subsection:cuspiness}, we
calculate some these cuspiness indicators for our sample, and
compare the results with those from our preferred methods.

Central cooling time and entropy are gas properties which are
well-established to differ between CC and NCC clusters. Both are determined
primarily by gas density and temperature, though cooling time is also
affected by metallicity, which we take to be 0.3 solar.  As a result, the
two properties are closely related. Cooling time (\tcool) is more directly
related to the cooling status of the system, so we use this for
preference. As discussed in Section~\ref{subsection:cooling_time} above,
our `central' cooling time is actually calculated at a radius 0.01\R.

It will be helpful for some of our analysis to adopt a threshold value for
\tcool\ to mark the transition between CC and NCC systems. Previous studies
in which central cooling time is used as a CC diagnostic have used a
variety of cooling time thresholds, ranging from 0.8 Gyr up to the age of
the Universe \citep{Peres1998,Bauer2005,Mittal2009}. To help motivate our
own choice, we note that some studies of the distribution of central
entropy in groups and clusters have shown the existence of {\it bimodality}
\citep{Cavagnolo2009,Sanderson2009,Mahdavi2013}.  Moreover, \cite{Cavagnolo2008} show
that systems with a central entropy lower than 30 keV cm$^{2}$ show
evidence for gas cooling at the cluster centre, in the form of optical
emission lines.

Although both the Cavagnolo and Mahdavi studies show the existence of
bimodality in the entropy distribution, the break between the two peaks
occurs at 30-50 keV cm$^{2}$ for \cite{Cavagnolo2009} but 70 keV cm$^{2}$
for \cite{Mahdavi2013}. However, the difference between these two values
can be explained by the difference in the radius at which the entropy has
been calculated. This is effectively the centre in the former case, but is
20 kpc for the latter. 

\begin{figure}
\centering
 \includegraphics[width=0.48\textwidth,keepaspectratio]{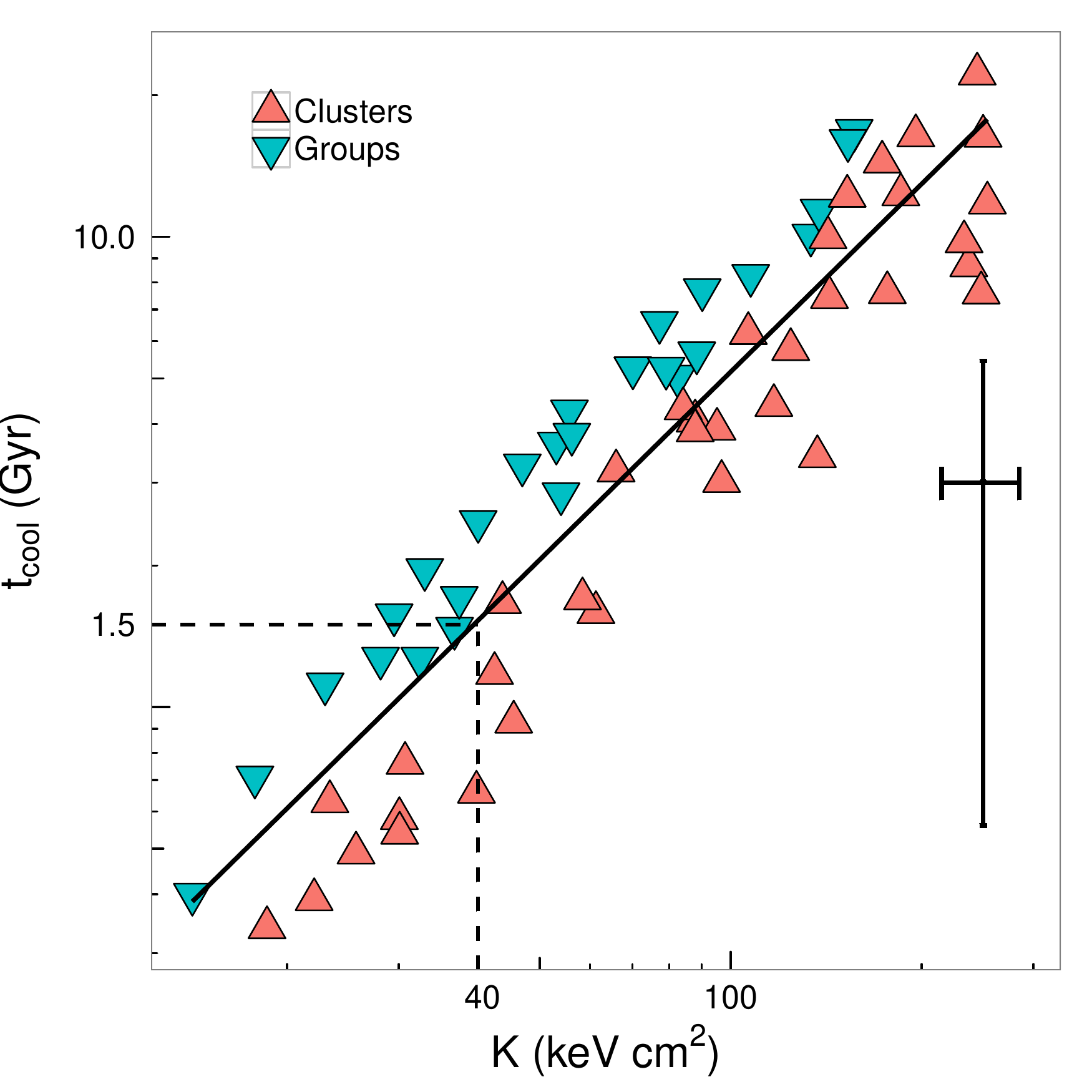}
    \caption{Relation between cooling time and entropy, both calculated at 0.01R$_{500}$, for our sample. Black solid line represents the best fit for all systems in the sample. Dashed lines mark the thresholds for K and \tcool\ used to separate CC from NCC systems. The error bars on the black point represent the median error for \tcool\ and K. These are $1\sigma$ errors.}
   \label{figTK}     
\end{figure}

Since our measurement is closer to the first of these, we adopt a cooling time threshold corresponding to
a central entropy of 40 keV cm$^{2}$, which lies within the 30-50 keV cm$^{2}$ interval from  \cite{Cavagnolo2009}. The tight correlation between our
cooling time and entropy values is shown in Figure~\ref{figTK}. Since
entropy scales as $T/n_e^{2/3}$, whilst cooling time scales (in the
bremsstrahlung regime) as $T^{1/2}/n_e \propto K^{3/2}/T$, we see that
there is some offset in the Figure between groups and clusters, such that
the gas in clusters has a rather shorter cooling time at given
entropy. Averaging over our sample, we adopt 1.5 Gyr as a sensible cooling
time threshold.

An important issue, highlighted in the recent study by \cite{McDonald2013},
is the distinction between the rate of current cooling and the amount of
gas which has been able to cool. \tcool\ is a measure of the former,
but for a cluster at high redshift less time has been available for
cooling to take effect. Since both current cooling and the accumulated
effects of cooling are of interest to us, we construct a further cool core
indicator, \TcoolTage, in which cooling time is divided by the age of
the Universe ($t_{\rm Uni}$) in our adopted cosmology, at the redshift
of the cluster. This represents the fraction of gas which
could have cooled in the lifetime of the cluster, in the absence of
AGN feedback. In practice, the impact of AGN feedback is believed to
suppress gas cooling by an order of magnitude \citep{McNamara2012}, but 
cannot prevent it altogether. In these circumstances, the integrated fraction of
a cluster's gas which could have cooled over its history should still scale
roughly with \TcoolTage, though the impact of cyclic AGN activity on the
cooling time in the core will introduce considerable scatter.

We calculate the threshold value for this parameter, separating CC from NCC systems, by dividing the threshold used for \tcool\ (1.5~Gyr) by the age of the Universe at the median redshift of systems from our sample (8.7~Gyr). This gives a threshold value for \TcoolTage\ of 0.17, which will be used below.

\section{Results}
\label{section:results}

\subsection{Cool core evolution}
\label{subsection:evolution}
\begin{figure*}\centering
\includegraphics[width=0.99\textwidth,keepaspectratio,trim={3cm 9cm 3cm 8cm}, clip=true]{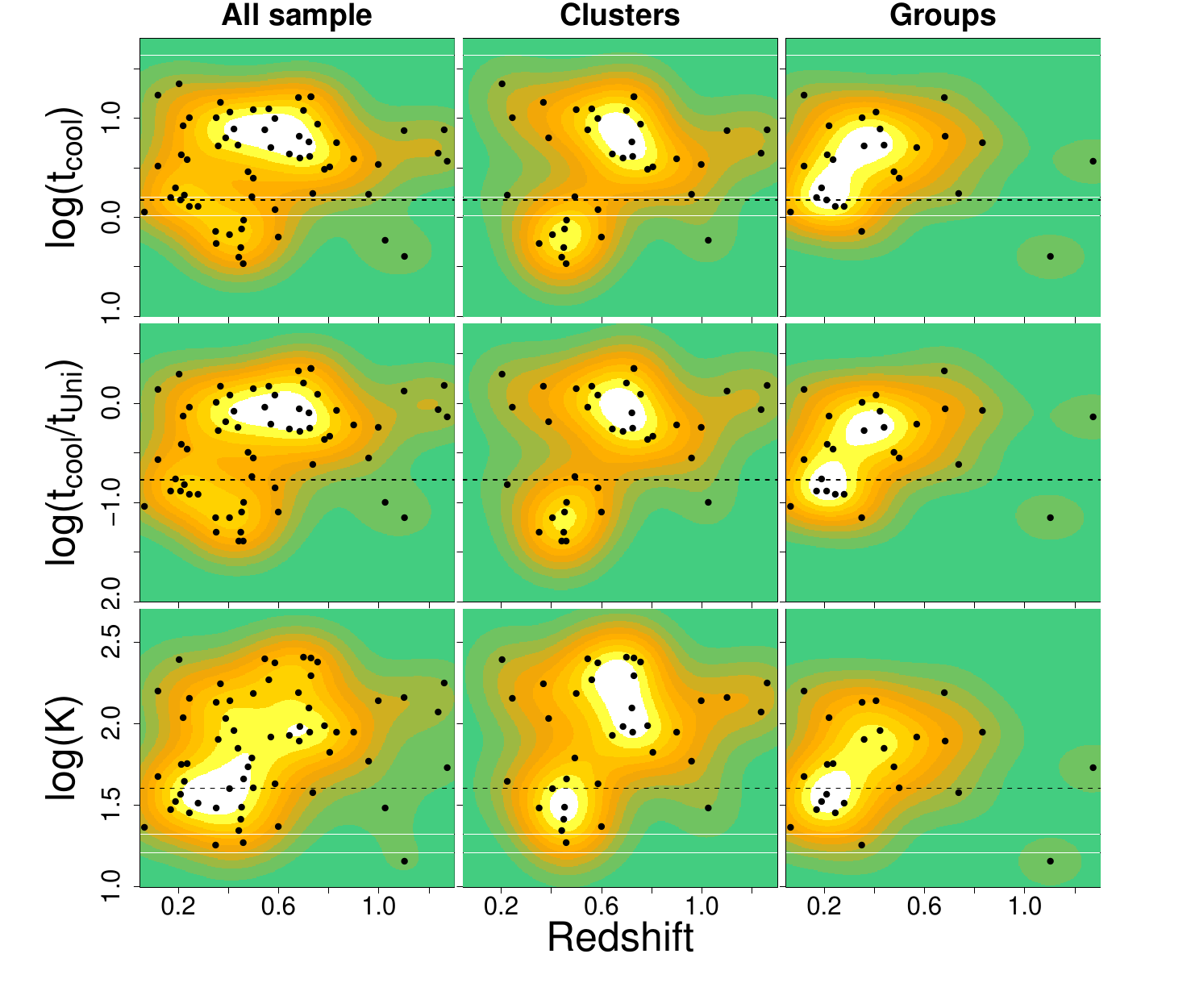}    
\caption{Distribution of different CC estimators with redshift: cooling time (top row), cooling time divided by the age of the Universe (middle row) and entropy (bottom row). For each parameter the distribution for all sample, clusters and groups is showed in the left, middle and right panel. Data points are showed as black dots and the contours represent number density contours. The dotted line represents the threshold between CCs and NCCs}
    \label{fig2}
\end{figure*}

The evolution of CC strength, as quantified by \tcool\ and
\TcoolTage, as well as entropy, all evaluated at radius 0.01\R,
is shown in Figure~\ref{fig2}.
For each parameter, we plot the results obtained when using the entire sample (left panel), a subsample which contains only clusters ($T \geq 3$~keV, middle panel) and one which contains only groups (right panel). This temperature cut allows us to compare the behaviour of evolutionary trends in the two mass regimes.

In each panel, black points represent the data, whilst the contoured colour scale
traces the smoothed density of points.
The black dotted line shows the threshold adopted for separating CC
from NCC systems (0.17 for \TcoolTage, 1.5 Gyr for cooling time and 40 keV cm$^2$ for
entropy). In each case, CC systems lie below the line.

A broadly similar pattern is seen in all three rows. Some
bimodality is apparent in the distribution for all three parameters. This
bimodality is more pronounced in the cluster sub-population, whilst in groups
the pattern is quite similar, but the CC and NCC peaks move closer together
and merge into a single elongated distribution.

Examining density plots such as Figure~\ref{fig2} is not a reliable way 
of establishing evolutionary trends. For example, the shape of the density
contours can be substantially modified by transformations of the axes
(plotting the cool core indicators in unlogged form, for example).
We have therefore tested for correlations of our CC indicators with redshift by
calculating the Spearman rank correlation coefficient. The results are
shown in the first three rows of Table~\ref{Spearman}. Values for our X-ray
selected sample occupy the left hand side of the Table. Corresponding
values for our `extended sample' will be discussed later, in 
Section~\ref{subsec:extended_results}.

\begin{table*}
\centering
\normalsize
\begin{minipage}{180mm}

    \caption{Statistical tests for redshift evolution of various cool core estimators. The correlation is quantified by Spearman's rank correlation coefficient which is given with associated p-value for each cool core parameter stated in the first column. Left hand side part of the table shows correlation test for X-ray selected sample presented in Table \ref{table:table2}, while the right hand side shows correlation for the extended sample which will be described in Section \ref{subsec:extended}. For each sample, correlation  is tested for the entire sample, clusters and groups. The first three rows present the correlation for cooling time normalized by the age of the cluster, cooling time and entropy, while last three rows present correlation for three cool core cuspiness parameters which are described in Section \ref{subsection:cuspiness}.   }
   \centering

\begin{tabular}{|l|r|r|r|r|r|r@{\hskip 1.4cm}|r|r|r|r|r|r|}
\cline{1-13}
& \multicolumn{6}{c|}{X-ray selected sample} & \multicolumn{6}{c|}{Extended sample}\\
\hline
  \multicolumn{1}{|c|}{Parameter} & \multicolumn{2}{c}{All sample} & \multicolumn{2}{c|}{Clusters} & \multicolumn{2}{c|@{\hskip 1.4cm}}{Groups} & \multicolumn{2}{c|}{All sample} & \multicolumn{2}{c|}{Clusters} & \multicolumn{2}{c|}{Groups} \\

 \multicolumn{1}{|c|}{} &
  \multicolumn{1}{|c|}{Coeff} &
  \multicolumn{1}{c|}{P-val} &
  \multicolumn{1}{c|}{Coeff} &
  \multicolumn{1}{c|}{P-val} &
  \multicolumn{1}{c|}{Coeff} &
  \multicolumn{1}{c@{\hskip 1.4cm}|}{P-val} &
  \multicolumn{1}{c|}{Coeff} &
  \multicolumn{1}{c|}{P-val} &
  \multicolumn{1}{c|}{Coeff} &
  \multicolumn{1}{c|}{P-val} &
  \multicolumn{1}{c|}{Coeff} &
  \multicolumn{1}{c|}{P-val} \\
\hline

  \TcoolTageb & 0.26 & 0.04 & 0.27 & 0.12 & 0.28 & 0.17 & 0.29 & 0.006 & 0.29 & 0.03 & 0.35 & 0.06\\
  \tcoolb & 0.07 & 0.58 & 0.12 & 0.49 & 0.14 & 0.49 & 0.11 & 0.35 & 0.14 & 0.30 & 0.19 & 0.32\\
  K & 0.26 & 0.04 & 0.22 & 0.20 & 0.17 & 0.41 & 0.29 & 0.006 & 0.24 & 0.07 & 0.22 & 0.25\\
  \csbb & -0.17 & 0.19 & -0.09 & 0.59 & -0.31 & 0.12 & -0.24 & 0.03 & -0.18 & 0.18 & -0.38 & 0.04\\
  \fcoreb & 0.04 & 0.74 & -0.07 & 0.70 & -0.15 & 0.45 & -0.06 & 0.57 & -0.16 & 0.24 & -0.26 & 0.17\\
  \fcb & -0.35 & 0.005 & -0.35 & 0.03 & -0.33 & 0.10 & -0.34 & 0.001 & -0.29 & 0.03 & -0.36 & 0.05\\
\hline
\label{Spearman}
\end{tabular}

\medskip

\end{minipage}
\end{table*}

The Table gives the values of the correlation coefficient for a trend
in each CC indicator with redshift. Being a rank correlation coefficient,
this is independent of any monotonic transformation of either axis.
For each coefficient, the chance probability (2-tailed) of obtaining a value
deviating from zero by this value or more is also quoted.

As can be seen, a significant trend ($p=0.04$) is apparent in 
\TcoolTage\ (and to a lesser extent in entropy) for both the full (cluster $+$ group)
sample, and for clusters alone. The group subsample shows a correlation coefficient
of similar size, but this is less significant, given the smaller number of
systems involved. However, the \tcool\ indicator shows {\it no} significant
trend with redshift.

\begin{figure}\centering
 \includegraphics[width=0.48\textwidth,keepaspectratio]{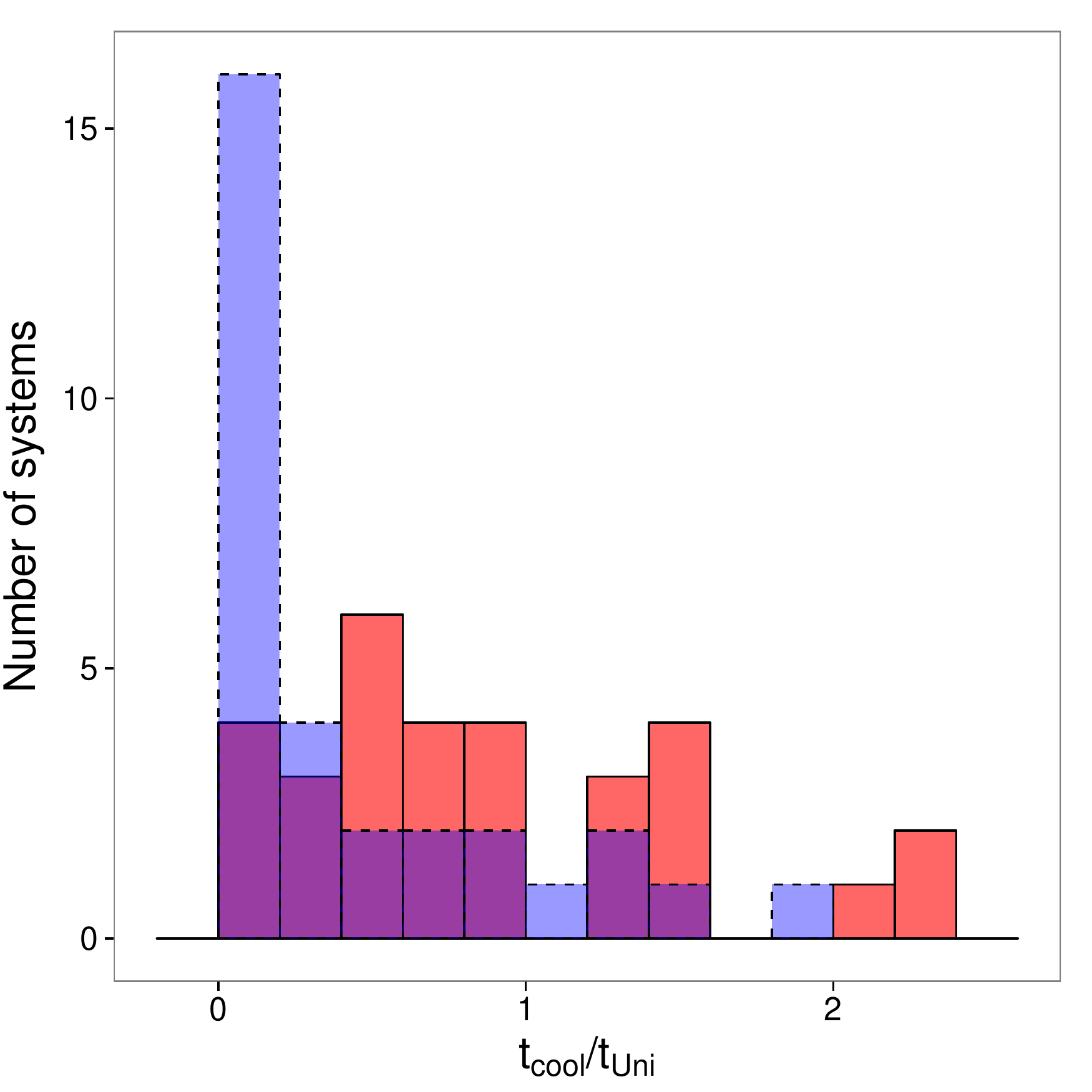}

    \caption{Distribution of \TcoolTage\ for the low (blue, dashed line) and high (red, solid line) redshift systems. The redshift threshold used to divide between these two subsamples is 0.5.}
    \label{fig3}
\end{figure}

As a further test, we examine the distribution of our two main CC indicators
across the sample at low and high redshift, and apply a Kolmogorov-Smirnov (K-S)
test to see whether they differ.
We choose a redshift cut at 0.5 to separate the low and high redshift
samples for this test, motivated by previous results in the literature
which report a change in the properties of CCs at redshifts greater
than 0.5 \citep{Vikhlinin2007}. However we have tested various
redshift thresholds and find similar results for any cut between 0.5
and 0.65. For \TcoolTage\ we find a highly significant difference ($p=0.009$)
between the distributions at high and low redshift. As shown in 
Figure~\ref{fig3}, our low redshift systems are more strongly concentrated
towards low values of \TcoolTage, confirming the redshift trend indicated 
by the Spearman rank analysis. Performing a similar analysis for \tcool\
we find a much weaker difference between the high and low redshift
distributions, though it can still be significant, depending on the 
value of the redshift cut. We will return to this with our
larger `extended sample' in Section~\ref{subsec:extended_results}
below.

Returning to the interpretation of our two main CC indicators as
representing {\it current} cooling (\tcool) and {\it accumulated}
cooling (\TcoolTage) in the core, our conclusion at this stage
seems to be that the latter is evolving, whilst the former is not.
However, before we can draw such a conclusion, we need to examine
the possibility that the trends we see could be driven primarily by
a changing composition in cluster richness with redshift, rather
than evolution in properties for clusters at a given richness.
Despite the rather similar behaviour in clusters and groups seen
in Figure~\ref{figTK}, it is well known that groups have gas properties
which differ systematically from richer clusters -- with flatter surface 
brightness profiles \citep{Ponman1999} and more compact central
cooling regions \citep{Rasmussen2007}.

\begin{figure}\centering
 \includegraphics[width=0.48\textwidth,keepaspectratio]{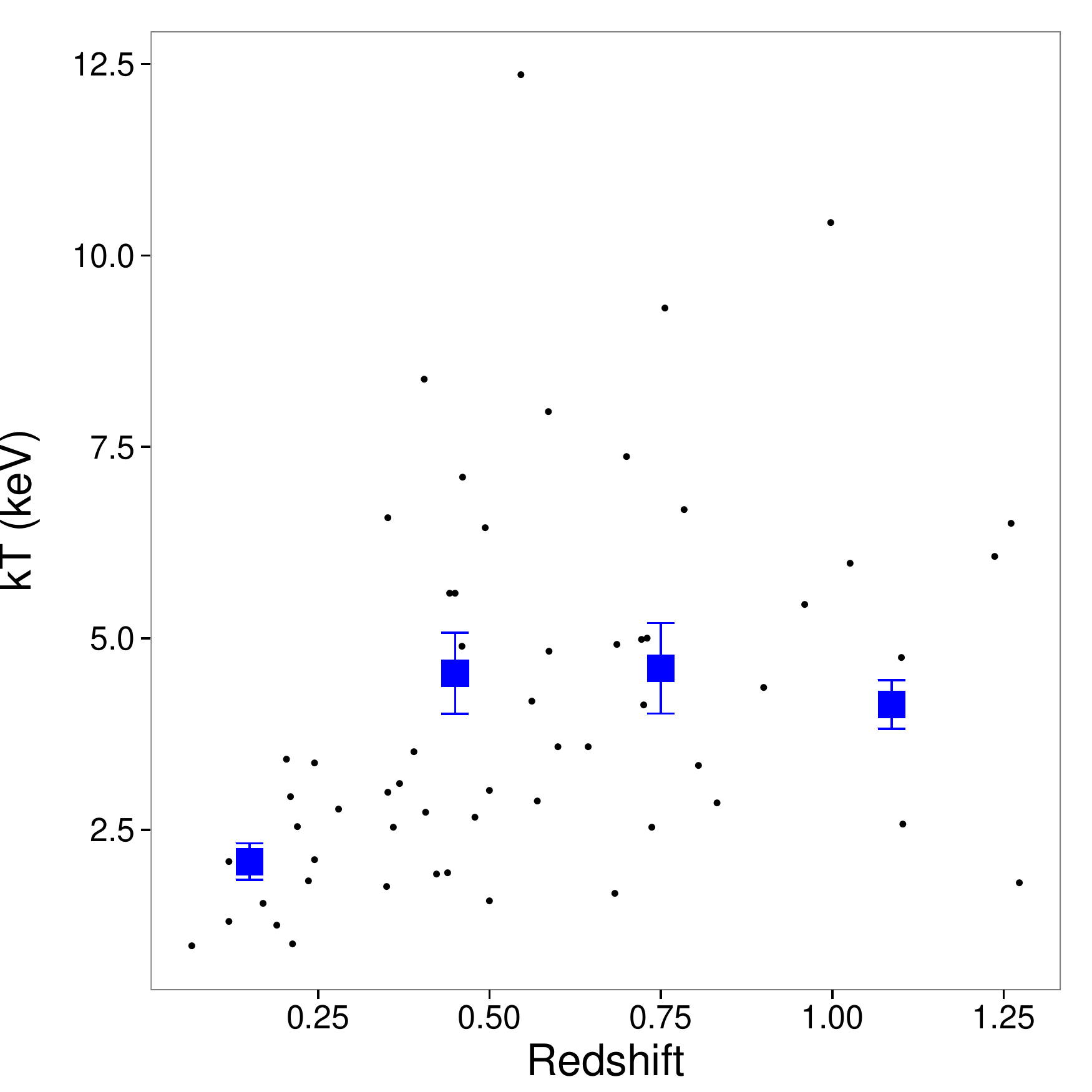}

    \caption{Relation between temperature and redshift for all systems in our sample. Black points marks individual systems while blue squares represent the mean temperature in four different redshift bins: 0-0.3, 0.3-0.6, 0.6-0.9, 0.9-1.27}

    \label{figTz}
\end{figure}

In Figure~\ref{figTz} we examine the distribution in system temperature
with redshift within our sample. As expected, the galaxy groups
($T<3$ keV), which are less luminous X-ray sources, are concentrated
towards lower redshifts. However, interestingly this effect is
largely confined to $z<0.35$, and above this redshift, the mean
temperature of our sample is essentially constant, at around 4.5~keV.
We have already seen that our conclusions about the trend in
\TcoolTage\ and the lack of evolution in \tcool\ apply even if we exclude
groups from our analysis. If we instead retain the full temperature
range, but exclude all systems with $z<0.35$, a positive correlation
(coefficient=0.21) remains, but its significance is reduced, due to
the smaller sample and reduced redshift baseline.
We conclude that our results are {\it not} being driven by 
redshift-dependent temperature biases in the sample.

\subsection{Cuspiness cool core indicators}
\label{subsection:cuspiness}
As we discussed earlier, most previous studies of cool core evolution
have been based on an analysis of central surface
brightness excess. We now apply some of these estimators to our
own sample, for comparison with our above findings based on
cooling time, and with results of earlier studies.
We use three CC estimators defined in the literature: surface brightness
concentration(\csb; \cite{Santos2008}), the core flux
ratio (\fcore; \cite{Maughan2012}) and the central excess
factor(\fc; \cite{Alshino2010}), for which we employ the same symbols
as the original authors.

The \csb\ parameter is defined as the ratio between the flux measured within circular apertures with radii of 40 kpc and
400 kpc, centred on the peak of the cluster X-ray emission. 
These radii were found by \cite{Santos2010} to optimize the separation between CC and NCC in a sample of simulated low redshift
clusters. They motivated the use of a fixed physical radius rather than a
fraction of the scale radius, \R, by the fact that cool cores are the
result of non-gravitational processes and therefore their sizes do not
scale self-similarly. In their study, \cite{Santos2010} used \csb\ to
divide the sample into strong (SCC), weak (WCC) and non cool core (NCC)
classes, with \csb$>0.155$, $0.075 \leq$\csb$\leq 0.155$, and \csb$<0.075$,
respectively.

Similar to the \csb\ parameter is the \fcore\ parameter, which is defined also as a flux ratio, but with aperture radii defined as fractions of $R_{500}$ instead of fixed physical sizes. Following \cite{Maughan2012}, \fcore\ is taken to be the ratio of flux within $0.15R_{500}$ to that within $R_{500}$. We add that while the definition of this parameter is similar to the one used by \cite{Maughan2012}, the way in which we calculate the fluxes is based only upon imaging data, whilst \cite{Maughan2012} calculate the unabsorbed flux from spectra extracted within each aperture. If the core flux is greater than half of the flux within $R_{500}$ (i.e. \fcore$>0.5$), the system is characterized as a CC.

While \csb\ and \fcore\ are simple parameters which do not require any modelling of the data, the \fc\ parameter of \citet{Alshino2010} quantifies the strength of a CC using the central excess in surface brightness profile above a fitted beta model with a fixed core radius of $0.105R_{500}$. This core radius was chosen by \cite{Alshino2010} to correspond to the observed size of cores seen in the group-scale emission of well-resolved low redshift groups of galaxies by \cite{Helsdon2000}. A CC is deemed to be present if the ratio (\fc) of the observed flux within $0.05R_{500}$ to the corresponding flux derived from the fitted beta model (with core radius of $0.105R_{500}$) is greater than unity.

\begin{figure}\centering

 \includegraphics[width=0.45\textwidth,keepaspectratio]{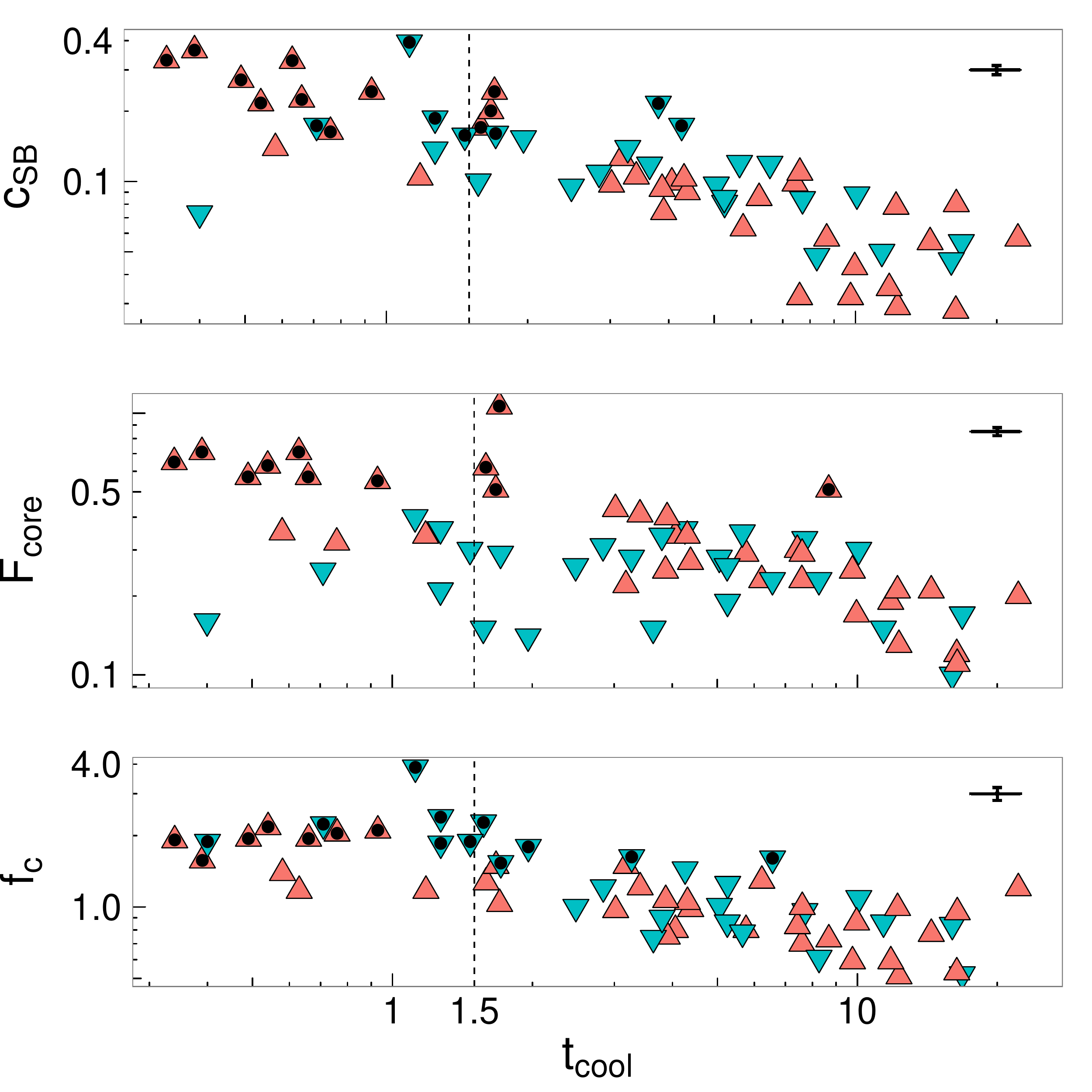}

    \caption{Correlation between \tcool\ and three cool core
estimators based on cuspiness in surface brightness: \csb, \fcore and \fc. In each panel, the symbol style and colour differentiates between groups (triangle point-down) and clusters (triangle point-up). Symbols which include a filled black circle are classified as CCs according to the surface brightness parameter represented on the y-axis. The dashed vertical line marks the threshold between CC and NCC for \tcool.}
    \label{figIndicators}

\end{figure}

All three of these CC indicators have been found by their proponents to show evolutionary trends, so we investigate their relationship with our \tcool\ indicator. Figure \ref{figIndicators} shows in each panel the correlation between \tcool\ and the three surface brightness based CC estimators. Different symbol styles and colours differentiate groups and clusters, and the presence of a central black point denotes systems characterized as CC according to the y-axis parameter. (For \csb\ we use the SCC criterion.) We have marked on the x-axis the value \tcool=1.5 which is our adopted CC threshold.

Note that, in contrast to our calculation of \tcool\ and \TcoolTage, no correction for any central AGN has been applied when calculating
the surface brightness cuspiness indicators. Hence clusters with a bright central AGN will be biased towards showing CC properties.
As we discuss later in Section~\ref{subsection:detbias}, the indications
are that AGN contamination is not a major problem in our sample.

Firstly, ignoring the distinction between groups and clusters, it can be seen that the best correlation with \tcool\ is found for \csb. We have calculated the Spearman coefficient for all parameters and find the highest coefficient for \csb\ (-0.82), closely followed by \fc\ (-0.77), while the lowest correlation is found for the \fcore\ parameter (Spearman coefficient of -0.60).

The correlation between \csb\ and \tcool\ is much stronger than that between \fcore\ and \tcool, although both parameters are defined as the flux ratio between the core and the bulk of the system, the only difference being in the sizes adopted for the inner and outer regions. Before drawing any conclusions about this discrepancy, we remind the reader that each CC indicator has been optimized to be applied to samples dominated by either clusters (\csb\ and \fcore) or groups (\fc), while our sample includes both types of system. Therefore we compare the performance of each parameter on the system class for which it has been designed.

Applying \csb\ and \fcore\ to just our cluster subsample, we see that both parameters give similar strong correlations: -0.85 (\fcore) and -0.89 (\csb). However, for the group subsample, there is a large discrepancy in the correlation coefficients: -0.64 for \csb\ and -0.17 for \fcore. The poor correlation seen for \fcore\ in the case of groups can be explained by the large size of the radius chosen to characterize the core region ($0.15R_{500}$). For clusters, this is approximately the size of the cool core, when it is present, whilst in groups cores are smaller, extending to a radius of typically only $0.1R_{500}$ \citep{Rasmussen2007}.

Comparing the symbols marked by black circles in Figure~\ref{figIndicators}
with the position of the vertical dashed line, we can examine the
fraction numbers of CC systems amongst groups and clusters identified by
each method. The \csb\ indicator shows excellent agreement with 
\tcool\ when applied to clusters, whilst for groups it identifies a
similar total number of CC systems, but not necessarily the same ones.
For \fcore, the clusters characterized as CC are again similar to 
those identified by \tcool, but not a single group is classified as a CC.
\fc\ identifies fewer CC clusters than \tcool, but includes some groups
with rather long cooling times as being CC systems. 

\begin{figure*}\centering

\includegraphics[width=0.99\textwidth,keepaspectratio,trim={3cm 9cm 3cm 8cm}, clip=true]{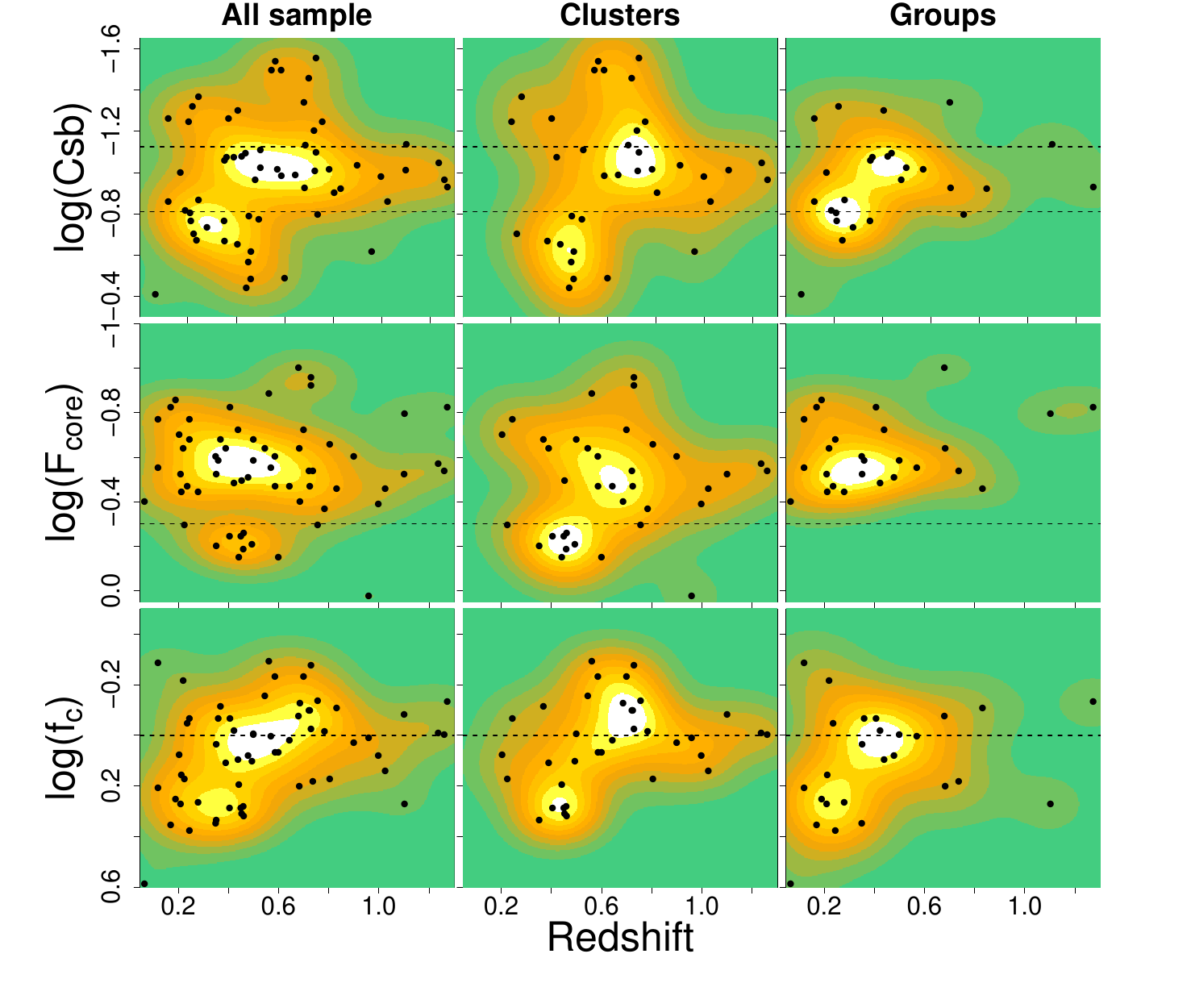}    
    \caption{Redshift distribution for three different cool core estimators defined in the literature based on the surface brightness excess: $c_{\rm SB}$, $F_{\rm core}$ and \fc. As in Figure~\ref{fig2}, left column of panels corresponds to all sample, middle panel to clusters and right one to groups. In each plot, the black horizontal line divides the sample into different classes according to their cool core strength. For $c_{\rm SB}$, the two lines at $c_{\rm SB}$=0.075 and $c_{\rm SB}$=0.155 divides sample into: NCCs, WCCs, and SCCs, while  a value for $F_{\rm core}$=0.5 and \fc=1 divides clusters into CCs and NCCs. For comparison with Figure \ref{fig2} we have used reversed axes for CC parameters so that CC systems lie at the bottom of each plot, as in Figure~\ref{fig2}.}

    \label{figIndEvol}

\end{figure*}

In Figure~\ref{figIndEvol} we plot the distribution of the surface brightness 
based CC estimators against redshift, in a similar fashion to Figure \ref{fig2}. 
Note that for these three estimators {\it high} values correspond to strong cool 
cores, in contrast to our three previous estimators. We have therefore flipped 
the y axis scales so that core dominance still increases downward on each plot. For 
the \csb\ plot, the two horizontal lines correspond to the two thresholds used
by \citet{Santos2008}, dividing clusters into SCC (bottom), WCC (middle) and 
NCC (top) classes.

The distributions for all three indicators show similarities with our
earlier cooling time and entropy based parameters. In particular, all
show some signs of bimodality, at least for clusters. 
In the case of \csb, there
is the wide variety in the CC strength at low redshifts,
while for redshifts greater than 0.6 the NCC and SCC classes largely disappear,
leaving only WCC systems. \fcore\ shows a similar pattern of narrowing
towards intermediate core strength at $z>0.7$, whilst \fc\ shows less
symmetrical behaviour. 

Results from Spearman rank tests for correlation with redshift are shown
in the bottom three rows of Table~\ref{Spearman}, and confirm the visual impression
from Figure~\ref{figIndEvol}. For the X-ray sample (left hand side of 
Table~\ref{Spearman}) only \fc\ shows a significant evolutionary trend.
This correlation (negative, due to the reversed sense of the indicator compared
to the physically based indicators shown in the first three rows of the Table)
is apparent for clusters and groups individually, as well as for the
combined sample.

\subsection{Systems with photometric redshift}
\label{subsection:photo-z}
The majority of the sample used in this study (presented in 
  Table \ref{table:table1} and Table \ref{table:table2}) consists of
  groups and clusters, detected as extended sources in X-ray images,
  whose nature is confirmed through spectroscopic redshifts of galaxy
  members. However, for 23$\%$ (14 out of 62) of the sample no spectroscopic
  redshift was available in the literature, and the redshift used in our
  analysis is photometric.

While the accuracy of cluster photometric redshifts is
  typically at a level of $\sim ∼ 0.02 - 0.05$ out to redshift of 1
  \citep{Bahcall2003, Koester2007,Pello2009, Takey2013}, which is
  perfectly adequate for our purposes, occasional `catastrophic errors'
  in photometric redshifts can be up to an order of magnitude higher
  \citep{Mullis2003,Koester2007,Pello2009}. Moreover, in the
  absence of spectroscopic confirmation that associated galaxies
  are really clustered in redshift, the identification of a cluster
  must be regarded as provisional.

We have therefore examined the effects of excluding the systems
  with photometric redshifts from our analysis. This produces no significant
  difference in our results. The nature of the trends seen
  do not change, but some become rather stronger. The most noticeable
  differences are found for the evolutionary trends in the cluster
  subsample for \TcoolTage\  (Spearman's rank coefficient of 0.40; p-val=0.02), 
  \tcool (coefficient=0.25; p-val=0.16)
  and K(coefficient=0.34; p-val=0.06), which can be compared with the values
  in Table~\ref{Spearman}.
  In addition, the trends seen
  in \csb for the full sample (coefficient=-0.25; p-val=0.07) and the group
  subsample (coefficient=-0.44; p-val=0.06) become more significant.

\section{Selection biases and AGN contamination}
Before drawing conclusions about the evolution of CCs in groups and clusters of galaxies we must consider first whether any differences
seen between the core properties of high and low redshift systems 
might simply result of the way in which our sample has been selected.

\begin{figure}\centering
 \includegraphics[width=0.45\textwidth,keepaspectratio]{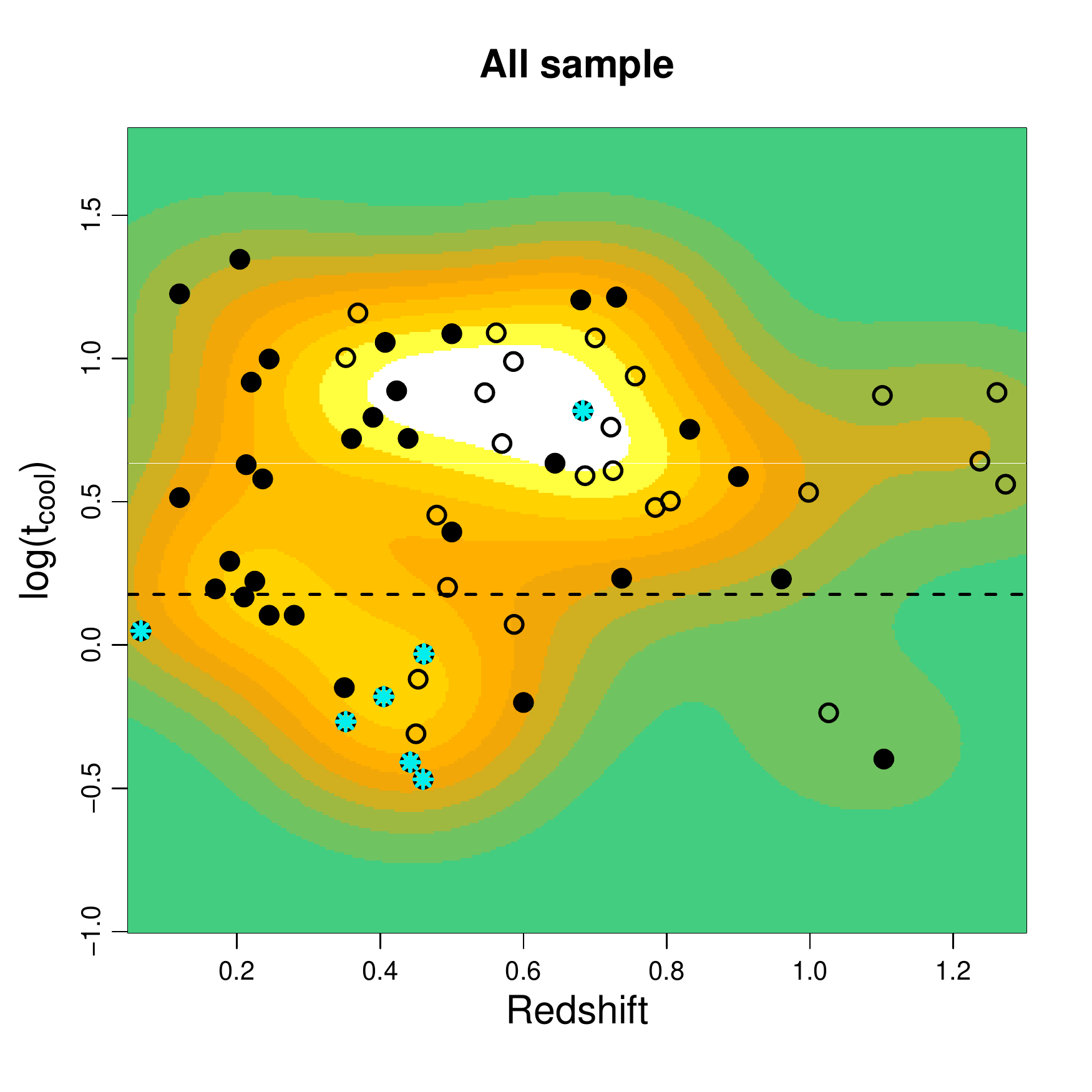}
    \caption{Same notations as in Figure \ref{fig2} but with different symbols representing serendipitous systems (filled circles) and target systems (open circles). Systems marked with a cyan asterisk are those which are contaminated by a central AGN that has been masked during our data analysis.  }
    \label{fig:PtSer}
\end{figure}

Our X-ray selected sample, constructed from extended sources
detected in Chandra archival observations which meet the criteria
mentioned in Section~\ref{secData}, contains two classes of systems: (i) groups and
clusters which represent the target of the Chandra observation,
and (ii) serendipitously detected sources. 
Figure~\ref{fig:PtSer} shows the \tcool\ distribution plot for the full
X-ray sample, marking targetted and serendipitous sources with open and filled 
symbols respectively. It can be seen that targetted sources account for the
majority of the sample at $z>0.7$.

The inclusion of deliberately targetted sources in our sample might introduce bias in favour of systems with a particular morphology or special properties, since these systems may have been observed because of these characteristics. While this kind of bias, known as {\it archival bias} affects only the non-serendipitous sources, a bias to which both types of systems are subject is {\it detection bias}. This is due to the effect of source properties on the efficiency with which they can be detected in an X-ray image. We now examine both these sources of bias in turn.

\subsection{Detection biases}
\label{subsection:detbias}

When constructing an X-ray selected sample of clusters, the probability that a system with a given flux and size will be included in it depends on the source detection efficiency and the ability to characterize the detected system as extended when compared to the telescope's point spread function.

As detection probability is a function of both the flux and spatial distribution of the X-ray emission, a different detection efficiency may be expected for sources with different intrinsic properties such as core size \citep{Eckert2011}, substructure, and the presence of intracluster point sources \citep{Vikhlinin1998a, Burenin2007}. For a given source flux, the detection probability may be increased by concentrating more of the flux in the core, until the concentration becomes so great that the cluster is rejected as appearing point-like.

Such an effect could, for example, account for the narrowing in 
core strength seen with the \csb\ indicator at high redshift, if
our detection method preferentially excludes systems
with very large and very small core radii.

One way to check this hypothesis is by answering the following
question: supposing that strong CC and NCC systems are common at high redshift, would we be able to detect such clusters with a flux value 
corresponding to our threshold limit of 100 soft band counts?

To answer this question we applied our detection algorithm to simulated observations of a high redshift CC and NCC cluster respectively. Observations were generated using the Chandra simulation software (MARX), which requires as input information about the system's spectral properties and its spatial distribution, in the form of a spectrum and values for beta model parameters, respectively. We base the properties on an observed high redshift cluster, but perturb its surface brightness distribution to generate an extreme CC and NCC system. Our template system is the cluster from our sample (Table~\ref{table:table1}) detected in CLJ1415.1+3612 field (CDGS57). This is a $\sim 6$ keV system at redshift 1.03. This provides the template for our input spectrum to MARX. For the spatial properties, we use the beta and normalization derived from our fit to the surface brightness profile of the CLJ1415.1+3612 cluster, but we perturb the core radius -- to $0.007R_{500}$ to represent a SCC and $0.3R_{500}$ for a NCC 
profile, where $R_{500}$ is the 
overdensity radius of our template system. These two values for core radius represent the median values for the size of core radii as a fraction of  $R_{500}$ for the low redshift ($z<0.3$) CC (\tcool$<1.5$~Gyr) and strongly NCC (for which we adopt \tcool$>7$~Gyr) systems in our sample. Having chosen the spectral and spatial parameters, we varied the exposure time of our simulations to obtain 100 soft band counts, which represents our threshold limit for source selection. 

So far, these simulated observations do not include any contribution from the background, which will degrade the source detection probability. To account for this, we added our simulated images to the observed image of our template cluster. Having the background level and spectral properties of a real detected system, we can now test if detection would still be possible in the case of CC and NCC cases. When we applied our detection and extension test procedure, we were able to reliably detect as extended sources both the CC system and the NCC one. This indicates that at our 100 count limit, the sample is not significantly affected by biases in detection efficiency due to the size of the core. Had we included much fainter sources in our sample, this would undoubtedly not have been the case. 

Another potential source of detection bias is the presence of intracluster point sources, especially central AGN which have a double influence on the detection efficiency. In the first case, point sources embedded in the intracluster medium can cause a positive bias, increasing the detection efficiency due to the central flux excess they add to the surface brightness distribution. On the other hand, there can be a negative bias if a bright AGN at the centre of a cluster dominates the cluster emission and leads to a misclassification of the cluster as a point source. \cite{Burenin2007} showed that the detection efficiency of a cluster varies in the presence of a central AGN according to the 
luminosity ratio between the AGN and the intracluster medium. The detection efficiency is raised if an AGN with a luminosity much less than that of the cluster is present. However, if the luminosity of the AGN dominates the cluster emission, the detection efficiency drops dramatically.

Our procedure for identifying central AGN was described in Section~\ref{subsection:SBP},
and 7 cases fell into our AGN `class 1', in which
we were able to remove the central point source and 
analyse the cluster containing it. These systems are flagged with asterisks
in Figure~\ref{fig:PtSer}. A strong
connection between the presence of a central AGN and CC status is apparent
-- most systems with a central point source are CCs.
\citep{Stott2012} showed that radio loud BCGs are more likely to
be found in more massive systems and at the centre of CCs. Also, based
on the observed correlation between the strength of CC and the radio
power of the central AGN \citep{Mittal2009}, we might expect that, at
least for clusters, high redshift systems dominated by strong AGN will
be SCCs. Is it possible that this has introduced a bias against
their inclusion in our sample?

The literature is limited in the number of X-ray studies of high redshift
clusters with dominant central AGN. Two which have been studied are
PKS1229-021 \citep{Russell2012} and 3C186 \citep{Siemiginowska2010}. Both 
lie at $z>1$ and have been reported to contain a strong CC. 
Since these two systems were observed with ACIS-S, they were not included
in our sample, which concentrated on ACIS-I observations. 
We have analysed the Chandra
data for these sources and checked into which of the previously mentioned 
AGN classes they would fall, had they been part of our sample. 
They would fall into our first AGN class -- sources with clear signs
of extension in which the central AGN does not dominate the total
flux. We conclude from this that, at least for massive systems
detected in observations with exposures of at least 70 ks like ours,
we are not strongly biased against CCs. This may not be the case for 
less massive systems.  

\begin{figure}\centering
 \includegraphics[width=0.45\textwidth,keepaspectratio]{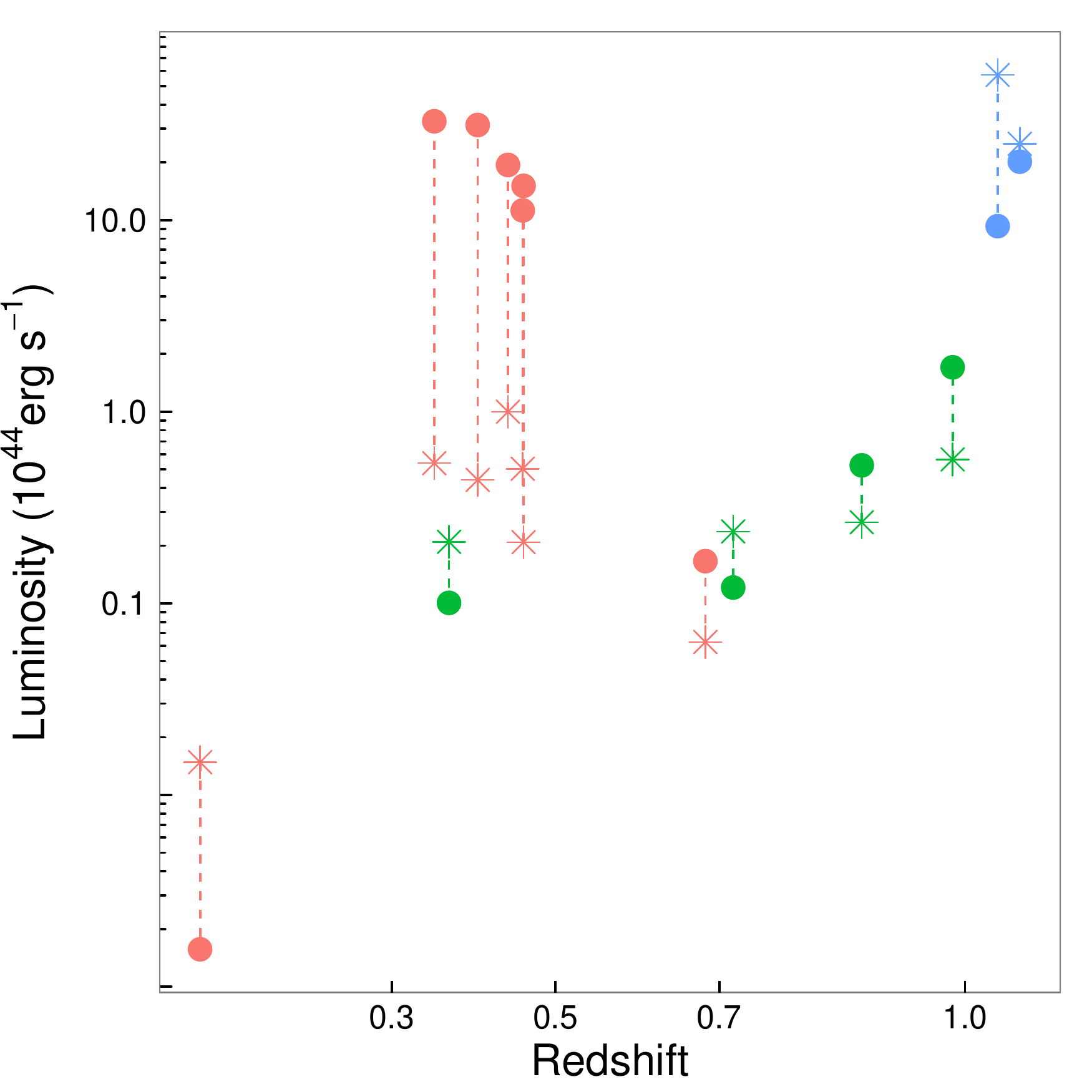}

    \caption{X-ray luminosity (0.5-7.0 keV) of the cluster (filled circles) and central point source 
(asterisks)  as a function of redshift for: (a) sources in our sample from which a
central point source that has been removed during the analysis (red); 
(b) extended sources detected in our fields and which have not been included 
into our sample because their X-ray flux is dominated by the central point source 
-- for these sources evidence for the existence of a cluster has been found 
in the literature (green); (c) the PKS1229-021 and 3C186 systems (blue). }

    \label{figAGN}

\end{figure}

To further examine the impact of central AGN on our results, we show
in Figure~\ref{figAGN} the X-ray luminosity of the cluster and AGN emission
in sources which appeared from our analysis to contain both point-like and 
extended components, and are confirmed from the literature to 
involve both an AGN and a cluster. 

Points marked in red correspond to the AGN (asterisk
symbols) and cluster (filled circles) contributions to the 7 systems
in which we were able to remove the AGN component and still perform
a useful analysis on the remaining cluster emission. The green points
correspond to clusters which were excluded from our sample, since the
remaining cluster component after removal of the central point source
did not leave enough signal/noise for a reliable analysis. 

Finally, we also mark (blue labelled symbols) the location of PKS1229-021 
and 3C186. The luminosities here have been estimated by fitting
a point source plus beta-model distribution to the X-ray surface
brightness distribution. For the green points, where the cluster contribution
is weak, the cluster luminosities should be regarded as rough
estimates.
 
Most of the systems (red points) in which we have been able to successfully remove
AGN contamination contain AGN which are less luminous than the cluster gas. 
The only exception to this is the lowest redshift system, which is
a nearby galaxy group ($T\approx 1$~keV) with a correspondingly large
X-ray extent. The clusters in which we
were unable to perform a useful analysis after removing the central AGN
(green points) have AGN which are brighter than the cluster, apart from the
two systems at $z=0.8$-1.0. These two are both observed at large off-axis
angles, where the instrument point spread function is broader, and 
the central AGN contaminates a region about $20\arcsec$ in diameter.

In general, our results suggest that the problem of AGN contamination
is a modest one in our sample. At low redshift ($z<0.5$), we find that 
about 19\% of our detected clusters contain central X-ray AGN, and in
most of these the AGN contributes less than 10\% of the cluster luminosity.
With the exception of PKS1229-021 and 3C186, which were not part of
our sample and were specially added to examine the case of powerful
central AGN at high redshift, there is little sign in Figure~\ref{figAGN}
that the luminosity of central AGN is increasing at redshifts
above 0.3, in which case only systems with cluster luminosities
$L_X \ltsim 10^{44}$~erg~s$^{-1}$
are likely to be lost from our sample due to AGN contamination.
The limited impact of AGN is confirmed by the results of Santos and
McDonald (private communication) who found the impact of central
X-ray point sources in their cluster samples to be modest.

\subsection{Archival biases}
The inclusion of targetted systems introduces biases which depend
upon the motivation of the observers who proposed these targets.
It is very difficult to decide how serious such biases might be, or in
which direction they might act, except that one would expect
{\it exceptional} objects to be especially popular targets.
The obvious way to avoid archival bias is by limiting the sample to
serendipitous sources, though the avoidance of targetted clusters
will introduce a certain bias in itself.
Although we might like to include in our
study only serendipitously detected systems, the lack of high redshift
serendipitous sources motivates us to include targetted systems in order
to improve the statistics available for evolutionary studies. 
It is clear from Figure~\ref{fig:PtSer} that including only serendipitous
sources, it will be difficult to draw conclusions about CC evolution.

We note from Figure~\ref{fig:PtSer} that most targetted sources at $z>0.7$ 
are WCC systems. This suggests that if an archival bias exists, it is towards systems 
with weak cool cores. This seems rather unlikely, since observers tend
to target interesting clusters, which would be expected to favour
dynamical disturbance (hence probably NCC) or strong AGN activity
(strong CC).

\subsection{Non X-ray selected clusters}
\label{subsec:extended}
The discussion above suggests that detection bias is unlikely to be a
serious problem for our survey, in which we require a minimum of 100
X-ray counts from each accepted source. AGN contamination does not
generally lead to the exclusion of luminous X-ray clusters from our
sample, but might affect systems with $L_X\ltsim 10^{44}$~erg~s$^{-1}$.
The influence of archival bias, especially at high
redshift, is difficult to assess due to the low number of high redshift
systems and the shortage of serendipitous ones. If we look at the
provenance of our high redshift systems we find that from 11 sources
detected at a redshift greater than 0.8, only four are serendipitous
systems. The other 7 systems represent the target of Chandra
follow-up observation of systems detected in earlier surveys at a variety
of wavelengths: near-infrared (1 system), Sunyaev-Zeldovich (SZ; 1 system), and two different ROSAT surveys
(WARPS; 1 system and RDCS; 4 systems). Since the majority of high
redshift sources come from ROSAT surveys, especially RDCS, we would
expect any bias in the RDCS sample to be reflected in our
sample. RDCS uses  a wavelet-based source detection algorithm
which is not expected to be substantially biased by the presence of 
a CC \citep{Rosati1995}. However, it is worth noting that the spatial
resolution of ROSAT is an order of magnitude poorer than that of Chandra.

In case there is some bias in X-ray properties arising from any of the
above factors, it is helpful to examine clusters selected in other
ways. To do this, and to improve our statistics at high redshift, we
added to our sample 24 systems with redshifts greater than 0.7, and
with at least 100 counts in the soft band, which result from Chandra
follow-up of groups and clusters selected from optical and SZ
surveys. These systems were not included in our initial sample for one
of three reasons: (a) they were observed for less than 70 ks, which
represents the lower limit adopted for our survey, (b) they were not
available in the archive at the time our sample was selected, or (c)
they were observed with the ACIS-S configuration, rather than ACIS-I.

\subsubsection{The South Pole Telescope (SPT) sample}
The South Pole Telescope survey \citep{Carlstrom2011} is a 2500 deg$^{2}$ 
survey that uses the distortion in the cosmic microwave background (CMB)
due to inverse Compton scattering of CMB photons by
electrons in the intracluster medium (Sunyaev Zeldovich effect) to detect galaxy clusters. From analysis of the first
720 deg$^{2}$ 224 galaxy cluster candidates have been found
\citep{Reichardt2013}. A significant number of SPT detected clusters
(52) have follow-up observations in the Chandra archive, and from these 
we have selected 17 clusters with redshifts greater than 0.7 and at least
100 soft band Chandra counts. Our SPT sample is presented in Table~\ref{SPT_RCS}. 
Detection of clusters using the SZ effect is not expected to be significantly biased 
by the dynamical state of the cluster or the presence of cool cores \citep{Motl2005}.

\subsubsection{The Red-Sequence Cluster Survey (RCS) sample}
The red-sequence method \citep{Gladders2000}
is a detection technique that exploits the observed tight correlation
between the colour and magnitude of the early-type galaxies in a
cluster. The RCS is a large optical imaging survey which uses the red
sequence method to detect clusters of galaxies out to redshift of 1. It
includes RCS1 \citep{Gladders2007} which covers an area of about 100 deg$^2$ and contains a sample of 429 cluster candidates, and
RCS2 which predicts the detection of ~30000 clusters from an area about 10 times larger \citep{Gilbank2011}. From these surveys, 21 clusters have been followed-up by
Chandra, from which we select only the 7 clusters at redshift greater than 0.7 and with at least 100 X-ray counts detected in ACIS-I observations. Since these clusters are optically selected, they are
free from any direct bias arising from their X-ray properties, including
the presence of AGN. Unlike SZ-detected clusters, which are invariably
massive systems, the RCS sample includes several high redshift groups.

\subsubsection{Results from the extended sample}
\label{subsec:extended_results}
The addition of 24 non-X-ray selected clusters doubles the number of high
redshift systems in our survey and creates what we refer to below as our 
{\it extended sample}.

\begin{figure*}\centering
\label{extended_sample}
 \includegraphics[width=0.98\textwidth,keepaspectratio,trim={2cm 10cm 2cm 10cm}, clip=true]{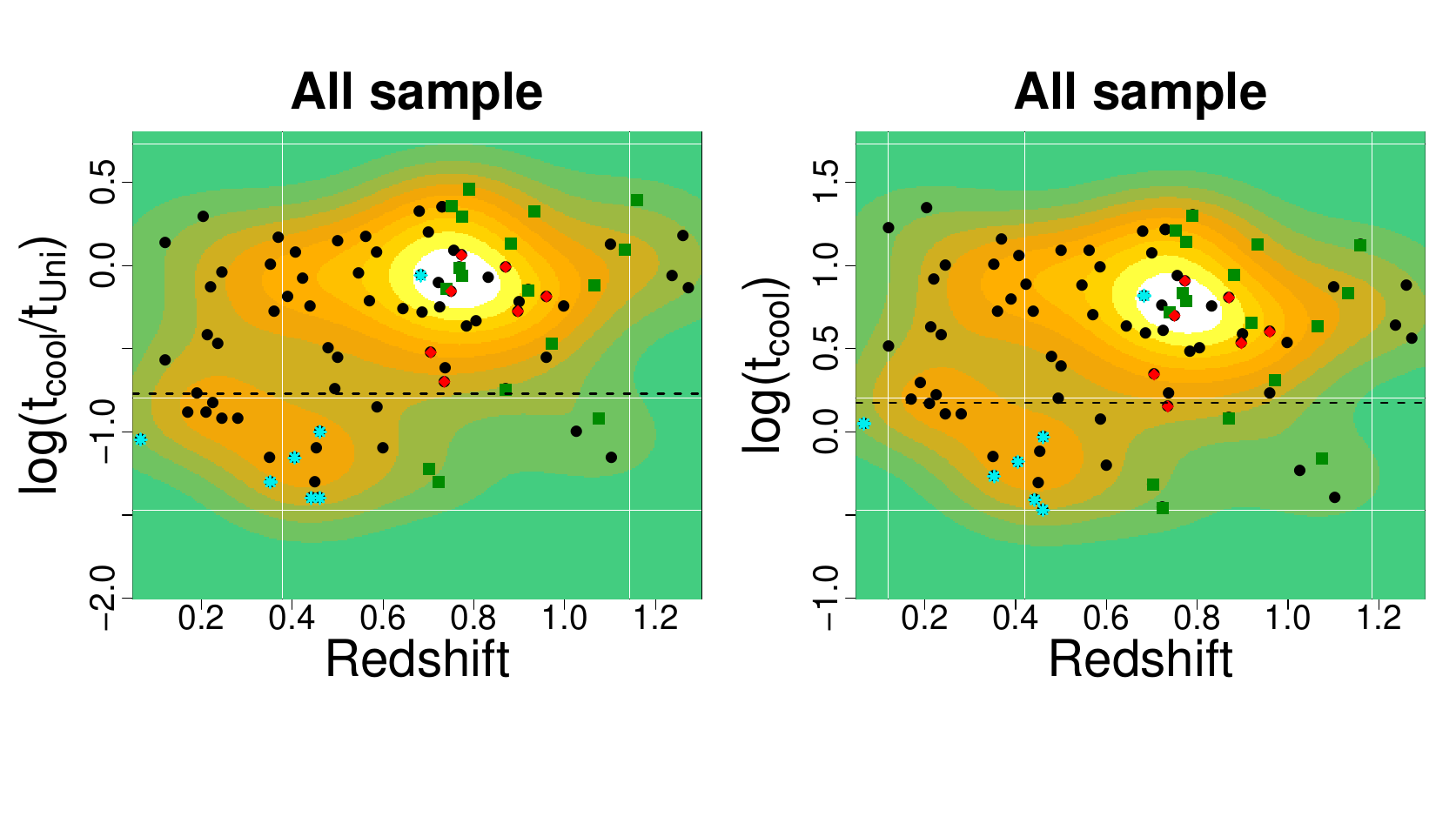}
    \caption{Evolution of \tcool and \TcoolTage for our extended sample, which includes the original X-ray selected sample to which we add 7 red-sequence selected systems (RCS sample) marked with red diamond symbols and 17 SZ selected systems (SPT sample) marked with green square symbols. AGN contaminated systems are marked with a cyan asterisk symbol. All other notations are the same as in Figure \ref{fig2}. }
    \label{fig:extended_sample}
\end{figure*}

Figure \ref{fig:extended_sample} shows the core evolution for our extended sample using
our two primary CC indicators: \tcool\ and \TcoolTage. SPT clusters are
shown in green, RCS clusters in red, and systems from which central
AGN have been removed are flagged with asterisks.
Examining the distribution of both \tcool\ and \TcoolTage\ in the extended sample, there
is an indication of some broadening of the high redshift distribution, especially towards
NCC systems. The spread in the NCC distributions at high redshift is
introduced by the existence of SPT clusters with very long cooling times.
This suggests that the shortage of such systems at high redshift in our
X-ray sample may be a selection effect. Possibly
NCC systems are underrepresented in the ROSAT surveys on which
most of our targetted high redshift observations were based.
It has already been noted in the context from Planck results
\citep{PlanckCollaboration2011} that SZ-selected clusters include a high
proportion of morphologically disturbed systems compared to X-ray selected
samples.

\begin{table*}
 \centering
    \caption{Non X-ray selected samples: 1) Clusters detected by the SZ effect using the South Pole Telescope and 2) Optically selected clusters detected based on the Red Sequence technique. }
           \label{SPT_RCS}
\small
\captionsetup[subfloat]{position=top,labelformat=empty}
\begin{subtable}[SPT sample]{.95\linewidth}
\caption{SPT sample}
        \begin{tabular}{{c@{\hskip 0.6cm \vspace{0.2 mm}} c@{\hskip 0.6cm \vspace{0.2 mm}} c@{\hskip 0.6cm \vspace{0.2 mm}} c@{\hskip 0.4cm \vspace{0.2 mm}} c@{\hskip 0.4cm \vspace{0.2 mm}} c@{\hskip 0.4cm \vspace{0.2 mm}} c@{\hskip 0.4cm \vspace{0.2 mm}} c@{\hskip 0.4cm \vspace{0.2 mm}} c@{\hskip 0.4cm \vspace{0.2 mm}} c@{\hskip 0.4cm \vspace{0.2 mm}}}}
	  \hline
   Field name & Ra & Dec & Redshft & Counts & R$_{500}$ & kT & \tcool & \TcoolTage & K \\
& (deg) &(deg) & & & (Mpc)  & (keV) & (Gyr) & & (keV cm$^{2}$) \\ [-0.0ex]
\hline
SPT-CLJ0001-5748 & 0.2500 & -57.8093 & 0.702 & 1226 & 0.981 & $ 8.01^{+3.61}_{-1.69}$ & $ 0.48 \pm 0.40$ & $ 0.06 \pm 0.05$ & $ 31.42 \pm 20.16$ \\
SPT-CLJ2043-5035 & 310.8242 & -50.5922 & 0.723 & 3957 & 0.797 & $ 5.26^{+0.27}_{-0.23}$ & $ 0.35 \pm 0.11$ & $ 0.05 \pm 0.02$ & $ 20.02 \pm 4.25$  \\
SPT-CLJ0324-6236 & 51.0483 & -62.5994 & 0.74 & 1249 & 0.888 & $ 6.55^{+1.45}_{-1.23}$ &$ 5.21 \pm 1.08$ & $ 0.72 \pm 0.15$ & $ 137.44 \pm 34.24$   \\
SPT-CLJ0014-4952 & 3.6921 & -49.8756 & 0.752 & 1600 & 0.951 & $ 7.56^{+1.80}_{-1.17}$ & $ 16.17 \pm 6.09$ & $ 2.26 \pm 0.85$ & $ 318.22 \pm 108.60$  \\
SPT-CLJ0528-5300 & 82.0216 & -52.9971 & 0.768 & 1203 & 0.777 & $ 5.14^{+1.09}_{-2.01}$ & $ 6.83 \pm 2.44$ & $ 0.97 \pm 0.34$ & $ 142.10 \pm 59.54$  \\
SPT-CLJ0000-5010 & 359.9323 & -50.1725 & 0.775 & 1447 & 0.925 & $ 7.98^{+2.00}_{-2.17}$ & $ 13.86 \pm 3.80$ & $ 1.97 \pm 0.54$ & $ 296.50 \pm 90.42$  \\
SPT-CLJ2337-5942 & 354.3574 & -59.7074 & 0.775 & 1205 & 0.986 & $ 8.26^{+3.77}_{-1.74}$ & $ 6.13 \pm 2.61$ & $ 0.87 \pm 0.37$ & $ 175.76 \pm 86.11$  \\
SPT-CLJ0449-4901 & 72.2773 & -49.0270 & 0.790 & 966 & 1.261 & $ 13.66^{+0.00}_{-5.25}$ &$ 20.00 \pm 10.30$ & $ 2.87 \pm 1.48$ & $ 504.29 \pm 245.82$   \\
SPT-CLJ0102-4915 & 15.7424 & -49.2742 & 0.870 &  48627 & 1.178 & $ 12.78^{+0.32}_{-0.34}$ & $ 1.21 \pm 0.74$ & $ 0.18 \pm 0.11$ & $ 75.59 \pm 27.75$  \\
SPT-CLJ0534-5005 & 83.4071 & -50.0965 & 0.881 & 342 & 0.702 & $ 2.88^{+0.76}_{-1.22}$ &$ 8.80 \pm 79.28$ & $ 1.35 \pm 12.18$ & $ 119.35 \pm 201.55$ \\
SPT-CLJ2034-5936 & 308.5370 & -59.6051 & 0.92 & 647 & 0.801 & $ 6.48^{+1.24}_{-0.76}$ &$ 4.51 \pm 1.23$ & $ 0.71 \pm 0.19$ & $ 124.16 \pm 33.51$ \\
SPT-CLJ2146-4632 & 326.6450 & -46.5495 & 0.933 & 1078 & 0.701 & $ 5.34^{+1.34}_{-1.06}$ & $ 13.30 \pm 6.53$ & $ 2.12 \pm 1.04$ & $ 226.65 \pm 93.38$ \\
SPT-CLJ0615-5746 & 93.9662 & -57.7800 & 0.972 & 16236 & 1.143 & $ 13.29^{+1.58}_{-0.94}$ & $ 2.05 \pm 0.94$ & $ 0.34 \pm 0.15$ & $ 110.20 \pm 33.73$  \\
SPT-CLJ0547-5345 & 86.6556 & -53.7606 & 1.066 & 1376 & 0.735 & $ 7.12^{+5.05}_{-2.08}$ & $ 4.32 \pm 2.38$ & $ 0.76 \pm 0.42$ & $ 127.83 \pm 86.22$   \\
SPT-CLJ2343-5411 & 355.6920 & -54.1850 & 1.075 & 1426 & 0.599 & $ 4.63^{+0.56}_{-1.13}$ & $ 0.69 \pm 0.58$ & $ 0.12 \pm 0.10$ & $ 28.78 \pm 16.77$ \\
SPT-CLJ0446-5849 & 71.5210 & -58.8308 & 1.16 & 281 & 1.001 & $ 10.85^{+5.72}_{-5.72}$ & $ 13.26 \pm 9.10$ & $ 2.46 \pm 1.69$ & $ 347.84 \pm 227.38$  \\
SPT-CLJ2106-5845 & 316.5226 & -58.7424 & 1.132 & 886 & 0.888 & $ 9.86^{+4.22}_{-2.14}$ & $ 6.79 \pm 2.89$ & $ 1.24 \pm 0.53$ & $ 210.16 \pm 99.20$ \\
\hline
        \end{tabular}

\end{subtable}

\hfill{}
\vspace{10 mm}
\begin{subtable}[RCS sample]{.95\linewidth}
\caption{RCS sample}
        \begin{tabular}{{c@{\hskip 0.6cm \vspace{0.2 mm}} c@{\hskip 0.6cm \vspace{0.2 mm}} c@{\hskip 0.6cm \vspace{0.2 mm}} c@{\hskip 0.4cm \vspace{0.2 mm}} c@{\hskip 0.4cm \vspace{0.2 mm}} c@{\hskip 0.4cm \vspace{0.2 mm}} c@{\hskip 0.4cm \vspace{0.2 mm}} c@{\hskip 0.4cm \vspace{0.2 mm}} c@{\hskip 0.4cm \vspace{0.2 mm}} c@{\hskip 0.4cm \vspace{0.2 mm}}}}
	  \hline

   Field name & Ra & Dec & Redshft & Counts & R$_{500}$ & kT & \tcool & \TcoolTage & K \\
& (deg) &(deg) & & & (Mpc)  & (keV) & (Gyr) & & (keV cm$^{2}$) \\ [-0.0ex]
\hline

RCS2327-0204 & 351.8647 & -02.0776 & 0.705 &  5144 & 1.199 & $ 10.71^{+1.84}_{-1.85}$ & $ 2.21 \pm 1.07$ & $ 0.30 \pm 0.14$ & $ 103.03 \pm 35.03$ \\
RCS1107-0523 & 166.8504 & -05.3890 & 0.735 &  896 & 0.694 & $ 3.97^{+1.09}_{-1.08}$ & $ 1.42 \pm 1.13$ & $ 0.20 \pm 0.16$ & $ 42.64 \pm 22.39$ \\
RCS1325+2858 & 201.6322 & +29.0586 & 0.75 & 110 & 0.393 & $ 1.43^{+3.33}_{-0.83}$ & $ 4.99 \pm 10.05$ & $ 0.70 \pm 1.40$ & $ 62.26 \pm 130.37$   \\
RCS0224-0002 & 36.1430 & -00.0406 & 0.773 & 758 & 0.614 & $ 3.39^{+1.96}_{-0.76}$ &$ 8.07 \pm 4.38$ & $ 1.15 \pm 0.62$ & $ 123.70 \pm 73.43$  \\
RCS1620+2929 & 245.0430 & +29.4898 & 0.870 & 181 & 0.630 & $ 2.81^{+1.21}_{-1.21}$ & $ 6.41 \pm 3.10$ & $ 0.98 \pm 0.47$ & $ 95.34 \pm 46.60$  \\
RCS2319+0038 & 349.9718 & +00.6370 & 0.897 & 1247 & 0.680 & $ 4.99^{+0.60}_{-0.63}$ & $ 3.41 \pm 1.31$ & $ 0.53 \pm 0.20$ & $ 87.70 \pm 24.13$ \\
RCS0439-2905 & 69.9075 & -29.0800 & 0.960 & 183 & 0.423 & $ 1.94^{+3.36}_{-0.38}$ & $ 3.99 \pm 3.46$ & $ 0.65 \pm 0.56$ & $ 58.08 \pm 72.00$ \\
\hline
\end{tabular}

\end{subtable}
\end{table*}

With the addition of 24 high redshift clusters and the moderation of 
any archival biases in our X-ray selected sample, the extended
sample forms a stronger basis for applying statistical tests
for cool core evolution. The Spearman rank tests for all six CC indicators
are given in the right hand half of Table~\ref{Spearman}. These results
confirm and strengthen the conclusions from the X-ray sample discussed
in Sections~\ref{subsection:evolution} and ~\ref{subsection:cuspiness}.
Using the full group$+$cluster sample, we see a highly significant ($p=0.006$)
correlation with redshift in \TcoolTage, but little trend in \tcool. These results
also apply to the group and cluster subsamples separately.
In terms of other indicators, as for the X-ray sample, we see evolutionary
trends in $K$ and \fc, but now also in \csb. All these trends imply
stronger cool cores at low redshift.

For our two main CC indicators, \tcool\ and \TcoolTage, we
conduct two further simple statistical tests which involve cutting
the extended sample into high and low redshift halves. The choice of the cut
redshift is arbitrary, and the results scatter with this choice, so
we present them for a series of cuts between $z=0.5$ and $z=0.7$.
For each split sample we calculate (a) the mean and standard deviation
for both CC indicators, and (b) a K-S test for consistency between 
the distribution seen in the high and low redshift samples.
Table~\ref{KS} shows the results. These broadly confirm the results
of the Spearman rank test; \TcoolTage\ is clearly evolving, wherever
the cut is placed, whilst differences in
\tcool\ between the low and high redshift subsamples are 
much weaker, though the cooling time does show a significant tendency 
to be somewhat shorter at low redshift.

\begin{table*}
     \caption{Comparison between the distribution of two cool core parameters (\tcool and \TcoolTage) at low and high redshift. Five threshold between 0.5 and 0.7 are chosen for redshift to divide the sample into low and high redshift subsamples. For each parameter and each redshift threshold the mean value of the parameter for the low and high redshift subsample is given together with the standard error on the mean. Also the p-value for a Kolmogorov-Smirnov test for similarity in the distribution for the low and high redshift subsamples is given.}
\label{KS}
\begin{minipage}{180mm}

\centering
\begin{tabular}{lccl@{\hskip 1.4cm}ccc}

\hline
 &\multicolumn{3}{c|@{\hskip 1.4cm}}{\tcoolb} & \multicolumn{3}{c|}{\TcoolTageb}\\
 \hline
 &\multicolumn{2}{c|}{Mean} & K-S &\multicolumn{2}{c|}{Mean} & K-S\\
  \multicolumn{1}{|c|}{Redshift} &
  \multicolumn{1}{c|}{low} &
  \multicolumn{1}{c|}{high} &
  \multicolumn{1}{c|@{\hskip 1.4cm}}{p-val} &
  \multicolumn{1}{c|}{low} &
  \multicolumn{1}{c|}{high} &
  \multicolumn{1}{c|}{p-val} \\
\hline

  0.5 & $4.79 \pm 0.98$ & $6.29 \pm 0.67$ & 0.03&  $0.46 \pm 0.09$ &$ 0.92 \pm 0.09$ & 0.005\\
  0.55 & $5.02 \pm 0.93 $&$ 6.23 \pm 0.69 $& 0.08&$  0.49 \pm 0.09 $& $0.92 \pm 0.10 $& 0.008\\
  0.6 &$ 5.24 \pm 0.86 $&$ 6.16 \pm 0.73$ & 0.13&$ 0.53 \pm 0.08 $&$ 0.92 \pm 0.11$ & 0.020\\
  0.65 & $5.10 \pm 0.83 $&$ 6.32 \pm 0.75$ & 0.08&  $0.52 \pm 0.08 $& $0.95 \pm 0.11$ & 0.016\\
  0.7 & $5.36 \pm 0.81$ & $6.14 \pm 0.77$ & 0.19&$0.56 \pm 0.08$ & $0.94 \pm 0.11$ & 0.070\\
\hline

\end{tabular}
\normalsize
\medskip

\end{minipage}
\end{table*}

\section{Discussion}

We conclude from the evidence presented above that most, but not all, of the 
cool core indicators
we have employed show evidence, confirmed by a number of statistical tests,
for evolution in the properties of cluster cores. Concentrating on our two
primary indicators, which are based on cooling time, we see significant evolution
in \TcoolTage\ evaluated at $r=0.01$\R, but 
at most a weak trend in the value of \tcool\ evaluated at
this radius. This behaviour is apparent for both our X-ray and extended cluster
samples, and it applies for clusters and groups separately and combined.
(See, for example, Table~\ref{Spearman}.)

There is no evidence here for a difference between the behaviour of groups
and clusters, such as was suggested by \cite{Alshino2010}. These authors
found, using XMM-Newton data for a sample of groups and clusters detected
in the XMM-LSS X-ray survey, that the cores in groups were actually {\it
 more} cuspy at high redshift, in contrast to the situation in
clusters. Using the same indicator as Alshino et al., the central excess
(\fc) above a standard beta-model fit, we find evolution towards {\it less}
prominent cores at high redshift in both groups and clusters, as can be
seen in Figure~\ref{figIndEvol} and Table~\ref{Spearman}. The reason for
this disagreement is unclear. The most relevant differences between the two
studies seems to be the angular resolution of the X-ray data and the
degree of uniformity of the survey.

Chandra has a much sharper point spread function than XMM, and so
our surface brightness profiles are subject to less instrumental
blurring. Although the PSF effects are modelled out during the profile
fitting in both studies, the work of Alshino et al. will be much more
vulnerable to any shortcomings in this process, since the impact of
blurring is greater for high redshift clusters.

The second relevant difference between our survey and the XMM-LSS survey on 
which the results of \cite{Alshino2010} are based, is that XMM-LSS is a more
uniform survey, with contiguous XMM-Newton exposures typically 10~ksec in duration,
whilst CDGS is based on Chandra exposures of widely varying depth (exposure
times ranging up to 4~Msec). This means that high redshift groups,
having low source flux, will be amongst the lowest significance sources
in XMM-LSS, but not necessarily in CDGS, especially since we have
imposed a minimum count threshold of 100 counts on all our sources.
A consequence is that the high redshift groups in the Alshino et al. survey will
be subject to strong selection effects, which may result in more
centrally peaked systems being preferentially detected. In contrast, the
simulations reported above in Section~\ref{subsection:detbias} establish
that no such significant bias should be present in our study. This seems
to us to be the most likely explanation for the contrary behaviour of
high redshift groups in the two studies.

In any case, we conclude that the combination of superior resolution
and the avoidance of systems close to the detection threshold, means that
the results from the present study regarding CC evolution in groups
should be more reliable than those reported by \cite{Alshino2010}.

We have considered the impact of systematic biases on our results, and conclude
that both detection bias and the effects of AGN contamination appear to be
modest. Archival bias, arising from the fact that many of our X-ray selected
clusters (especially those at high redshift) were deliberately targetted 
for Chandra observations, is of greater concern. We addressed this by adding
a further 24 clusters at $z>0.7$ selected from SZ and optical surveys.
These show a somewhat wider range in core properties than our high redshift X-ray 
sample, however the main thrust of our conclusions on core
evolution are unchanged by the addition of these clusters to the sample.

Following the initial indications reported by \cite{Vikhlinin2007}, subsequent studies
of the cuspiness of the profiles of X-ray selected samples dominated by
clusters ($T>3$~keV) by \cite{Santos2008,Santos2010} and \citet{Maughan2012} have
found a reduction in the fraction of clusters hosting strong central
surface brightness cusps at high ($z>0.5$) redshift. Our results are consistent
with the existence of such a trend, which we have shown extends also to galaxy groups.

Using cool core indicators based on cooling time, a more nuanced picture
emerges, which can be usefully compared with the study of \cite{McDonald2013,McDonald2014}. 
This examined the X-ray properties of a sample of 80
SZ-detected (hence rather rich) clusters and, like the present study,
explored a variety of different CC indicators. The use of a SZ-selected
sample reduces direct selection bias arising from
the X-ray properties of the clusters. Indirect biases are still possible --
for example, dense core gas does enhance the SZ signal, and radio-bright AGN may also increase the probability of cluster detection.
\cite{McDonald2013} conclude that both are minor effects.

\cite{McDonald2013} also find no evolutionary trend in cooling time calculated
within the core (in their case at a radius of $0.012$\R). They do
not compute \TcoolTage, but they do calculate \csb\ and also find that this
evolves towards increasing cuspiness, as do two other indicators: the 
logarithmic density slope at $0.04$\R, and the mass cooling rate integrated
within a cooling radius, which itself depends upon the age of the Universe
at the redshift of the cluster. These results are highly consistent with our
own and suggests that these trends are rather robust against the method
of cluster selection (X-ray vs SZ) and the mass range considered (our sample
extends to considerably lower masses).

What do these results imply about the evolution of cluster cores?
In the first place, it is clear that these cores do not follow the
self-similar evolution seen in the outer regions of clusters. Here
the gas density at a given scaled radius (e.g. \R) scales with
the critical density of the Universe, and hence as $E(z)^2$, whilst
from the virial theorem the characteristic temperature is related to 
cluster mass via $T\propto (M E(z))^{2/3}$. If the core gas followed 
the same scaling laws, then the cooling time at a given scaled radius 
would (in the approximation that thermal bremsstrahlung dominates) scale as 
\tcool $\propto \frac{n_e T}{n_e^2 \Lambda(T)} \propto \frac{T^{1/2}}{n_e} \propto T^{1/2}E(z)^{-2}$~~,
where the cooling function, $\Lambda(T)$, scales as $T^{1/2}$ for bremsstrahlung
emission. This implies that cooling times should be significantly {\it shorter}
at high redshift, which is clearly inconsistent with our observations.

\cite{Voit2011b} has proposed an interesting model for the thermal state
of cluster cores whereby there exists a critical line in the radius-entropy
plane, $K(r) \approx 5 r_{\rm kpc}^{2/3}$, along which conductive heat transfer
can balance radiative cooling. Above this line, cooling is subdominant, and
the gas entropy drops inwards according to the
$K \propto r^{1.1}$ relation predicted by simple models of gas accretion and
shock heating. However, once this steeper radial trend intersects the
conductive balance line the gas cools and ultimately becomes thermally
unstable, and its entropy profile within this radius
is determined by feedback processes, probably associated with a central AGN,
which prevent catastrophic cooling. 

The radial entropy profiles reported in a sample of low redshift groups
and clusters by \cite{Panagoulia2014} seems to accord remarkably well
with this model (though these authors seem not to have noticed this),
and suggest that the entropy follows the conductive
equilibrium line inwards in cool core systems, once the steeper
outer entropy profile hits the critical equilibrium line. This might
be explained if AGN feedback were able to prevent the entropy from falling
much below the conductive balance value.

What evolutionary behaviour in cores would be predicted by such a model?
The conductive balance $K(r)$ line is independent of redshift, but the
radius at which we measure entropy ($0.01$\R) will evolve -- for 
temperature $T$, \R\ scales as $T^{1/2}E(z)^{-1}$, so the entropy at
$0.01$\R\ will scale as $T^{1/3}E(z)^{-2/3}$, and hence for a given temperature
should be {\it lower} at high redshift (though not as much lower as for self-similar
evolution). In practice (see Table~\ref{Spearman}) we see the reverse --
somewhat higher entropies at high redshift. This suggests that the
conductive equilibrium model in its simplest form cannot account for
the evolutionary trend we observe.

A more recent development of the model by \cite{Voit2014} demonstrates that
low redshift cool core clusters have cooling time profiles 
(which closely follow from entropy profiles -- see Figure~\ref{figTK})
that are bracketed by the conductive balance locus (at high \tcool) and a 
lower \tcool\ limit set by the point at which thermal instability
causes gas to generate cool clouds which can precipitate onto a
central galaxy, causing AGN feedback which heats and mixes the core
gas, limiting further cooling. This lower `precipitation line'
corresponds to the locus along which the
cooling time is approximately 10 times the free fall
time, which in turn is set by the gravitational potential. This will not
evolve strongly with redshift, which might again lead to an expectation that
\tcool($0.01$\R) would be smaller at high redshift, due to the smaller
value of \R\ (at a given system temperature). However, the model predicts
that cooling time profiles will be distributed between the conductive balance 
and precipitation lines in a way which depends on details of the AGN feedback
process, such as the duty cycle. This leaves open the possibility that
the average cooling time over a sample of CC systems might evolve very
little. Detailed entropy profiles for a sample of high redshift CC systems
are required to explore the viability of the model. This may have to await
the next generation of X-ray observatories.

In terms of our two main CC indicators, the fact that \TcoolTage\ is decreasing
with time follows directly from the fact that \tcool\ is {\it not} evolving, since
the age of the Universe (obviously) increases with time. 
The reason that \tcool\ does not evolve to any great extent must be connected to
the processes which break self-similarity in cluster cores:
cooling, conductive heat transfer and the feedback processes
which prevent runaway cooling. It is well known that cooling
in cluster cores is suppressed well below the naive rate derived from the observed
X-ray luminosity. Nonetheless, some cooling does take place and, for example,
star formation in central galaxies within cool cores implies that gas
can cool at a rate 1-10\% of the uncontrolled value, most likely due
to countervailing AGN heating \citep{ODea2008}. If some of this gas is
able to accumulate within the cluster core, outside 0.01\R, it might explain
why cuspiness indicators like \csb\ evolve with time, in addition to 
quantities like \TcoolTage. This is essentially the explanation
proposed by \cite{McDonald2013} to account for their results, and in
\cite{McDonald2014} they show that pressure tends to rise over time
within CC clusters, which they take to be a result of a build-up of gas.

\begin{figure}\centering

 \includegraphics[width=0.43\textwidth,keepaspectratio,trim={0cm 0cm 0cm 2.5cm}, clip=true]{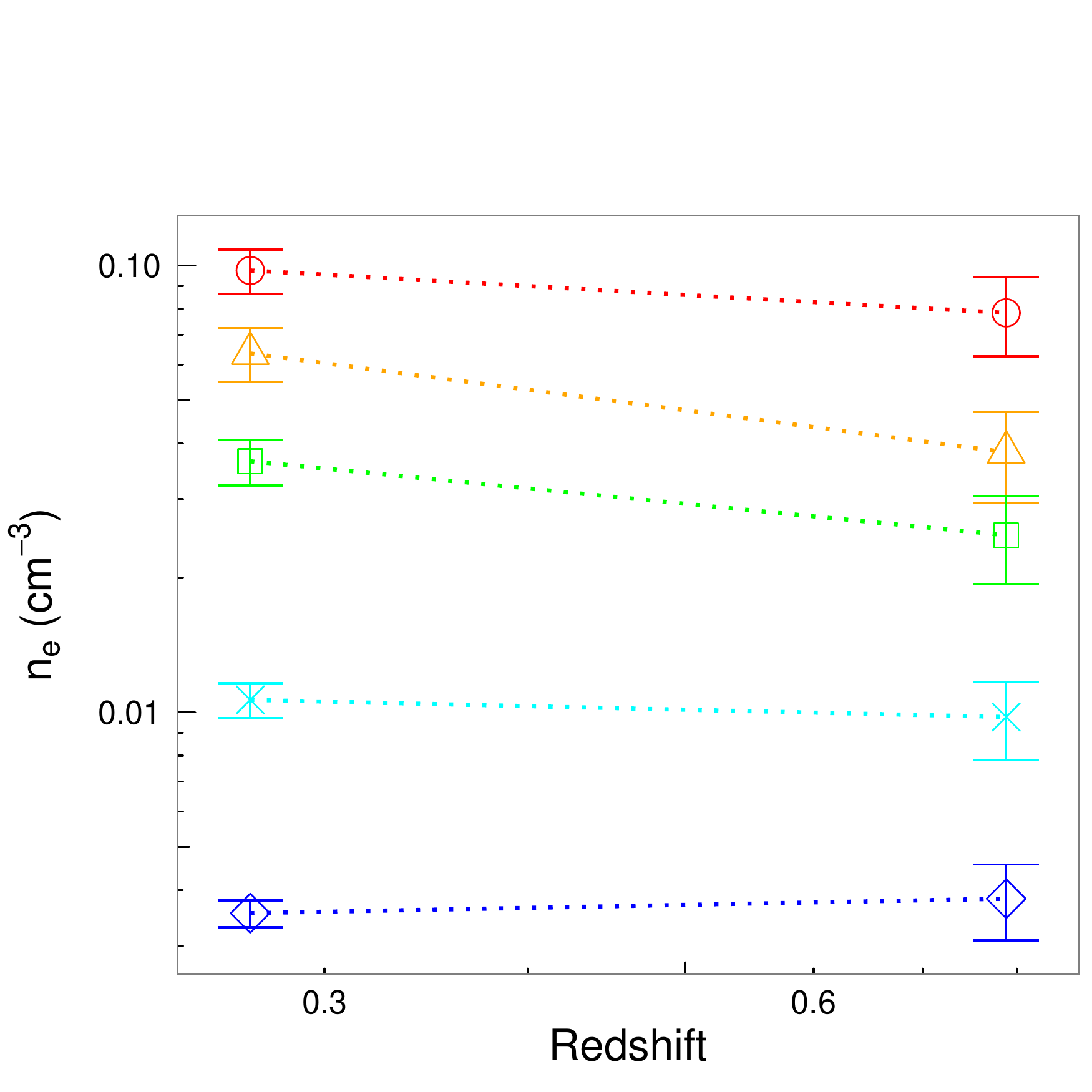}
    \caption{ Evolution of mean density for cool core systems within the extended sample,
calculated in two redshift bins for six different radii: 0.01\R (circle), 0.02\R (triangle), 
0.04\R (square), 0.1\R (cross) and 0.2\R (diamond). Each point in the plot represents the mean 
density of the sample, with associated standard error.  }
    \label{densevoln}
\end{figure}

We can assess the evidence for such a rise in gas density in the outer core
in our clusters by examining the evolution of gas density in CC systems
at a range of different radii. For this purpose, we restrict ourselves to
clusters. Groups have lower gas densities than clusters over most of their 
radial range as a result of the action of feedback processes \citep{Ponman1999},
and since groups are concentrated at low redshift
this difference will swamp any evolutionary trends if they are included.
We are interested only in CC systems here, so we include only systems
which have \tcool$(0.01\R)<1.5$~Gyr.
Figure~\ref{densevoln} shows that the mean gas density for this subsample, derived from our
analytical deprojection analysis, as a function of redshift for several different
overdensity radii. This confirms that density increases more strongly
with time immediately outside 0.01\R. At larger radii ($>0.1$\R) this evolution reverses
as it tends towards self-similar behaviour.

\section{Conclusion}

We have presented in this paper a study of the
evolution of cluster cores, based on a sample of 62 X-ray selected systems with
temperatures between 1-12 keV and redshifts up to 1.3. We have
investigated the existence of evolutionary trends in the entire
sample, as well as in the subsamples of 26 groups ($T< 3$~keV) and
36 clusters ($T\geq3$~keV) separately.  Our main results can be
summarised as follows:

\begin{itemize}

\item[--] Six different parameters have been used to
quantify the strength of cool cores, and different evolutionary trends are
found for CC strength, depending on the CC estimator used. This behaviour
is found for the entire sample and the subsamples of clusters and groups
separately.

\item[--] For the entire sample of 62 systems, we find a
decrease in the fraction of CC with redshift using the \TcoolTage, K and
\fc\ indicators, a weak evolution for \csb, and no significant evolution
for \tcool\ and \fcore. 

\item[--] Groups and clusters show similar trends
irrespective of the parameter used to characterize CCs, although the
statistical significance of the trends found for groups is lower than that
for clusters. These trends are similar to those seen for the entire
sample.

\item[--] In particular, a clear reduction in the fraction
of cool cores at high redshift is found for both groups and clusters when
the \fc\ indicator is used. This is inconsistent with the
opposite trends for groups and clusters reported by \cite{Alshino2010}
using this estimator.

\item[--] The impact on our results of a variety of different
biases was investigated. Bias due to the impact of core properties
on the ability to detect an extended X-ray source in our Chandra data
appears to be modest, as do biases arising from the 
presence of a central AGN. The impact of archival bias, arising from
the agenda of observers who targetted the non-serendipitous systems
which dominate our sample at high redshift, is potentially more serious.
Its impact was tested by adding 24 non-X-ray selected systems
at $z>0.7$ to generate an extended sample.
In general, the effect of adding these systems is to strengthen the trends seen when
using only the X-ray selected sample. The most noticeable difference is
for \csb, which shows a more pronounced evolutionary trend in the extended
sample. Removal from our sample of systems with photometric, rather
than spectroscopic, redshifts also leave our findings substantially unchanged. 

\item[--] A reasonable interpretation of our results is
that, in both groups and clusters, the cooling time of gas in the inner
core is held at an approximately constant value by AGN feedback. However,
cooling gas accumulates in the outer core, driving an increase in the
cuspiness of CC systems with time. We find evidence for this behaviour in
the evolution of gas density as a function of radius, within CC systems.

\end{itemize}

\section{Acknowledgements}

We thank Nathan Slack for the use of his Bayesian source extension
software, and Michael McDonald, Mark Voit, Joanna Santos and Kathy Romer for useful
discussions of cooling in cluster cores. We thank the anonymous referee for stimulating a number of clarifications
in the text. AP acknowledges financial support from
the School of Physics and Astronomy at the University of Birmingham.

\bibliography{CDGS-CC_evolution}

\begin{thebibliography}{136}
\expandafter\ifx\csname natexlab\endcsname\relax\def\natexlab#1{#1}\fi

\bibitem[{{Allen} {et~al}\mbox{.}(2008){Allen}, {Rapetti}, {Schmidt},
  {Ebeling}, {Morris}, \& {Fabian}}]{Allen2008}
{Allen} S.~W., {Rapetti} D.~A., {Schmidt} R.~W., {Ebeling} H., {Morris} R.~G.,
  {Fabian} A.~C., 2008, MNRAS, 383, 879

\bibitem[{{Alshino} {et~al}\mbox{.}(2010){Alshino}, {Ponman}, {Pacaud}, \&
  {Pierre}}]{Alshino2010}
{Alshino} A., {Ponman} T., {Pacaud} F., {Pierre} M., 2010, MNRAS, 407, 2543

\bibitem[{{Anders} \& {Grevesse}(1989)}]{Anders1989}
{Anders} E., {Grevesse} N., 1989, Geochim. Cosmochim. Acta, 53, 197

\bibitem[{{Bahcall} {et~al}\mbox{.}(2003){Bahcall}, {McKay}, {Annis}, {Kim},
  {Dong}, {Hansen}, {Goto}, {Gunn}, {Miller}, {Nichol}, {Postman}, {Schneider},
  {Schroeder}, {Voges}, {Brinkmann}, \& {Fukugita}}]{Bahcall2003}
{Bahcall} N.~A. {et~al.}, 2003, ApJ, 148, 243

\bibitem[{{Barkhouse} {et~al}\mbox{.}(2006){Barkhouse}, {Green}, {Vikhlinin},
  {Kim}, {Perley}, {Cameron}, {Silverman}, {Mossman}, {Burenin}, {Jannuzi},
  {Kim}, {Smith}, {Smith}, {Tananbaum}, \& {Wilkes}}]{Barkhouse2006}
{Barkhouse} W.~A. {et~al.}, 2006, ApJ, 645, 955

\bibitem[{{Bauer} {et~al}\mbox{.}(2005){Bauer}, {Fabian}, {Sanders}, {Allen},
  \& {Johnstone}}]{Bauer2005}
{Bauer} F.~E., {Fabian} A.~C., {Sanders} J.~S., {Allen} S.~W., {Johnstone}
  R.~M., 2005, MNRAS, 359, 1481

\bibitem[{{B{\^i}rzan} {et~al}\mbox{.}(2004){B{\^i}rzan}, {Rafferty},
  {McNamara}, {Wise}, \& {Nulsen}}]{Birzan2004}
{B{\^i}rzan} L., {Rafferty} D.~A., {McNamara} B.~R., {Wise} M.~W., {Nulsen}
  P.~E.~J., 2004, ApJ, 607, 800

\bibitem[{{Blanton} {et~al}\mbox{.}(2011){Blanton}, {Randall}, {Clarke},
  {Sarazin}, {McNamara}, {Douglass}, \& {McDonald}}]{Blanton2011}
{Blanton} E.~L., {Randall} S.~W., {Clarke} T.~E., {Sarazin} C.~L., {McNamara}
  B.~R., {Douglass} E.~M., {McDonald} M., 2011, ApJ, 737, 99

\bibitem[{{Boehringer} {et~al}\mbox{.}(1993){Boehringer}, {Voges}, {Fabian},
  {Edge}, \& {Neumann}}]{Boehringer1993}
{Boehringer} H., {Voges} W., {Fabian} A.~C., {Edge} A.~C., {Neumann} D.~M.,
  1993, MNRAS, 264, L25

\bibitem[{{Bonamente} {et~al}\mbox{.}(2012){Bonamente}, {Landry}, {Maughan},
  {Giles}, {Joy}, \& {Nevalainen}}]{Bonamente2012}
{Bonamente} M., {Landry} D., {Maughan} B., {Giles} P., {Joy} M., {Nevalainen}
  J., 2012, MNRAS, 177

\bibitem[{{Brada{\v c}} {et~al}\mbox{.}(2008){Brada{\v c}}, {Allen}, {Treu},
  {Ebeling}, {Massey}, {Morris}, {von der Linden}, \&
  {Applegate}}]{Bradavc2008}
{Brada{\v c}} M., {Allen} S.~W., {Treu} T., {Ebeling} H., {Massey} R., {Morris}
  R.~G., {von der Linden} A., {Applegate} D., 2008, ApJ, 687, 959

\bibitem[{{Burenin} {et~al}\mbox{.}(2007){Burenin}, {Vikhlinin}, {Hornstrup},
  {Ebeling}, {Quintana}, \& {Mescheryakov}}]{Burenin2007}
{Burenin} R.~A., {Vikhlinin} A., {Hornstrup} A., {Ebeling} H., {Quintana} H.,
  {Mescheryakov} A., 2007, ApJS, 172, 561

\bibitem[{{Burns} {et~al}\mbox{.}(2008){Burns}, {Hallman}, {Gantner}, {Motl},
  \& {Norman}}]{Burns2008}
{Burns} J.~O., {Hallman} E.~J., {Gantner} B., {Motl} P.~M., {Norman} M.~L.,
  2008, ApJ, 675, 1125

\bibitem[{{Carlstrom} {et~al}\mbox{.}(2011){Carlstrom}, {Ade}, {Aird},
  {Benson}, {Bleem}, {Busetti}, {Chang}, {Chauvin}, {Cho}, {Crawford},
  {Crites}, {Dobbs}, {Halverson}, {Heimsath}, {Holzapfel}, {Hrubes}, {Joy},
  {Keisler}, {Lanting}, {Lee}, {Leitch}, {Leong}, {Lu}, {Lueker}, {Luong-van},
  {McMahon}, {Mehl}, {Meyer}, {Mohr}, {Montroy}, {Padin}, {Plagge}, {Pryke},
  {Ruhl}, {Schaffer}, {Schwan}, {Shirokoff}, {Spieler}, {Staniszewski},
  {Stark}, {Tucker}, {Vanderlinde}, {Vieira}, \& {Williamson}}]{Carlstrom2011}
{Carlstrom} J.~E. {et~al.}, 2011, PASP, 123, 568

\bibitem[{{Cash}(1979)}]{Cash1979}
{Cash} W., 1979, ApJ, 228, 939

\bibitem[{{Cavagnolo} {et~al}\mbox{.}(2008){Cavagnolo}, {Donahue}, {Voit}, \&
  {Sun}}]{Cavagnolo2008}
{Cavagnolo} K.~W., {Donahue} M., {Voit} G.~M., {Sun} M., 2008, ApJ, 682, 821

\bibitem[{{Cavagnolo} {et~al}\mbox{.}(2009){Cavagnolo}, {Donahue}, {Voit}, \&
  {Sun}}]{Cavagnolo2009}
{Cavagnolo} K.~W., {Donahue} M., {Voit} G.~M., {Sun} M., 2009, ApJS, 182, 12

\bibitem[{{Cavaliere} \& {Fusco-Femiano}(1976)}]{Cavaliere1976}
{Cavaliere} A., {Fusco-Femiano} R., 1976, A\&A, 49, 137

\bibitem[{{Chen} {et~al}\mbox{.}(2007){Chen}, {Reiprich}, {B{\"o}hringer},
  {Ikebe}, \& {Zhang}}]{Chen2007}
{Chen} Y., {Reiprich} T.~H., {B{\"o}hringer} H., {Ikebe} Y., {Zhang} Y.-Y.,
  2007, A\&A, 466, 805

\bibitem[{{Croston} {et~al}\mbox{.}(2005){Croston}, {Hardcastle}, \&
  {Birkinshaw}}]{Croston2005}
{Croston} J.~H., {Hardcastle} M.~J., {Birkinshaw} M., 2005, MNRAS, 357, 279

\bibitem[{{De Luca} \& {Molendi}(2004)}]{DeLuca2004}
{De Luca} A., {Molendi} S., 2004, A\&A, 419, 837

\bibitem[{{Dickey} \& {Lockman}(1990)}]{Dickey1990}
{Dickey} J.~M., {Lockman} F.~J., 1990, ARA\&A, 28, 215

\bibitem[{{Donahue} {et~al}\mbox{.}(1999){Donahue}, {Voit}, {Scharf}, {Gioia},
  {Mullis}, {Hughes}, \& {Stocke}}]{Donahue1999}
{Donahue} M., {Voit} G.~M., {Scharf} C.~A., {Gioia} I.~M., {Mullis} C.~R.,
  {Hughes} J.~P., {Stocke} J.~T., 1999, ApJ, 527, 525

\bibitem[{{Dong} {et~al}\mbox{.}(2010){Dong}, {Rasmussen}, \&
  {Mulchaey}}]{Dong2010}
{Dong} R., {Rasmussen} J., {Mulchaey} J.~S., 2010, ApJ, 712, 883

\bibitem[{{Dressler} \& {Gunn}(1992)}]{Dressler1992}
{Dressler} A., {Gunn} J.~E., 1992, ApJ, 78, 1

\bibitem[{{Ebeling} {et~al}\mbox{.}(2007){Ebeling}, {Barrett}, {Donovan}, {Ma},
  {Edge}, \& {van Speybroeck}}]{Ebeling2007}
{Ebeling} H., {Barrett} E., {Donovan} D., {Ma} C.-J., {Edge} A.~C., {van
  Speybroeck} L., 2007, ApJ, 661, L33

\bibitem[{{Ebeling} {et~al}\mbox{.}(2010){Ebeling}, {Edge}, {Mantz}, {Barrett},
  {Henry}, {Ma}, \& {van Speybroeck}}]{Ebeling2010}
{Ebeling} H., {Edge} A.~C., {Mantz} A., {Barrett} E., {Henry} J.~P., {Ma}
  C.~J., {van Speybroeck} L., 2010, MNRAS, 407, 83

\bibitem[{{Eckert} {et~al}\mbox{.}(2011){Eckert}, {Molendi}, \&
  {Paltani}}]{Eckert2011}
{Eckert} D., {Molendi} S., {Paltani} S., 2011, A\&A, 526, A79

\bibitem[{{Edge} {et~al}\mbox{.}(1992){Edge}, {Stewart}, \&
  {Fabian}}]{Edge1992}
{Edge} A.~C., {Stewart} G.~C., {Fabian} A.~C., 1992, MNRAS, 258, 177

\bibitem[{{Fabian} {et~al}\mbox{.}(2000){Fabian}, {Sanders}, {Ettori},
  {Taylor}, {Allen}, {Crawford}, {Iwasawa}, {Johnstone}, \&
  {Ogle}}]{Fabian2000}
{Fabian} A.~C. {et~al.}, 2000, MNRAS, 318, L65

\bibitem[{{Feruglio} {et~al}\mbox{.}(2008){Feruglio}, {Fiore}, {La Franca},
  {Sacchi}, {Puccetti}, {Comastri}, {Berta}, {Brusa}, {Franceschini},
  {Gruppioni}, {Mathur}, {Matute}, {Mignoli}, {Pozzi}, {Vignali}, \&
  {Zamorani}}]{Feruglio2008}
{Feruglio} C. {et~al.}, 2008, A\&A, 488, 417

\bibitem[{{Finoguenov} {et~al}\mbox{.}(2009){Finoguenov}, {Connelly}, {Parker},
  {Wilman}, {Mulchaey}, {Saglia}, {Balogh}, {Bower}, \&
  {McGee}}]{Finoguenov2009}
{Finoguenov} A. {et~al.}, 2009, ApJ, 704, 564

\bibitem[{{Finoguenov} {et~al}\mbox{.}(2007){Finoguenov}, {Guzzo}, {Hasinger},
  {Scoville}, {Aussel}, {B{\"o}hringer}, {Brusa}, {Capak}, {Cappelluti},
  {Comastri}, {Giodini}, {Griffiths}, {Impey}, {Koekemoer}, {Kneib},
  {Leauthaud}, {Le F{\`e}vre}, {Lilly}, {Mainieri}, {Massey}, {McCracken},
  {Mobasher}, {Murayama}, {Peacock}, {Sakelliou}, {Schinnerer}, {Silverman},
  {Smol{\v c}i{\'c}}, {Taniguchi}, {Tasca}, {Taylor}, {Trump}, \&
  {Zamorani}}]{Finoguenov2007}
{Finoguenov} A. {et~al.}, 2007, ApJ, 172, 182

\bibitem[{{Finoguenov} {et~al}\mbox{.}(2001){Finoguenov}, {Reiprich}, \&
  {B{\"o}hringer}}]{Finoguenov2001}
{Finoguenov} A., {Reiprich} T.~H., {B{\"o}hringer} H., 2001, A\&A, 368, 749

\bibitem[{{Gastaldello} {et~al}\mbox{.}(2007){Gastaldello}, {Buote},
  {Humphrey}, {Zappacosta}, {Bullock}, {Brighenti}, \&
  {Mathews}}]{Gastaldello2007}
{Gastaldello} F., {Buote} D.~A., {Humphrey} P.~J., {Zappacosta} L., {Bullock}
  J.~S., {Brighenti} F., {Mathews} W.~G., 2007, ApJ, 669, 158

\bibitem[{{Gastaldello} {et~al}\mbox{.}(2009){Gastaldello}, {Buote}, {Temi},
  {Brighenti}, {Mathews}, \& {Ettori}}]{Gastaldello2009}
{Gastaldello} F., {Buote} D.~A., {Temi} P., {Brighenti} F., {Mathews} W.~G.,
  {Ettori} S., 2009, ApJ, 693, 43

\bibitem[{{Gilbank} {et~al}\mbox{.}(2011){Gilbank}, {Gladders}, {Yee}, \&
  {Hsieh}}]{Gilbank2011}
{Gilbank} D.~G., {Gladders} M.~D., {Yee} H.~K.~C., {Hsieh} B.~C., 2011, AJ,
  141, 94

\bibitem[{{Gitti} {et~al}\mbox{.}(2011){Gitti}, {Nulsen}, {David}, {McNamara},
  \& {Wise}}]{Gitti2011}
{Gitti} M., {Nulsen} P.~E.~J., {David} L.~P., {McNamara} B.~R., {Wise} M.~W.,
  2011, ApJ, 732, 13

\bibitem[{{Gitti} {et~al}\mbox{.}(2010){Gitti}, {O'Sullivan}, {Giacintucci},
  {David}, {Vrtilek}, {Raychaudhury}, \& {Nulsen}}]{Gitti2010}
{Gitti} M., {O'Sullivan} E., {Giacintucci} S., {David} L.~P., {Vrtilek} J.,
  {Raychaudhury} S., {Nulsen} P.~E.~J., 2010, ApJ, 714, 758

\bibitem[{{Gladders} \& {Yee}(2000)}]{Gladders2000}
{Gladders} M.~D., {Yee} H.~K.~C., 2000, AJ, 120, 2148

\bibitem[{{Gladders} {et~al}\mbox{.}(2007){Gladders}, {Yee}, {Majumdar},
  {Barrientos}, {Hoekstra}, {Hall}, \& {Infante}}]{Gladders2007}
{Gladders} M.~D., {Yee} H.~K.~C., {Majumdar} S., {Barrientos} L.~F., {Hoekstra}
  H., {Hall} P.~B., {Infante} L., 2007, ApJ, 655, 128

\bibitem[{{Gonzalez} {et~al}\mbox{.}(2005){Gonzalez}, {Tran}, {Conbere}, \&
  {Zaritsky}}]{Gonzalez2005}
{Gonzalez} A.~H., {Tran} K.-V.~H., {Conbere} M.~N., {Zaritsky} D., 2005, ApJ,
  624, L73

\bibitem[{{Hao} {et~al}\mbox{.}(2010){Hao}, {McKay}, {Koester}, {Rykoff},
  {Rozo}, {Annis}, {Wechsler}, {Evrard}, {Siegel}, {Becker}, {Busha}, {Gerdes},
  {Johnston}, \& {Sheldon}}]{Hao2010}
{Hao} J. {et~al.}, 2010, ApJ, 191, 254

\bibitem[{{Helsdon} \& {Ponman}(2000)}]{Helsdon2000}
{Helsdon} S.~F., {Ponman} T.~J., 2000, MNRAS, 315, 356

\bibitem[{{Henley} \& {Shelton}(2010)}]{Henley2010a}
{Henley} D.~B., {Shelton} R.~L., 2010, ApJS, 187, 388

\bibitem[{{Henning} {et~al}\mbox{.}(2009){Henning}, {Gantner}, {Burns}, \&
  {Hallman}}]{Henning2009}
{Henning} J.~W., {Gantner} B., {Burns} J.~O., {Hallman} E.~J., 2009, ApJ, 697,
  1597

\bibitem[{{Hlavacek-Larrondo} {et~al}\mbox{.}(2012){Hlavacek-Larrondo},
  {Fabian}, {Edge}, {Ebeling}, {Sanders}, {Hogan}, \&
  {Taylor}}]{Hlavacek-Larrondo2012}
{Hlavacek-Larrondo} J., {Fabian} A.~C., {Edge} A.~C., {Ebeling} H., {Sanders}
  J.~S., {Hogan} M.~T., {Taylor} G.~B., 2012, MNRAS, 421, 1360

\bibitem[{{Holden} {et~al}\mbox{.}(2001){Holden}, {Stanford}, {Rosati},
  {Squires}, {Tozzi}, {Fosbury}, {Papovich}, {Eisenhardt}, {Elston}, \&
  {Spinrad}}]{Holden2001}
{Holden} B.~P. {et~al.}, 2001, AJ, 122, 629

\bibitem[{{Holden} {et~al}\mbox{.}(2002){Holden}, {Stanford}, {Squires},
  {Rosati}, {Tozzi}, {Eisenhardt}, \& {Spinrad}}]{Holden2002}
{Holden} B.~P., {Stanford} S.~A., {Squires} G.~K., {Rosati} P., {Tozzi} P.,
  {Eisenhardt} P., {Spinrad} H., 2002, AJ, 124, 33

\bibitem[{{Hsieh} {et~al}\mbox{.}(2005){Hsieh}, {Yee}, {Lin}, \&
  {Gladders}}]{Hsieh2005}
{Hsieh} B.~C., {Yee} H.~K.~C., {Lin} H., {Gladders} M.~D., 2005, ApJ, 158, 161

\bibitem[{{Huang} {et~al}\mbox{.}(2009){Huang}, {Morokuma}, {Fakhouri},
  {Aldering}, {Amanullah}, {Barbary}, {Brodwin}, {Connolly}, {Dawson}, {Doi},
  {Faccioli}, {Fadeyev}, {Fruchter}, {Goldhaber}, {Gladders}, {Hennawi},
  {Ihara}, {Jee}, {Kowalski}, {Konishi}, {Lidman}, {Meyers}, {Moustakas},
  {Perlmutter}, {Rubin}, {Schlegel}, {Spadafora}, {Suzuki}, {Takanashi}, \&
  {Yasuda}}]{Huang2009}
{Huang} X. {et~al.}, 2009, ApJ, 707, L12

\bibitem[{{Hudson} {et~al}\mbox{.}(2010){Hudson}, {Mittal}, {Reiprich},
  {Nulsen}, {Andernach}, \& {Sarazin}}]{Hudson2010}
{Hudson} D.~S., {Mittal} R., {Reiprich} T.~H., {Nulsen} P.~E.~J., {Andernach}
  H., {Sarazin} C.~L., 2010, A\&A, 513, A37

\bibitem[{{Humphrey} {et~al}\mbox{.}(2009){Humphrey}, {Liu}, \&
  {Buote}}]{Humphrey2009}
{Humphrey} P.~J., {Liu} W., {Buote} D.~A., 2009, ApJ, 693, 822

\bibitem[{{Kleinmann} {et~al}\mbox{.}(1988){Kleinmann}, {Hamilton}, {Keel},
  {Wynn-Williams}, {Eales}, {Becklin}, \& {Kuntz}}]{Kleinmann1988}
{Kleinmann} S.~G., {Hamilton} D., {Keel} W.~C., {Wynn-Williams} C.~G., {Eales}
  S.~A., {Becklin} E.~E., {Kuntz} K.~D., 1988, ApJ, 328, 161

\bibitem[{{Knobel} {et~al}\mbox{.}(2012){Knobel}, {Lilly}, {Iovino}, {Kova{\v
  c}}, {Bschorr}, {Presotto}, {Oesch}, {Kampczyk}, {Carollo}, {Contini},
  {Kneib}, {Le Fevre}, {Mainieri}, {Renzini}, {Scodeggio}, {Zamorani},
  {Bardelli}, {Bolzonella}, {Bongiorno}, {Caputi}, {Cucciati}, {de la Torre},
  {de Ravel}, {Franzetti}, {Garilli}, {Lamareille}, {Le Borgne}, {Le Brun},
  {Maier}, {Mignoli}, {Pello}, {Peng}, {Perez Montero}, {Silverman}, {Tanaka},
  {Tasca}, {Tresse}, {Vergani}, {Zucca}, {Barnes}, {Bordoloi}, {Cappi},
  {Cimatti}, {Coppa}, {Koekemoer}, {L{\'o}pez-Sanjuan}, {McCracken}, {Moresco},
  {Nair}, {Pozzetti}, \& {Welikala}}]{Knobel2012}
{Knobel} C. {et~al.}, 2012, ApJ, 753, 121

\bibitem[{{Koester} {et~al}\mbox{.}(2007){Koester}, {McKay}, {Annis},
  {Wechsler}, {Evrard}, {Bleem}, {Becker}, {Johnston}, {Sheldon}, {Nichol},
  {Miller}, {Scranton}, {Bahcall}, {Barentine}, {Brewington}, {Brinkmann},
  {Harvanek}, {Kleinman}, {Krzesinski}, {Long}, {Nitta}, {Schneider},
  {Sneddin}, {Voges}, \& {York}}]{Koester2007}
{Koester} B.~P. {et~al.}, 2007, ApJ, 660, 239

\bibitem[{{Kotov} \& {Vikhlinin}(2006)}]{Kotov2006}
{Kotov} O., {Vikhlinin} A., 2006, ApJ, 641, 752

\bibitem[{{Ma} {et~al}\mbox{.}(2013){Ma}, {McNamara}, \& {Nulsen}}]{Ma2013}
{Ma} C.-J., {McNamara} B.~R., {Nulsen} P.~E.~J., 2013, ApJ, 763, 63

\bibitem[{{Mahdavi} {et~al}\mbox{.}(2013){Mahdavi}, {Hoekstra}, {Babul},
  {Bildfell}, {Jeltema}, \& {Henry}}]{Mahdavi2013}
{Mahdavi} A., {Hoekstra} H., {Babul} A., {Bildfell} C., {Jeltema} T., {Henry}
  J.~P., 2013, ApJ, 767, 116

\bibitem[{{Manners} {et~al}\mbox{.}(2003){Manners}, {Johnson}, {Almaini},
  {Willott}, {Gonzalez-Solares}, {Lawrence}, {Mann}, {Perez-Fournon}, {Dunlop},
  {McMahon}, {Oliver}, {Rowan-Robinson}, \& {Serjeant}}]{Manners2003}
{Manners} J.~C. {et~al.}, 2003, MNRAS, 343, 293

\bibitem[{{Maughan} {et~al}\mbox{.}(2012){Maughan}, {Giles}, {Randall},
  {Jones}, \& {Forman}}]{Maughan2012}
{Maughan} B.~J., {Giles} P.~A., {Randall} S.~W., {Jones} C., {Forman} W.~R.,
  2012, MNRAS, 2419

\bibitem[{{McCarthy} {et~al}\mbox{.}(2008){McCarthy}, {Babul}, {Bower}, \&
  {Balogh}}]{McCarthy2008}
{McCarthy} I.~G., {Babul} A., {Bower} R.~G., {Balogh} M.~L., 2008, MNRAS, 386,
  1309

\bibitem[{{McCarthy} {et~al}\mbox{.}(2004){McCarthy}, {Balogh}, {Babul},
  {Poole}, \& {Horner}}]{McCarthy2004a}
{McCarthy} I.~G., {Balogh} M.~L., {Babul} A., {Poole} G.~B., {Horner} D.~J.,
  2004, ApJ, 613, 811

\bibitem[{{McDonald} {et~al}\mbox{.}(2014){McDonald}, {Benson}, {Vikhlinin},
  {Aird}, {Allen}, {Bautz}, {Bayliss}, {Bleem}, {Bocquet}, {Brodwin},
  {Carlstrom}, {Chang}, {Cho}, {Clocchiatti}, {Crawford}, {Crites}, {de Haan},
  {Dobbs}, {Foley}, {Forman}, {George}, {Gladders}, {Gonzalez}, {Halverson},
  {Hlavacek-Larrondo}, {Holder}, {Holzapfel}, {Hrubes}, {Jones}, {Keisler},
  {Knox}, {Lee}, {Leitch}, {Liu}, {Lueker}, {Luong-Van}, {Mantz}, {Marrone},
  {McMahon}, {Meyer}, {Miller}, {Mocanu}, {Mohr}, {Murray}, {Padin}, {Pryke},
  {Reichardt}, {Rest}, {Ruhl}, {Saliwanchik}, {Saro}, {Sayre}, {Schaffer},
  {Shirokoff}, {Spieler}, {Stalder}, {Stanford}, {Staniszewski}, {Stark},
  {Story}, {Stubbs}, {Vanderlinde}, {Vieira}, {Williamson}, {Zahn}, \&
  {Zenteno}}]{McDonald2014}
{McDonald} M. {et~al.}, 2014, ApJ, 794, 67

\bibitem[{{McDonald} {et~al}\mbox{.}(2013){McDonald}, {Benson}, {Vikhlinin},
  {Stalder}, {Bleem}, {de Haan}, {Lin}, {Aird}, {Ashby}, {Bautz}, {Bayliss},
  {Bocquet}, {Brodwin}, {Carlstrom}, {Chang}, {Cho}, {Clocchiatti}, {Crawford},
  {Crites}, {Desai}, {Dobbs}, {Dudley}, {Foley}, {Forman}, {George},
  {Gettings}, {Gladders}, {Gonzalez}, {Halverson}, {High}, {Holder},
  {Holzapfel}, {Hoover}, {Hrubes}, {Jones}, {Joy}, {Keisler}, {Knox}, {Lee},
  {Leitch}, {Liu}, {Lueker}, {Luong-Van}, {Mantz}, {Marrone}, {McMahon},
  {Mehl}, {Meyer}, {Miller}, {Mocanu}, {Mohr}, {Montroy}, {Murray},
  {Nurgaliev}, {Padin}, {Plagge}, {Pryke}, {Reichardt}, {Rest}, {Ruel}, {Ruhl},
  {Saliwanchik}, {Saro}, {Sayre}, {Schaffer}, {Shirokoff}, {Song}, {{\v
  S}uhada}, {Spieler}, {Stanford}, {Staniszewski}, {Stark}, {Story}, {van
  Engelen}, {Vanderlinde}, {Vieira}, {Williamson}, {Zahn}, \&
  {Zenteno}}]{McDonald2013}
{McDonald} M. {et~al.}, 2013, ApJ, 774, 23

\bibitem[{{McNamara} \& {Nulsen}(2007)}]{McNamara2007}
{McNamara} B.~R., {Nulsen} P.~E.~J., 2007, ARA\&A, 45, 117

\bibitem[{{McNamara} \& {Nulsen}(2012)}]{McNamara2012}
{McNamara} B.~R., {Nulsen} P.~E.~J., 2012, New Journal of Physics, 14, 055023

\bibitem[{{McNamara} {et~al}\mbox{.}(2000){McNamara}, {Wise}, {Nulsen},
  {David}, {Sarazin}, {Bautz}, {Markevitch}, {Vikhlinin}, {Forman}, {Jones}, \&
  {Harris}}]{McNamara2000}
{McNamara} B.~R. {et~al.}, 2000, ApJ, 534, L135

\bibitem[{{Mehrtens} {et~al}\mbox{.}(2012){Mehrtens}, {Romer}, {Hilton},
  {Lloyd-Davies}, {Miller}, {Stanford}, {Hosmer}, {Hoyle}, {Collins}, {Liddle},
  {Viana}, {Nichol}, {Stott}, {Dubois}, {Kay}, {Sahl{\'e}n}, {Young}, {Short},
  {Christodoulou}, {Watson}, {Davidson}, {Harrison}, {Baruah}, {Smith},
  {Burke}, {Mayers}, {Deadman}, {Rooney}, {Edmondson}, {West}, {Campbell},
  {Edge}, {Mann}, {Sabirli}, {Wake}, {Benoist}, {da Costa}, {Maia}, \&
  {Ogando}}]{Mehrtens2012}
{Mehrtens} N. {et~al.}, 2012, MNRAS, 423, 1024

\bibitem[{Mittal {et~al}\mbox{.}(2009)Mittal, Hudson, Reiprich, \&
  Clarke}]{Mittal2009}
Mittal R., Hudson D.~S., Reiprich T.~H., Clarke T., 2009, A\&A, 501, 835

\bibitem[{{Mohr} {et~al}\mbox{.}(1999){Mohr}, {Mathiesen}, \&
  {Evrard}}]{Mohr1999}
{Mohr} J.~J., {Mathiesen} B., {Evrard} A.~E., 1999, ApJ, 517, 627

\bibitem[{{Morita} {et~al}\mbox{.}(2006){Morita}, {Ishisaki}, {Yamasaki},
  {Ota}, {Kawano}, {Fukazawa}, \& {Ohashi}}]{Morita2006}
{Morita} U., {Ishisaki} Y., {Yamasaki} N.~Y., {Ota} N., {Kawano} N., {Fukazawa}
  Y., {Ohashi} T., 2006, PASJ, 58, 719

\bibitem[{{Motl} {et~al}\mbox{.}(2005){Motl}, {Hallman}, {Burns}, \&
  {Norman}}]{Motl2005}
{Motl} P.~M., {Hallman} E.~J., {Burns} J.~O., {Norman} M.~L., 2005, ApJ, 623,
  L63

\bibitem[{{Mullis} {et~al}\mbox{.}(2003){Mullis}, {McNamara}, {Quintana},
  {Vikhlinin}, {Henry}, {Gioia}, {Hornstrup}, {Forman}, \&
  {Jones}}]{Mullis2003}
{Mullis} C.~R. {et~al.}, 2003, ApJ, 594, 154

\bibitem[{{Mushotzky} \& {Loewenstein}(1997)}]{Mushotzky1997}
{Mushotzky} R.~F., {Loewenstein} M., 1997, ApJ, 481, L63

\bibitem[{{Neumann} \& {Arnaud}(1999)}]{Neumann1999}
{Neumann} D.~M., {Arnaud} M., 1999, A\&A, 348, 711

\bibitem[{{O'Dea} {et~al}\mbox{.}(2008){O'Dea}, {Baum}, {Privon}, {Noel-Storr},
  {Quillen}, {Zufelt}, {Park}, {Edge}, {Russell}, {Fabian}, {Donahue},
  {Sarazin}, {McNamara}, {Bregman}, \& {Egami}}]{ODea2008}
{O'Dea} C.~P. {et~al.}, 2008, ApJ, 681, 1035

\bibitem[{{O'Hara} {et~al}\mbox{.}(2006){O'Hara}, {Mohr}, {Bialek}, \&
  {Evrard}}]{O'Hara2006}
{O'Hara} T.~B., {Mohr} J.~J., {Bialek} J.~J., {Evrard} A.~E., 2006, ApJ, 639,
  64

\bibitem[{{Osmond} \& {Ponman}(2004)}]{Osmond2004}
{Osmond} J.~P.~F., {Ponman} T.~J., 2004, MNRAS, 350, 1511

\bibitem[{{O'Sullivan} {et~al}\mbox{.}(2011{\natexlab{a}}){O'Sullivan},
  {Giacintucci}, {David}, {Gitti}, {Vrtilek}, {Raychaudhury}, \&
  {Ponman}}]{OSullivan2011a}
{O'Sullivan} E., {Giacintucci} S., {David} L.~P., {Gitti} M., {Vrtilek} J.~M.,
  {Raychaudhury} S., {Ponman} T.~J., 2011{\natexlab{a}}, ApJ, 735, 11

\bibitem[{{O'Sullivan} {et~al}\mbox{.}(2011{\natexlab{b}}){O'Sullivan},
  {Worrall}, {Birkinshaw}, {Trinchieri}, {Wolter}, {Zezas}, \&
  {Giacintucci}}]{OSullivan2011}
{O'Sullivan} E., {Worrall} D.~M., {Birkinshaw} M., {Trinchieri} G., {Wolter}
  A., {Zezas} A., {Giacintucci} S., 2011{\natexlab{b}}, MNRAS, 416, 2916

\bibitem[{{Panagoulia} {et~al}\mbox{.}(2014){Panagoulia}, {Fabian}, \&
  {Sanders}}]{Panagoulia2014}
{Panagoulia} E.~K., {Fabian} A.~C., {Sanders} J.~S., 2014, MNRAS, 438, 2341

\bibitem[{{Pell{\'o}} {et~al}\mbox{.}(2009){Pell{\'o}}, {Rudnick}, {De Lucia},
  {Simard}, {Clowe}, {Jablonka}, {Milvang-Jensen}, {Saglia}, {White},
  {Arag{\'o}n-Salamanca}, {Halliday}, {Poggianti}, {Best}, {Dalcanton},
  {Dantel-Fort}, {Fort}, {von der Linden}, {Mellier}, {Rottgering}, \&
  {Zaritsky}}]{Pello2009}
{Pell{\'o}} R. {et~al.}, 2009, A\&A, 508, 1173

\bibitem[{{Peres} {et~al}\mbox{.}(1998){Peres}, {Fabian}, {Edge}, {Allen},
  {Johnstone}, \& {White}}]{Peres1998}
{Peres} C.~B., {Fabian} A.~C., {Edge} A.~C., {Allen} S.~W., {Johnstone} R.~M.,
  {White} D.~A., 1998, MNRAS, 298, 416

\bibitem[{{Perlman} {et~al}\mbox{.}(2002){Perlman}, {Horner}, {Jones},
  {Scharf}, {Ebeling}, {Wegner}, \& {Malkan}}]{Perlman2002}
{Perlman} E.~S., {Horner} D.~J., {Jones} L.~R., {Scharf} C.~A., {Ebeling} H.,
  {Wegner} G., {Malkan} M., 2002, ApJ, 140, 265

\bibitem[{{Planck Collaboration} {et~al}\mbox{.}(2011){Planck Collaboration},
  {Ade}, {Aghanim}, {Arnaud}, {Ashdown}, {Aumont}, {Baccigalupi}, {Balbi},
  {Banday}, {Barreiro}, {Bartelmann}, {Bartlett}, {Battaner}, {Battye},
  {Benabed}, {Beno{\^i}t}, {Bernard}, {Bersanelli}, {Bhatia}, {Bock},
  {Bonaldi}, {Bond}, {Borrill}, {Bouchet}, {Brown}, {Bucher}, {Burigana},
  {Cabella}, {Cantalupo}, {Cardoso}, {Carvalho}, {Catalano}, {Cay{\'o}n},
  {Challinor}, {Chamballu}, {Chary}, {Chiang}, {Chiang}, {Chon}, {Christensen},
  {Churazov}, {Clements}, {Colafrancesco}, {Colombi}, {Couchot}, {Coulais},
  {Crill}, {Cuttaia}, {da Silva}, {Dahle}, {Danese}, {Davis}, {de Bernardis},
  {de Gasperis}, {de Rosa}, {de Zotti}, {Delabrouille}, {Delouis},
  {D{\'e}sert}, {Dickinson}, {Diego}, {Dolag}, {Dole}, {Donzelli}, {Dor{\'e}},
  {D{\"o}rl}, {Douspis}, {Dupac}, {Efstathiou}, {Eisenhardt}, {En{\ss}lin},
  {Feroz}, {Finelli}, {Flores-Cacho}, {Forni}, {Fosalba}, {Frailis},
  {Franceschi}, {Fromenteau}, {Galeotta}, {Ganga}, {G{\'e}nova-Santos},
  {Giard}, {Giardino}, {Giraud-H{\'e}raud}, {Gonz{\'a}lez-Nuevo},
  {Gonz{\'a}lez-Riestra}, {G{\'o}rski}, {Grainge}, {Gratton}, {Gregorio},
  {Gruppuso}, {Harrison}, {Hein{\"a}m{\"a}ki}, {Henrot-Versill{\'e}},
  {Hern{\'a}ndez-Monteagudo}, {Herranz}, {Hildebrandt}, {Hivon}, {Hobson},
  {Holmes}, {Hovest}, {Hoyland}, {Huffenberger}, {Hurier}, {Hurley-Walker},
  {Jaffe}, {Jones}, {Juvela}, {Keih{\"a}nen}, {Keskitalo}, {Kisner}, {Kneissl},
  {Knox}, {Kurki-Suonio}, {Lagache}, {Lamarre}, {Lasenby}, {Laureijs},
  {Lawrence}, {Le Jeune}, {Leach}, {Leonardi}, {Li}, {Liddle}, {Lilje},
  {Linden-V{\o}rnle}, {L{\'o}pez-Caniego}, {Lubin}, {Mac{\'{\i}}as-P{\'e}rez},
  {MacTavish}, {Maffei}, {Maino}, {Mandolesi}, {Mann}, {Maris}, {Marleau},
  {Mart{\'{\i}}nez-Gonz{\'a}lez}, {Masi}, {Matarrese}, {Matthai}, {Mazzotta},
  {Mei}, {Meinhold}, {Melchiorri}, {Melin}, {Mendes}, {Mennella}, {Mitra},
  {Miville-Desch{\^e}nes}, {Moneti}, {Montier}, {Morgante}, {Mortlock},
  {Munshi}, {Murphy}, {Naselsky}, {Nati}, {Natoli}, {Netterfield},
  {N{\o}rgaard-Nielsen}, {Noviello}, {Novikov}, {Novikov}, {Olamaie},
  {Osborne}, {Pajot}, {Pasian}, {Patanchon}, {Pearson}, {Perdereau}, {Perotto},
  {Perrotta}, {Piacentini}, {Piat}, {Pierpaoli}, {Piffaretti}, {Plaszczynski},
  {Pointecouteau}, {Polenta}, {Ponthieu}, {Poutanen}, {Pratt}, {Pr{\'e}zeau},
  {Prunet}, {Puget}, {Rachen}, {Reach}, {Rebolo}, {Reinecke}, {Renault},
  {Ricciardi}, {Riller}, {Ristorcelli}, {Rocha}, {Rosset},
  {Rubi{\~n}o-Mart{\'{\i}}n}, {Rusholme}, {Saar}, {Sandri}, {Santos},
  {Saunders}, {Savini}, {Schaefer}, {Scott}, {Seiffert}, {Shellard}, {Smoot},
  {Stanford}, {Starck}, {Stivoli}, {Stolyarov}, {Stompor}, {Sudiwala},
  {Sunyaev}, {Sutton}, {Sygnet}, {Taburet}, {Tauber}, {Terenzi}, {Toffolatti},
  {Tomasi}, {Torre}, {Tristram}, {Tuovinen}, {Valenziano}, {Vibert}, {Vielva},
  {Villa}, {Vittorio}, {Wade}, {Wandelt}, {Weller}, {White}, {White}, {Yvon},
  {Zacchei}, \& {Zonca}}]{PlanckCollaboration2011}
{Planck Collaboration} {et~al.}, 2011, A\&A, 536, A8

\bibitem[{{Ponman} {et~al}\mbox{.}(1999){Ponman}, {Cannon}, \&
  {Navarro}}]{Ponman1999}
{Ponman} T.~J., {Cannon} D.~B., {Navarro} J.~F., 1999, Nature, 397, 135

\bibitem[{{Poole} {et~al}\mbox{.}(2006){Poole}, {Fardal}, {Babul}, {McCarthy},
  {Quinn}, \& {Wadsley}}]{Poole2006}
{Poole} G.~B., {Fardal} M.~A., {Babul} A., {McCarthy} I.~G., {Quinn} T.,
  {Wadsley} J., 2006, MNRAS, 373, 881

\bibitem[{{Pratt} {et~al}\mbox{.}(2009){Pratt}, {Croston}, {Arnaud}, \&
  {B{\"o}hringer}}]{Pratt2009}
{Pratt} G.~W., {Croston} J.~H., {Arnaud} M., {B{\"o}hringer} H., 2009, A\&A,
  498, 361

\bibitem[{{Rafferty} {et~al}\mbox{.}(2006){Rafferty}, {McNamara}, {Nulsen}, \&
  {Wise}}]{Rafferty2006}
{Rafferty} D.~A., {McNamara} B.~R., {Nulsen} P.~E.~J., {Wise} M.~W., 2006, ApJ,
  652, 216

\bibitem[{{Randall} {et~al}\mbox{.}(2009){Randall}, {Jones}, {Markevitch},
  {Blanton}, {Nulsen}, \& {Forman}}]{Randall2009}
{Randall} S.~W., {Jones} C., {Markevitch} M., {Blanton} E.~L., {Nulsen}
  P.~E.~J., {Forman} W.~R., 2009, ApJ, 700, 1404

\bibitem[{{Rasmussen} \& {Ponman}(2007)}]{Rasmussen2007}
{Rasmussen} J., {Ponman} T.~J., 2007, MNRAS, 380, 1554

\bibitem[{{Reichardt} {et~al}\mbox{.}(2013){Reichardt}, {Stalder}, {Bleem},
  {Montroy}, {Aird}, {Andersson}, {Armstrong}, {Ashby}, {Bautz}, {Bayliss},
  {Bazin}, {Benson}, {Brodwin}, {Carlstrom}, {Chang}, {Cho}, {Clocchiatti},
  {Crawford}, {Crites}, {de Haan}, {Desai}, {Dobbs}, {Dudley}, {Foley},
  {Forman}, {George}, {Gladders}, {Gonzalez}, {Halverson}, {Harrington},
  {High}, {Holder}, {Holzapfel}, {Hoover}, {Hrubes}, {Jones}, {Joy}, {Keisler},
  {Knox}, {Lee}, {Leitch}, {Liu}, {Lueker}, {Luong-Van}, {Mantz}, {Marrone},
  {McDonald}, {McMahon}, {Mehl}, {Meyer}, {Mocanu}, {Mohr}, {Murray}, {Natoli},
  {Padin}, {Plagge}, {Pryke}, {Rest}, {Ruel}, {Ruhl}, {Saliwanchik}, {Saro},
  {Sayre}, {Schaffer}, {Shaw}, {Shirokoff}, {Song}, {Spieler}, {Staniszewski},
  {Stark}, {Story}, {Stubbs}, {{\v S}uhada}, {van Engelen}, {Vanderlinde},
  {Vieira}, {Vikhlinin}, {Williamson}, {Zahn}, \& {Zenteno}}]{Reichardt2013}
{Reichardt} C.~L. {et~al.}, 2013, ApJ, 763, 127

\bibitem[{{Romer} {et~al}\mbox{.}(2000){Romer}, {Nichol}, {Holden}, {Ulmer},
  {Pildis}, {Merrelli}, {Adami}, {Burke}, {Collins}, {Metevier}, {Kron}, \&
  {Commons}}]{Romer2000}
{Romer} A.~K. {et~al.}, 2000, ApJS, 126, 209

\bibitem[{{Rosati} {et~al}\mbox{.}(1995){Rosati}, {Della Ceca}, {Burg},
  {Norman}, \& {Giacconi}}]{Rosati1995}
{Rosati} P., {Della Ceca} R., {Burg} R., {Norman} C., {Giacconi} R., 1995, ApJ,
  445, L11

\bibitem[{{Rosati} {et~al}\mbox{.}(1999){Rosati}, {Stanford}, {Eisenhardt},
  {Elston}, {Spinrad}, {Stern}, \& {Dey}}]{Rosati1999}
{Rosati} P., {Stanford} S.~A., {Eisenhardt} P.~R., {Elston} R., {Spinrad} H.,
  {Stern} D., {Dey} A., 1999, AJ, 118, 76

\bibitem[{{Rosati} {et~al}\mbox{.}(2004){Rosati}, {Tozzi}, {Ettori},
  {Mainieri}, {Demarco}, {Stanford}, {Lidman}, {Nonino}, {Borgani}, {Della
  Ceca}, {Eisenhardt}, {Holden}, \& {Norman}}]{Rosati2004}
{Rosati} P. {et~al.}, 2004, AJ, 127, 230

\bibitem[{{Rossetti} \& {Molendi}(2010)}]{Rossetti2010}
{Rossetti} M., {Molendi} S., 2010, A\&A, 510, A83

\bibitem[{{Rumbaugh} {et~al}\mbox{.}(2013){Rumbaugh}, {Kocevski}, {Gal},
  {Lemaux}, {Lubin}, {Fassnacht}, \& {Squires}}]{Rumbaugh2013}
{Rumbaugh} N., {Kocevski} D.~D., {Gal} R.~R., {Lemaux} B.~C., {Lubin} L.~M.,
  {Fassnacht} C.~D., {Squires} G.~K., 2013, ApJ, 763, 124

\bibitem[{{Russell} {et~al}\mbox{.}(2012){Russell}, {Fabian}, {Taylor},
  {Sanders}, {Blundell}, {Crawford}, {Johnstone}, \& {Belsole}}]{Russell2012}
{Russell} H.~R., {Fabian} A.~C., {Taylor} G.~B., {Sanders} J.~S., {Blundell}
  K.~M., {Crawford} C.~S., {Johnstone} R.~M., {Belsole} E., 2012, MNRAS, 422,
  590

\bibitem[{{Russell} {et~al}\mbox{.}(2013){Russell}, {McNamara}, {Edge},
  {Hogan}, {Main}, \& {Vantyghem}}]{Russell2013}
{Russell} H.~R., {McNamara} B.~R., {Edge} A.~C., {Hogan} M.~T., {Main} R.~A.,
  {Vantyghem} A.~N., 2013, MNRAS, 432, 530

\bibitem[{{Samuele} {et~al}\mbox{.}(2011){Samuele}, {McNamara}, {Vikhlinin}, \&
  {Mullis}}]{Samuele2011}
{Samuele} R., {McNamara} B.~R., {Vikhlinin} A., {Mullis} C.~R., 2011, ApJ, 731,
  31

\bibitem[{{Sanderson} {et~al}\mbox{.}(2009){Sanderson}, {O'Sullivan}, \&
  {Ponman}}]{Sanderson2009}
{Sanderson} A.~J.~R., {O'Sullivan} E., {Ponman} T.~J., 2009, MNRAS, 395, 764

\bibitem[{{Sanderson} {et~al}\mbox{.}(2006){Sanderson}, {Ponman}, \&
  {O'Sullivan}}]{Sanderson2006}
{Sanderson} A.~J.~R., {Ponman} T.~J., {O'Sullivan} E., 2006, MNRAS, 372, 1496

\bibitem[{{Santos} {et~al}\mbox{.}(2008){Santos}, {Rosati}, {Tozzi},
  {B{\"o}hringer}, {Ettori}, \& {Bignamini}}]{Santos2008}
{Santos} J.~S., {Rosati} P., {Tozzi} P., {B{\"o}hringer} H., {Ettori} S.,
  {Bignamini} A., 2008, A\&A, 483, 35

\bibitem[{{Santos} {et~al}\mbox{.}(2010){Santos}, {Tozzi}, {Rosati}, \&
  {B{\"o}hringer}}]{Santos2010}
{Santos} J.~S., {Tozzi} P., {Rosati} P., {B{\"o}hringer} H., 2010, A\&A, 521,
  A64

\bibitem[{{Santos} {et~al}\mbox{.}(2012){Santos}, {Tozzi}, {Rosati}, {Nonino},
  \& {Giovannini}}]{Santos2012}
{Santos} J.~S., {Tozzi} P., {Rosati} P., {Nonino} M., {Giovannini} G., 2012,
  A\&A, 539, A105

\bibitem[{{Schmidt} \& {Allen}(2007)}]{Schmidt2007}
{Schmidt} R.~W., {Allen} S.~W., 2007, MNRAS, 379, 209

\bibitem[{Semler {et~al}\mbox{.}(2012)Semler, ?uhada, Aird, Ashby, Bautz,
  Bayliss, Bazin, Bocquet, Benson, Bleem, Brodwin, Carlstrom, Chang, Cho,
  Clocchiatti, Crawford, Crites, de~Haan, Desai, Dobbs, Dudley, Foley, George,
  Gladders, Gonzalez, Halverson, Harrington, High, Holder, Holzapfel, Hoover,
  Hrubes, Jones, Joy, Keisler, Knox, Lee, Leitch, Liu, Lueker, Luong-Van,
  Mantz, Marrone, McDonald, McMahon, Mehl, Meyer, Mocanu, Mohr, Montroy,
  Murray, Natoli, Padin, Plagge, Pryke, Reichardt, Rest, Ruel, Ruhl,
  Saliwanchik, Saro, Sayre, Schaffer, Shaw, Shirokoff, Song, Spieler, Stalder,
  Staniszewski, Stark, Story, Stubbs, van Engelen, Vanderlinde, Vieira,
  Vikhlinin, Williamson, Zahn, \& Zenteno}]{Semler2012}
Semler D.~R. {et~al.}, 2012, ApJ, 761, 183

\bibitem[{{Siemiginowska} {et~al}\mbox{.}(2010){Siemiginowska}, {Burke},
  {Aldcroft}, {Worrall}, {Allen}, {Bechtold}, {Clarke}, \&
  {Cheung}}]{Siemiginowska2010}
{Siemiginowska} A., {Burke} D.~J., {Aldcroft} T.~L., {Worrall} D.~M., {Allen}
  S., {Bechtold} J., {Clarke} T., {Cheung} C.~C., 2010, ApJ, 722, 102

\bibitem[{{Slack} \& {Ponman}(2014)}]{Slack2014}
{Slack} N.~W., {Ponman} T.~J., 2014, MNRAS, 439, 102

\bibitem[{{Snowden} {et~al}\mbox{.}(1998){Snowden}, {Egger}, {Finkbeiner},
  {Freyberg}, \& {Plucinsky}}]{Snowden1998}
{Snowden} S.~L., {Egger} R., {Finkbeiner} D.~P., {Freyberg} M.~J., {Plucinsky}
  P.~P., 1998, ApJ, 493, 715

\bibitem[{{Snowden} {et~al}\mbox{.}(2008){Snowden}, {Mushotzky}, {Kuntz}, \&
  {Davis}}]{Snowden2008}
{Snowden} S.~L., {Mushotzky} R.~F., {Kuntz} K.~D., {Davis} D.~S., 2008, A\&A,
  478, 615

\bibitem[{{Song} {et~al}\mbox{.}(2012){Song}, {Zenteno}, {Stalder}, {Desai},
  {Bleem}, {Aird}, {Armstrong}, {Ashby}, {Bayliss}, {Bazin}, {Benson},
  {Bertin}, {Brodwin}, {Carlstrom}, {Chang}, {Cho}, {Clocchiatti}, {Crawford},
  {Crites}, {de Haan}, {Dobbs}, {Dudley}, {Foley}, {George}, {Gettings},
  {Gladders}, {Gonzalez}, {Halverson}, {Harrington}, {High}, {Holder},
  {Holzapfel}, {Hoover}, {Hrubes}, {Joy}, {Keisler}, {Knox}, {Lee}, {Leitch},
  {Liu}, {Lueker}, {Luong-Van}, {Marrone}, {McDonald}, {McMahon}, {Mehl},
  {Meyer}, {Mocanu}, {Mohr}, {Montroy}, {Natoli}, {Nurgaliev}, {Padin},
  {Plagge}, {Pryke}, {Reichardt}, {Rest}, {Ruel}, {Ruhl}, {Saliwanchik},
  {Saro}, {Sayre}, {Schaffer}, {Shaw}, {Shirokoff}, {{\v S}uhada}, {Spieler},
  {Stanford}, {Staniszewski}, {Stark}, {Story}, {Stubbs}, {van Engelen},
  {Vanderlinde}, {Vieira}, {Williamson}, \& {Zahn}}]{Song2012}
{Song} J. {et~al.}, 2012, ApJ, 761, 22

\bibitem[{{Stanford} {et~al}\mbox{.}(1997){Stanford}, {Elston}, {Eisenhardt},
  {Spinrad}, {Stern}, \& {Dey}}]{Stanford1997}
{Stanford} S.~A., {Elston} R., {Eisenhardt} P.~R., {Spinrad} H., {Stern} D.,
  {Dey} A., 1997, AJ, 114, 2232

\bibitem[{{Stern} {et~al}\mbox{.}(2010){Stern}, {Jimenez}, {Verde}, {Stanford},
  \& {Kamionkowski}}]{Stern2010}
{Stern} D., {Jimenez} R., {Verde} L., {Stanford} S.~A., {Kamionkowski} M.,
  2010, ApJ, 188, 280

\bibitem[{{Stott} {et~al}\mbox{.}(2012){Stott}, {Hickox}, {Edge}, {Collins},
  {Hilton}, {Harrison}, {Romer}, {Rooney}, {Kay}, {Miller}, {Sahl{\'e}n},
  {Lloyd-Davies}, {Mehrtens}, {Hoyle}, {Liddle}, {Viana}, {McCarthy}, {Schaye},
  \& {Booth}}]{Stott2012}
{Stott} J.~P. {et~al.}, 2012, MNRAS, 422, 2213

\bibitem[{{Struble} \& {Rood}(1987)}]{Struble1987}
{Struble} M.~F., {Rood} H.~J., 1987, ApJ, 63, 543

\bibitem[{{Sun} {et~al}\mbox{.}(2007){Sun}, {Jones}, {Forman}, {Vikhlinin},
  {Donahue}, \& {Voit}}]{Sun2007}
{Sun} M., {Jones} C., {Forman} W., {Vikhlinin} A., {Donahue} M., {Voit} M.,
  2007, ApJ, 657, 197

\bibitem[{{Sun} {et~al}\mbox{.}(2009){Sun}, {Voit}, {Donahue}, {Jones},
  {Forman}, \& {Vikhlinin}}]{Sun2009}
{Sun} M., {Voit} G.~M., {Donahue} M., {Jones} C., {Forman} W., {Vikhlinin} A.,
  2009, ApJ, 693, 1142

\bibitem[{{Sutherland} \& {Dopita}(1993)}]{Sutherland1993}
{Sutherland} R.~S., {Dopita} M.~A., 1993, ApJ, 88, 253

\bibitem[{{Szokoly} {et~al}\mbox{.}(2004){Szokoly}, {Bergeron}, {Hasinger},
  {Lehmann}, {Kewley}, {Mainieri}, {Nonino}, {Rosati}, {Giacconi}, {Gilli},
  {Gilmozzi}, {Norman}, {Romaniello}, {Schreier}, {Tozzi}, {Wang}, {Zheng}, \&
  {Zirm}}]{Szokoly2004}
{Szokoly} G.~P. {et~al.}, 2004, ApJ, 155, 271

\bibitem[{{Takey} {et~al}\mbox{.}(2013){Takey}, {Schwope}, \&
  {Lamer}}]{Takey2013}
{Takey} A., {Schwope} A., {Lamer} G., 2013, A\&A, 558, A75

\bibitem[{{Tanaka} {et~al}\mbox{.}(2008){Tanaka}, {Finoguenov}, {Kodama},
  {Morokuma}, {Rosati}, {Stanford}, {Eisenhardt}, {Holden}, \&
  {Mei}}]{Tanaka2008}
{Tanaka} M. {et~al.}, 2008, A\&A, 489, 571

\bibitem[{{Tran} {et~al}\mbox{.}(2009){Tran}, {Saintonge}, {Moustakas}, {Bai},
  {Gonzalez}, {Holden}, {Zaritsky}, \& {Kautsch}}]{Tran2009a}
{Tran} K.-V.~H., {Saintonge} A., {Moustakas} J., {Bai} L., {Gonzalez} A.~H.,
  {Holden} B.~P., {Zaritsky} D., {Kautsch} S.~J., 2009, ApJ, 705, 809

\bibitem[{{Tran} {et~al}\mbox{.}(2005){Tran}, {van Dokkum}, {Illingworth},
  {Kelson}, {Gonzalez}, \& {Franx}}]{Tran2005a}
{Tran} K.-V.~H., {van Dokkum} P., {Illingworth} G.~D., {Kelson} D., {Gonzalez}
  A., {Franx} M., 2005, ApJ, 619, 134

\bibitem[{{Vikhlinin} {et~al}\mbox{.}(2007){Vikhlinin}, {Burenin}, {Forman},
  {Jones}, {Hornstrup}, {Murray}, \& {Quintana}}]{Vikhlinin2007}
{Vikhlinin} A., {Burenin} R., {Forman} W.~R., {Jones} C., {Hornstrup} A.,
  {Murray} S.~S., {Quintana} H., 2007, in Heating versus Cooling in Galaxies
  and Clusters of Galaxies, {B{\"o}hringer} H., {Pratt} G.~W., {Finoguenov} A.,
  {Schuecker} P., eds., p.~48

\bibitem[{{Vikhlinin} {et~al}\mbox{.}(2006){Vikhlinin}, {Kravtsov}, {Forman},
  {Jones}, {Markevitch}, {Murray}, \& {Van Speybroeck}}]{Vikhlinin2006}
{Vikhlinin} A., {Kravtsov} A., {Forman} W., {Jones} C., {Markevitch} M.,
  {Murray} S.~S., {Van Speybroeck} L., 2006, ApJ, 640, 691

\bibitem[{{Vikhlinin} {et~al}\mbox{.}(2001){Vikhlinin}, {Markevitch}, {Forman},
  \& {Jones}}]{Vikhlinin2001}
{Vikhlinin} A., {Markevitch} M., {Forman} W., {Jones} C., 2001, ApJ, 555, L87

\bibitem[{{Vikhlinin} {et~al}\mbox{.}(1998){Vikhlinin}, {McNamara}, {Forman},
  {Jones}, {Quintana}, \& {Hornstrup}}]{Vikhlinin1998a}
{Vikhlinin} A., {McNamara} B.~R., {Forman} W., {Jones} C., {Quintana} H.,
  {Hornstrup} A., 1998, ApJ, 502, 558

\bibitem[{{Voit}(2005)}]{Voit2005}
{Voit} G.~M., 2005, Reviews of Modern Physics, 77, 207

\bibitem[{{Voit}(2011)}]{Voit2011b}
{Voit} G.~M., 2011, ApJ, 740, 28

\bibitem[{{Voit} {et~al}\mbox{.}(2014){Voit}, {Donahue}, {Bryan}, \&
  {McDonald}}]{Voit2014}
{Voit} G.~M., {Donahue} M., {Bryan} G.~L., {McDonald} M., 2014, ArXiv e-prints

\bibitem[{{Wen} \& {Han}(2011)}]{Wen2011}
{Wen} Z.~L., {Han} J.~L., 2011, ApJ, 734, 68

\bibitem[{{Wen} {et~al}\mbox{.}(2012){Wen}, {Han}, \& {Liu}}]{Wen2012}
{Wen} Z.~L., {Han} J.~L., {Liu} F.~S., 2012, ApJ, 199, 34

\bibitem[{{Yang} {et~al}\mbox{.}(2004){Yang}, {Mushotzky}, {Steffen}, {Barger},
  \& {Cowie}}]{Yang2004}
{Yang} Y., {Mushotzky} R.~F., {Steffen} A.~T., {Barger} A.~J., {Cowie} L.~L.,
  2004, AJ, 128, 1501

\end{thebibliography}

\end{document}